\documentclass[12pt]{article}

\usepackage{jheppub}

\usepackage{pgfplots}
\usepackage{pgfplotstable}
\usepackage{pgfkeys}
\allowdisplaybreaks
\usepackage{ytableau}
\usepackage{comment}
\usepackage[vcentermath]{youngtab}
\usepackage{bm}
\usepackage{ulem}

\newcommand\be{\begin{equation}}
\newcommand\ee{\end{equation}}

\usepackage{braket}
\usepackage{subfigure}

\title{Kac-Moody algebras from M5-giants}
\abstract{
We examine the giant graviton expansions of the Higgs indices for the 3d $\mathcal{N}=4$ $U(N)$ ADHM theories with $l$ fundamental hypermultiplets. 
The indices for the M5-brane giant gravitons of wrapping number $m$ appearing in the expansions 
consist of the contributions that generalize the characters of the W-algebra $\mathcal{W}(\mathfrak{gl}(m))$ 
and those which realize the characters of the affine Kac-Moody algebras of type $A_{l-1}$. 
Also we confirm that the inverse giant graviton expansions of the resulting M5-brane indices consistently reproduce the Higgs indices. 
}

\author[a]{Hirotaka Hayashi,}
\emailAdd{h.hayashi@tokai.ac.jp}
\affiliation[a]{Department of Physics, School of Science, Tokai University,\\
4-1-1 Kitakaname, Hiratsuka-shi, Kanagawa 259-1292, Japan}

\author[b,c]{Tomoki Nosaka}
\emailAdd{nosaka@simis.cn}
\affiliation[b]{
Center for Mathematics and Interdisciplinary Sciences, Fudan University, Shanghai 200433, China
}
\affiliation[c]{
Shanghai Institute for Mathematics and Interdisciplinary Sciences,\\
Block A, International Innovation Plaza, No.~657 Songhu Road, Yangpu District, Shanghai, China
}
\author[d]{and Tadashi Okazaki}
\emailAdd{tokazaki@seu.edu.cn}
\affiliation[d]{
Shing-Tung Yau Center of Southeast University,\\
Yifu Architecture Building, No.2 Sipailou, Xuanwu district, Nanjing, Jiangsu, 210096, China}

\begin{document}
\maketitle

\ytableausetup{boxsize=1.5mm}

\section{Introduction and summary}

In this paper we study the so called $U(N)$ ADHM theory with $l$ flavors, which is the ${\cal N}=4$ $U(N)$ supersymmetric Yang-Mills theory 
coupled with an adjoint hypermultiplet and $l$ fundamental hypermultiplets \cite{deBoer:1996mp,deBoer:1996ck}. 
The theory describes a stack of $N$ M2-branes probing an $l$-centered Taub-NUT space 
which is asymptotically $\mathbb{C}^2/\mathbb{Z}_l$ space with the $A_{l-1}$ singularity 
so that the space transverse to the M2-branes is $\mathbb{C}^4/\mathbb{Z}_l$ $=$ $\mathbb{C}^2\times (\mathbb{C}^2/\mathbb{Z}_l)$ \cite{Benini:2009qs,Bashkirov:2010kz}. 
In the near horizon limit for large $N$ and finite $l$, the theory is holographically dual to M-theory on $AdS_4\times S^7/\mathbb{Z}_l$, 
where the $S^7$ is embedded into the $\mathbb{C}^4$ 
\begin{align}
|z_1|^2+|z_2|^2+|z_3|^2+|z_4|^2=1
\end{align}
and the $\mathbb{Z}_l$ acts on the four complex coordinates on the $\mathbb{C}^4$ as
\begin{align}
\label{orbifold_Z_l}
(z_1,z_2,z_3,z_4)&\mapsto 
(z_1,e^{\frac{2\pi i}{l}}z_2,z_3,e^{\frac{2\pi i}{l}}z_4). 
\end{align}

The spectrum of the BPS local operators in the large $N$ limit of the effective theories describing a stack of $N$ coincident branes in string/M-theory is related to
that of the Kaluza-Klein (KK) modes on the holographically dual supergravity backgrounds. 
For the BPS local operators with relatively large conformal dimension, 
the dual BPS states in the supergravity side are described by the giant gravitons which behave as extended branes \cite{McGreevy:2000cw,Myers:1999ps,Grisaru:2000zn,Hashimoto:2000zp}. 
In the near horizon geometry $AdS_4\times S^7/\mathbb{Z}_l$, 
there exist giant gravitons as M5-branes wrapping a supersymmetric $5$-cycle in $S^7/\mathbb{Z}_l$ 
as the intersection of holomorphic surfaces in $\mathbb{C}^4/\mathbb{Z}_l$ with the $S^7/\mathbb{Z}_l$ \cite{Mikhailov:2000ya}. 
The fact that the $\mathbb{C}^4$ background factorizes into $\mathbb{C}^2$ and $\mathbb{C}^2/\mathbb{Z}_l$ 
leads to the distinguished BPS spectrum of the M5-branes. 

Accordingly, the large $N$ limits of the supersymmetric indices of the effective theories are related to the gravity indices for the KK modes in the near horizon geometry. On the other hand, the finite $N$ indices should contain the contributions from the giant gravitons whose angular momenta are of order $N$. 
The giant graviton expansions are manipulated by expanding the ratios of the finite $N$ supersymmetric indices to the large $N$ indices 
with respect to the fugacities coupled to the angular momenta carried by the giant gravitons \cite{Arai:2019xmp,Arai:2020qaj,Gaiotto:2021xce}. 
Remarkably, the emerging expansion coefficients can be viewed as the indices for the theories on the giant gravitons 
so that one can also derive the indices for the associated world-volume theories upon an appropriate change of variables through the giant graviton expansions. 

In \cite{Hayashi:2024aaf} we analyzed the giant graviton expansions of the Coulomb and Higgs indices for the M2-brane superconformal field theories (SCFTs).\footnote{
See \cite{Gaiotto:2021xce,Arai:2020uwd,Beccaria:2023cuo} for the relevant analyses of the giant graviton expansions of the supersymmetric indices for the M2-brane SCFTs.
} 
When $l=1$, the ADHM theory has the equivalent Higgs and Coulomb branches 
and the theory is IR equivalent to the $U(N)_1\times U(N)_{-1}$ ABJM theory \cite{Aharony:2008ug} with opposite Chern-Simons level $+1$ and $-1$, 
both of which describe a stack of $N$ M2-branes probing flat space. 
For $l>1$ the ADHM theory has the Higgs branch of supersymmetric vacua as the moduli space of $N$ $SU(l)$ instantons,  
which is distinguished from the Coulomb branch, the $N$-th symmetric product of the A-type singularity. 
We found that the unflavored vacuum character of the affine Kac-Moody algebra $\widehat{\mathfrak{su}}(l)_1$ at level $1$ 
emerges from the index of a single M5-brane associated with the leading term in the giant graviton expansion of the Higgs index of the $U(N)$ ADHM theory with $l$ flavors. 
The aim of this paper is to promote the analysis to higher order terms in the giant graviton expansion to obtain the flavored indices for multiple $m$ M5-branes. 
We find that the indices of a stack of $m$ M5-branes are subdivided to two kinds of contributions. 
One is the $1/4$-BPS index which can be obtained by taking the twisted limit \cite{Hayashi:2024aaf} of the supersymmetric index of a stack of $m$ M5-branes. 
It reduces to the unrefined index \cite{Kim:2012qf,Beem:2014kka}, that is the vacuum character of the W-algebra $\mathcal{W}(\mathfrak{gl}(m))$ in the special fugacity limit.
The other involves the affine Kac-Moody algebraic structure that 
will describe the spectrum of the $m$ M5-branes on the $l$-centered Taub-NUT space in M-theory 
or the I-brane configuration consisting of $m$ D4-branes intersecting with $l$ D6-branes in Type IIA string theory 
\cite{MR1302318,MR1441880,Vafa:1994rv,Itzhaki:2005tu,Dijkgraaf:2007sw,Dijkgraaf:2007fe,Tan:2008wp,Witten:2009at} (see also \cite{Ohlsson:2012yn,Lambert:2018mfb,Gustavsson:2022jpo}). 
We find that the indices admit two interesting special fugacity limits that shed light on the structure of the current algebra. 
One leads to a product of the $m$ vacuum characters of $\widehat{\mathfrak{su}}(l)_1$, 
which can be viewed as the unflavored limit of the indices contributed from the spectrum of the excitations on the I-brane. 
Another limit gives rise to the vacuum characters of $\widehat{\mathfrak{su}}(l)_m$ at level $m$, 
which can be found from the full M5-brane indices by keeping the terms with fixed combination of the fugacities. 
The emergence of the affine Kac-Moody algebra $\widehat{\mathfrak{su}}(l)_m$ is quite similar to the proposal \cite{Belavin:2011pp,Belavin:2012aa,Nishioka:2011jk} 
(also see \cite{Bonelli:2011jx,Belavin:2011tb,Bonelli:2011kv,Wyllard:2011mn,Ito:2011mw,Alfimov:2011ju,Foda:2019msm,Manabe:2020etw})
that $m$ M5-branes on $\mathbb{C}^2/\mathbb{Z}_l$ with $\Omega$-deformation generates the current algebra. 
Moreover, we confirm that the giant graviton expansion of the resulting M5-brane indices can be viewed as the inverse relation 
in such a way that the expansion coefficients are given by the M2-brane giant graviton indices transformed from the Higgs indices upon a change of variables. 
This strongly indicates that the Higgs indices are dual to the resulting M5-brane indices. 

\subsection{Future works}

\begin{itemize}

\item In this work, we have found the special fugacity limits of the indices for M5-branes 
appearing in the giant graviton expansions coincide with the characters of the affine Kac-Moody algebras, 
but further mathematical verification of our results and exact closed-form expressions for the indices of M5-branes are still highly desirable. 

\item One obvious extension is to survey the giant graviton expansions of the indices for the other M2-brane SCFTs, 
including the theories with non-unitary gauge groups and the quiver Chern-Simons theories \cite{Imamura:2008nn,Imamura:2008dt}. 

\item It would be interesting to explore the giant graviton expansions of the indices in the presence of the defect operators. 
The expansions of such defect indices have been examined for 4d $\mathcal{N}=4$ SYM theory in \cite{Imamura:2024lkw,Beccaria:2024oif,Imamura:2024pgp,Beccaria:2024dxi,Beccaria:2024lbt,Hatsuda:2024uwt,Imamura:2024zvw}. 
On the other hand, several exact results for the M2-brane SCFTs are presented in \cite{Crew:2020psc,Hayashi:2024jof,Hayashi:2025guk} for the line defect indices 
and in \cite{Okazaki:2025rjl} for the boundary conditions. 
We hope to elaborate on the expansions of the defect indices in future work. 

\item The Higgs index of $U(N)$ ADHM theory with $l$ flavors coincide, up to an overall factor, with the instanton partition function of 5d ${\cal N}=1$ $SU(l)$ pure Yang-Mills theory with the instanton number $N$ (see e.g.~\cite{Hayashi:2024jof}).
These 5d partition functions are known to satisfy non-linear recursion relations with respect to the instanton number \cite{Kim:2019uqw}, which arises from the blowup equation \cite{MR2183121}.
It would be interesting to provide analytic derivation of our proposals for the giant graviton indices through the recursion relations.
\end{itemize}

\subsection{Structure}
The organization of the paper is straightforward. 
In section \ref{sec_review} we make a brief review of the supersymmetric indices for the $U(N)$ ADHM theory with $l$ flavors and the giant graviton expansions. 
In section \ref{app:howtodeterminehatcalFm} we explain the methods and working hypotheses to determine the explicit expansion coefficients in the giant graviton indices.
In section \ref{sec_l=2} we analyze the single sum giant graviton expansion of the Higgs index for the $U(N)$ ADHM theory with $l=2$ flavors. 
In section \ref{sec_general_l} we examine the single sum giant graviton expansions of the Higgs indices for the theories with higher $l$. 
We propose the conjectured relations of the indices for a stack of $m$ M5-branes on the $l$-centered Taub-NUT space 
resulting from the giant graviton expansions and the characters of the affine Kac-Moody algebras.
The explicit expressions for the giant graviton indices for various flavors $l$ and wrapping numbers $m$ are listed in Appendix \ref{app:listofFlm}, which are written in terms of the function $F_m^{(l)}(y_\alpha;x_1;x_2)$ defined by \eqref{Fml}.

\section{ADHM theory with $l$ flavors}
\label{sec_review}
In this section we briefly review the supersymmetric indices for the $U(N)$ ADHM theory.\footnote{See \cite{Okazaki:2019ony,Hayashi:2022ldo} for the details including the definition and convention. }
The $U(N)$ ADHM theory with $l$ flavors is a 3d $\mathcal{N}=4$ supersymmetric gauge theory 
of gauge group $U(N)$, an adjoint hypermultiplet and $l$ fundamental hypermultiplets \cite{deBoer:1996mp,deBoer:1996ck}. 
The supersymmetric index of $U(N)$ ADHM theory is given explicitly by the supersymmetric localization formula \cite{Kim:2009wb,Imamura:2011su} as
\begin{align}
&I^{U(N)\text{ADHM-}[l]}(t,x,y_\alpha,z;q)\nonumber \\
&=\frac{1}{N!}
\frac{(q^{\frac{1}{2}}t^2;q)_\infty^N}{(q^{\frac{1}{2}}t^{-2};q)_\infty^N}\sum_{m_1,\cdots,m_N\in\mathbb{Z}}\oint\prod_{i=1}^N\frac{ds_i}{2\pi is_i}\nonumber \\
&\quad\times \prod_{i<j} \Bigl(1-q^{\frac{|m_i-m_j|}{2}}\Bigl(\frac{s_i}{s_j}\Bigr)^{\pm 1}\Bigr)
\frac{(q^{\frac{1}{2}+\frac{|m_i-m_j|}{2}}t^2(\frac{s_i}{s_j})^{\mp 1};q)_\infty
}{
(q^{\frac{1}{2}+\frac{|m_i-m_j|}{2}}t^{-2}(\frac{s_i}{s_j})^{\pm 1};q)_\infty
}\nonumber \\
&\quad\times \frac{
(q^{\frac{3}{4}}t^{-1}x^{\mp 1};q)_\infty^N
}{
(q^{\frac{1}{4}}tx^{\pm 1};q)_\infty^N
}
\prod_{i<j}
\frac{
(q^{\frac{3}{4}+\frac{|m_i-m_j|}{2}}t^{-1}(x\frac{s_i}{s_j})^{\mp 1};q)_\infty
}{
(q^{\frac{1}{4}+\frac{|m_i-m_j|}{2}}t(x\frac{s_i}{s_j})^{\pm 1};q)_\infty
}
\frac{
(q^{\frac{3}{4}+\frac{|m_i-m_j|}{2}}t^{-1}(x^{-1}\frac{s_i}{s_j})^{\mp 1};q)_\infty
}{
(q^{\frac{1}{4}+\frac{|m_i-m_j|}{2}}t(x^{-1}\frac{s_i}{s_j})^{\pm 1};q)_\infty
}\nonumber \\
&\quad\times \prod_{i=1}^N\prod_{\alpha=1}^l\frac{
(q^{\frac{3}{4}+\frac{|m_i|}{2}}t^{-1}(y_\alpha s_i)^{\mp 1};q)_\infty
}{
(q^{\frac{1}{4}+\frac{|m_i|}{2}}t(y_\alpha s_i)^{\pm 1};q)_\infty
}
q^{\frac{l}{4}\sum_{i=1}^N|m_i|}t^{-l\sum_{i=1}^N|m_i|}z^{l\sum_{i=1}^Nm_i}, 
\label{fullSCI}
\end{align}
where $y_{\alpha}$ are the fugacities for the $SU(l)$ flavor symmetry with $\prod_{\alpha}y_{\alpha}=1$.
Here and hereafter we use the symbol $\pm$ appearing in the powers of variables as an abbreviation of repeated factors with $+$ and $-$. For example, $(x^{\pm};q)_{\infty}$ $=$ $(x;q)_{\infty}(x^{-1};q)_{\infty}$.

For $l=1$ the ADHM theory is a self-mirror theory with enhanced $\mathcal{N}=8$ supersymmetry. 
For $l>1$ the theory has the Higgs branch as $N$ $SU(l)$ instanton moduli space, which is distinguished from the Coulomb branch. 

\subsection{Higgs indices}
The supersymmetric indices for 3d $\mathcal{N}=4$ supersymmetric field theories admit two special fugacity limits 
by keeping the ratio of two variables $q$ and $t$ and sending $q$ to zero, 
the Coulomb and Higgs limits \cite{Razamat:2014pta}, 
in which they enumerate the Coulomb and Higgs branch local operators as the Coulomb and Higgs indices, \textit{a.k.a.} refined Hilbert series \cite{Benvenuti:2010pq,Hanany:2012dm,Cremonesi:2013lqa}
for the Coulomb and Higgs branches respectively. 
In the Higgs limit
\begin{align}
q,t^{-1}\rightarrow 0\quad \text{with}\quad \mathfrak{t}=q^{\frac{1}{4}}t\quad \text{fixed},
\end{align}
the supersymmetric index \eqref{fullSCI} reduces to the Higgs index
\begin{align}
&{\cal I}^{U(N)\text{ADHM-}[l]}_H(x,y_\alpha;\mathfrak{t})\nonumber \\
&=
\frac{1}{N!}
\frac{(1-\mathfrak{t}^2)^N}{(1-x^{\pm 1}\mathfrak{t})^N}
\oint\prod_{i=1}^N\frac{ds_i}{2\pi is_i}
\frac{
\prod_{i<j}
(1-(\frac{s_i}{s_j})^{\pm 1})(1-\mathfrak{t}^2(\frac{s_i}{s_j})^{\pm 1})
}{
\prod_{i<j} (1-\mathfrak{t}x(\frac{s_i}{s_j})^{\pm 1}) (1-\mathfrak{t}x^{-1}(\frac{s_i}{s_j})^{\pm 1})
}\nonumber \\
&\quad\times \prod_{i=1}^N\prod_{\alpha=1}^l\frac{1}{1-\mathfrak{t}(y_\alpha s_i)^{\pm 1}}.
\label{IH}
\end{align}
In the large $N$ limit, the Higgs index was found to be given as \cite{Crew:2020psc}
\begin{align}
{\cal I}^{U(\infty)\text{ADHM-[}l]}_H(x,y_\alpha;\mathfrak{t})=
\prod_{a=1}^\infty
\frac{1}{1-(x^{\pm 1}\mathfrak{t})^a}
\prod_{\alpha,\beta=1}^l
\prod_{a,b=1}^\infty 
\frac{1}{1-(x^{-1}\mathfrak{t})^a(x\mathfrak{t})^b\frac{y_\alpha}{y_\beta}}.
\label{IADHMNflinfty}
\end{align}
On the other hand, the closed-form expressions for the Higgs index for finite $N$ can also be obtained 
by several methods, including the Hall-Littlewood expansions and the {Jeffrey-Kirwan (JK) residue formula (see e.g.~\cite{Hayashi:2024jof}). 

\subsection{Giant graviton expansions}
It follows that 
the coefficients of the small $\mathfrak{t}$ expansion of the Higgs index for finite $N$ 
deviate from those of the large $N$ index \eqref{IADHMNflinfty} at ${\cal O}(\mathfrak{t}^{N+1})$. 
In general we observe that the finite $N$ corrections have the following structure
\begin{align}
\frac{
{\cal I}^{U(N)\text{ADHM-[}l]}_H(x,y_\alpha;\mathfrak{t})
}{
{\cal I}^{U(\infty)\text{ADHM-[}l]}_H(x,y_\alpha;\mathfrak{t})
}
=
1+\sum_{\substack{m_1,m_2\ge 0\\ (m_1,m_2)\neq (0,0)}}
(x^{-1}\mathfrak{t})^{m_1 N}
(x\mathfrak{t})^{m_2 N}
\hat{{\cal F}}^{(l)}_{m_1,m_2}.
\label{doublesum}
\end{align}
The expansion describes the finite $N$ corrections to the Higgs index when $|\mathfrak{t}|$ is small and $x,y_\alpha$ are kept of ${\cal O}(\mathfrak{t}^0)$.
Such expansions of the ratios of the finite $N$ supersymmetric indices to the large $N$ indices are called the \textit{giant graviton expansions}  \cite{Arai:2019xmp,Arai:2020qaj,Gaiotto:2021xce}. 
Remarkably, the expansion coefficients in the giant graviton expansions can be viewed as the index for the effective theory on the giant gravitons. 

We can also consider a different expansion where finite $N$ corrections take simpler form.
For this purpose we define new variables $x_1,x_2$ as
\begin{align}
x_1=x^{-1}\mathfrak{t},\quad
x_2=x\mathfrak{t}.
\end{align}
Then, we expand the left-hand side of \eqref{doublesum} to an arbitrary but finite order in $x_2$ first, and then study the finite $N$ corrections to this truncated series in terms of the parameter $|x_1|<1$.
In this way, we find that the finite $N$ corrections are given by the same series but only the terms with $(m_1,m_2)=(m,0)$ being kept.
Namely,
\begin{align}
\frac{
{\cal I}^{U(N)\text{ADHM-[}l]}_H(x,y_\alpha;\mathfrak{t})
}{
{\cal I}^{U(\infty)\text{ADHM-[}l]}_H(x,y_\alpha;\mathfrak{t})
}
\biggr|_{\text{single sum}}
=
1+\sum_{m=1}^\infty x_1^{mN} \hat{{\cal F}}_m^{(l)}(y_{\alpha};x_1;x_2).
\label{singlesum}
\end{align}
with
\begin{align}
\hat{{\cal F}}^{(l)}_m(y_\alpha;x_1;x_2)=\hat{{\cal F}}^{(l)}_{m,0}.
\end{align}
We shall call this expansion \textit{single sum expansion}.
In the remaining part of this paper we always consider the single sum expansion and hence omit the subscript ``$|_{\text{single sum}}$''.

Some preliminary analyses of the single sum expansions of the Higgs indices of the $U(N)$ ADHM theory were made for $l=1$ in \cite{Gaiotto:2021xce} and for $l>1$ in \cite{Hayashi:2024aaf}.
In particular, it was found that the first coefficient $\hat{{\cal F}}_1^{(l)}(y_\alpha;x_1;x_2)$ is given for $y_\alpha=1$ as \cite{Hayashi:2024aaf}
\begin{align}
\hat{{\cal F}}^{(l)}_1({y_{\alpha}=1;}x_1;x_2)=\frac{
\chi_{\widehat{\mathfrak{su}}(l)_{1}}(x_2)
}{(x_1^{-1};x_2)_\infty},
\label{hatcalFm1y1}
\end{align}
where
\begin{align}
\chi_{\widehat{\mathfrak{su}}(l)_{1}}(q)=\frac{1}{(q;q)_\infty^{l-1}}\sum_{\substack{m_1,\cdots,m_l\in\mathbb{Z}\\ (m_1+\cdots+m_l=0)}}q^{\frac{1}{2}\sum_{i=1}^lm_i^2}
\label{chisul1q}
\end{align}
is the unflavored vacuum character of the 
affine Kac-Moody algebra $\widehat{\mathfrak{su}}(2)_1$ at level $1$.
For the first few expansion coefficients of $\chi_{\widehat{\mathfrak{su}}(l)_1}(q)$, see e.g.~Appendix B in \cite{Hayashi:2024aaf}. 
Furthermore, for $l=1$ one can derive the double sum expansion \eqref{doublesum} together with the explicit expressions of all the coefficients $\hat{{\cal F}}_{m_1,m_2}^{(l)}$ analytically by using the equivalence of the Higgs index to the Coulomb index \cite{Gaiotto:2021xce,Hayashi:2024aaf}.
In particular, the coefficients $\hat{{\cal F}}_m^{(l=1)}(x_1;x_2)=\hat{{\cal F}}_{m,0}^{(l=1)}$ which appear in the single sum expansion are given as
\begin{align}
\hat{{\cal F}}^{(l=1)}_m(x_1;x_2)=\prod_{k=1}^m\frac{1}{(x_1^{-k};x_2)_\infty}.
\end{align}

We also note that when we set $x_2=0$, the Higgs index \eqref{IH} for any $l$ reduces to the half BPS index \cite{Bhattacharyya:2007sa}
\begin{align}
{\cal I}^{U(N)\text{ADHM-}[l]}_H(y_\alpha;x_1;0)
={\cal I}^{U(N)\text{ADHM-}[l]}_{\frac{1}{2}\text{BPS},H}(x_1)
=\prod_{n=1}^N\frac{1}{1-x_1^n},
\end{align}
for which the coefficients of the single sum expansion are given explicitly as \cite{Gaiotto:2021xce}
\begin{align}
\hat{{\cal F}}_m^{(l)}(y_\alpha;x_1;0)=\prod_{n=1}^m\frac{1}{1-x_1^{-n}}.
\end{align}
These half BPS indices turn out to be the universal overall factors and play a crucial role in the determination of the $x_2$ dependence of the giant graviton indices.

\section{Determination of expansion coefficients}
\label{app:howtodeterminehatcalFm}

In this section we explain how we can determine the coefficients $\hat{{\cal F}}_m^{(l)}(y_\alpha;x_1;x_2)$ of the single sum expansion \eqref{singlesum}.
Readers who are interested in our main conjectures rather than the supporting evidence of the conjectures may skip this section for their first reading.

Note that in the single sum expansion we can study the coefficients of $x_2^j$ in the finite $N$ corrections separately for different $j$'s.
The expansion \eqref{singlesum} suggests that when $N$ is sufficiently large, the finite $N$ correction is given only by the first term $x_1^N\hat{\cal F}_1^{(l)}(y_\alpha;x_1;x_2)$.
Hence we can determine $\hat{{\cal F}}_1^{(l)}(y_\alpha;x_1;x_2)$ by expanding the left-hand side of \eqref{singlesum} with sufficiently large $N$ so that all the expansion coefficients in $x_1$ saturate.
Once we determine the coefficient of $x_2^j$ in $\hat{\cal F}_1^{(l)}(y_\alpha;x_1;x_2)$ in this way, we can determine the coefficient of $x_2^j$ in $\hat{{\cal F}}_2^{(l)}(y_\alpha;x_1;x_2)$ by the same analysis after subtracting $x_1^N\hat{{\cal F}}_1^{(l)}(y_\alpha;x_1;x_2)$ from the both sides of \eqref{singlesum}.
By repeating this analysis we can determine the coefficient of $x_2^j$ in $\hat{{\cal F}}_m^{(l)}(y_\alpha;x_1;x_2)$ sequentially in $m$.

After several trials with sufficiently large $N$ ($N\le 14$), we observe that the expansion coefficients have the following structure
\begin{align}
\hat{{\cal F}}^{(l)}_m(y_\alpha;x_1;x_2)=\biggl(\prod_{n=1}^m\frac{1}{1-x_1^{-n}}\biggr)\biggl(1+\sum_{j=1}^\infty x_1^{-mj}f^{(l)}_{m,j}(y_\alpha;x_1)x_2^j\biggr),
\label{Fhatansatz}
\end{align}
where $f^{(l)}_{m,j}(y_\alpha;x_1)$ are some $(mj)$-th order polynomials in $x_1$
\begin{align}
f^{(l)}_{m,j}(y_\alpha;x_1)=1+\sum_{a=1}^{mj}f_{m,j;a}^{(l)}(y_\alpha)x_1^a,
\end{align}
with $f_{m,j;a}^{(l)}(y_\alpha)$ being some finite Laurent polynomial in $y_\alpha$ with non-negative coefficients.
We further observe that the coefficients $f^{(l)}_{m,j;a}(y_\alpha)$ of the polynomial $f_{m,j}^{(l)}(y_\alpha;x_1)$ show the following saturating behavior
\begin{subequations}
\label{fmjsaturation}
\begin{align}
&f_{m,j;a}^{(l)}(y_\alpha)=f_{m-1,j;a}^{(l)}(y_\alpha),\quad (m\ge 3,a=1,\cdots, m+j-3),\\
&f_{2,j;a}^{(l)}(y_\alpha)=f_{1,j;a}^{(l)}(y_\alpha),\quad \Bigl(a=1,\cdots, \Bigl\lfloor\frac{j}{2}\Bigr\rfloor\Bigr),\\
&f_{m,j;mj-a}^{(l)}(y_\alpha)=f_{m-1,j;(m-1)j-a}^{(l)}(y_\alpha),\quad (a=0,1,\cdots,m-j-1),\\
&f_{m,j;a}^{(l)}(y_\alpha)=f_{m,j-1;a}^{(l)}(y_\alpha),\quad (a=1,2,\cdots,j-1).
\end{align}
\end{subequations}
The saturating property \eqref{fmjsaturation} enables us to determine the coefficient polynomials $f^{(l)}_{m,j}(y_\alpha;x_1)$ for higher order $j$ in $x_2$ and wrapping number $m$ only from relatively a few data of ${\cal I}^{U(N)\text{ADHM-}[l]}_H(x,y_\alpha;\mathfrak{t})$.

Let us show some examples for how the above strategy works: the determination of $f_{m,j}^{(l=2)}(x_1)$ with $y_\alpha=1$ from the exact expressions of ${\cal I}^{U(N)\text{ADHM-}[2]}_H(x,1;\mathfrak{t})$ with $N\le 1,2,3$, which are given as
{\fontsize{9pt}{1pt}\selectfont
\begin{subequations}
\begin{align}
&{\cal I}^{U(1)\text{ADHM-}[2]}_H(x,1;\mathfrak{t})= \frac{1 + x_1 x_2}{(1 - x_1) (1 - x_2) (1 - x_1 x_2)^2},\\
&{\cal I}^{U(2)\text{ADHM-}[2]}_H(x,1;\mathfrak{t})\nonumber \\
&=
\frac{
1}{ (1 - x_1) (1 - x_1^2) (1 - x_2) (1 - x_2^2) (1 - x_1 x_2)^2 (1 - x_1^2 x_2)^2 (1 - x_1 x_2^2)^2
}
\Bigl[
1
+x_1 (2+x_1)    x_2\nonumber \\
&\quad -x_1 (-1-2 x_1+2 x_1^2)    x_2^2
-x_1^2 (2+x_1+2 x_1^2)    x_2^3
-x_1^3 (2+x_1+2 x_1^2)    x_2^4
+x_1^4 (-2+2 x_1+x_1^2)    x_2^5\nonumber \\
&\quad +x_1^5 (1+2 x_1)    x_2^6
+x_1^7    x_2^7
\Bigr],\\
&{\cal I}^{U(3)\text{ADHM-}[2]}_H(x,1;\mathfrak{t})\nonumber \\
&=
\frac{
1}{
(1 - x_1) (1 - x_1^2) (1 - x_1^3) (1 - x_2) (1 - x_2^2) (1 - x_2^3) (1 - x_1 x_2)^2 (1 - x_1^2 x_2)^2 (1 - x_1 x_2^2)^2 (1 - x_1^3 x_2)^2 (1 - x_1 x_2^3)^2
}\nonumber \\
&\quad\times \Bigl[
1
+x_1 (2+2 x_1+x_1^2)    x_2
-x_1 (-2-6 x_1-3 x_1^2+2 x_1^4)    x_2^2
-x_1 (-1-3 x_1-6 x_1^2-x_1^3+3 x_1^4+4 x_1^5+2 x_1^6)    x_2^3\nonumber \\
&\quad +x_1^3 (1+3 x_1-6 x_1^2-11 x_1^3-7 x_1^4-x_1^5+x_1^6)    x_2^4
+x_1^2 (1+x_1) (-2-x_1-5 x_1^2-3 x_1^3-10 x_1^4+x_1^6+2 x_1^7)    x_2^5\nonumber \\
&\quad
+x_1^3 (-4-11 x_1-13 x_1^2-6 x_1^3+x_1^4+3 x_1^5+7 x_1^6+4 x_1^7+2 x_1^8)    x_2^6\nonumber \\
&\quad
+x_1^3 (-2-7 x_1-10 x_1^2+x_1^3+14 x_1^4+14 x_1^5+8 x_1^6+7 x_1^7+3 x_1^8+x_1^9)    x_2^7\nonumber \\
&\quad
-x_1^4 (1-x_1-3 x_1^2-14 x_1^3-20 x_1^4-14 x_1^5-3 x_1^6-x_1^7+x_1^8)    x_2^8\nonumber \\
&\quad
-x_1^4 (-1-3 x_1-7 x_1^2-8 x_1^3-14 x_1^4-14 x_1^5-x_1^6+10 x_1^7+7 x_1^8+2 x_1^9)    x_2^9\nonumber \\
&\quad
-x_1^5 (-2-4 x_1-7 x_1^2-3 x_1^3-x_1^4+6 x_1^5+13 x_1^6+11 x_1^7+4 x_1^8)    x_2^{10}\nonumber \\
&\quad
-x_1^6 (1+x_1) (-2-x_1+10 x_1^3+3 x_1^4+5 x_1^5+x_1^6+2 x_1^7)    x_2^{11}
+x_1^7 (1-x_1-7 x_1^2-11 x_1^3-6 x_1^4+3 x_1^5+x_1^6)    x_2^{12}\nonumber \\
&\quad
+x_1^9 (-2-4 x_1-3 x_1^2+x_1^3+6 x_1^4+3 x_1^5+x_1^6)    x_2^{13}
+x_1^{11} (-2+3 x_1^2+6 x_1^3+2 x_1^4)    x_2^{14}
+x_1^{13} (1+2 x_1+2 x_1^2)    x_2^{15}\nonumber \\
&\quad
+x_1^{16}    x_2^{16}
\Bigr].
\end{align}
\end{subequations}
}
First let us consider the coefficients $f^{(l=2)}_{m,j}(y_\alpha=1;x_1)$ with $j=1$, where the coefficient $f_{m,1}^{(l=2)}(y_\alpha=1;x_1)$ with $m=1$ is given by expanding \eqref{hatcalFm1y1} as
\begin{align}
f^{(l=2)}_{1,1}(y_\alpha=1;x_1)=1+3x_1.
\end{align}
Taking into account the saturating property \eqref{fmjsaturation}, let us denote the unknown coefficients as
\begin{subequations}
\begin{align}
&f^{(l=2)}_{2,1}(y_\alpha=1;x_1)=1+a_1x_1+3x_1^2,\\
&f^{(l=2)}_{3,1}(y_\alpha=1;x_1)=1+a_1x_1+a_1x_1^2+3x_1^3,\\
&\quad \vdots.\nonumber 
\end{align}
\end{subequations}
Then, by subtracting the right-hand side of \eqref{singlesum} with the summation over the wrapping number truncated for $m\le 3$ from the left hand side, we obtain 
\begin{subequations}
\begin{align}
N&=1:\cr
0&=x_1^2\Bigl[(4-a_1)x_1^2+(4-a_1)x_1^3+(8-2 a_1)x_1^4+(4-a_1  )x_1^5+(4-a_1   )x_1^6\nonumber \\
&\qquad\quad +(-3+a_1)x_1^8+\cdots\Bigr],\\
N&=2:\cr
0&=x_1^4\Bigl[(4-a_1)x_1^2+(4-a_1)x_1^3+(8-2 a_1)x_1^4+(8-2 a_1)x_1^5+(8-2 a_1 )x_1^6\nonumber \\
&\qquad\quad +(4-a_1)x_1^7+(4-a_1)x_1^8+\cdots\Bigr],\\
N&=3:\cr
0&=x_1^6\Bigl[(4-a_1)x_1^2+(4-a_1)x_1^3+(8-2 a_1)x_1^4+(8-2 a_1)x_1^5+(12-3 a_1)x_1^6\nonumber \\
&\qquad\quad +(8-2 a_1)x_1^7+(8-2 a_1)x_1^8+\cdots\Bigr].
\end{align}
\end{subequations}
Note that the higher expansion coefficients vary as $N$ increases, which is because we have omitted the contributions from higher wrapping number $m$.
Conversely, when the coefficients are saturated in $N$, we can regard them as the exact result without truncation of $m$ and use them as the equations to determine $f_{m,j}^{(l=2)}(y_\alpha=1;x_1)$.
In the current case, from the coefficients of $x_1^{2N+\nu}$ with $\nu\le 5$ which saturate at $N=2$, we can determine $a_1$ as $a_1=4$.

Next let us consider the coefficients $f^{(l=2)}_{m,j}(y_\alpha=1;x_1)$ with $j=2$, where the coefficient $f_{m,2}^{(l=2)}(y_\alpha=1;x_1)$ with $m=1$ is given by expanding \eqref{hatcalFm1y1} as
\begin{align}
f^{(l=2)}_{1,2}(y_\alpha=1;x_1)=1+4x_1+4x_1^2.
\end{align}
Taking into account the saturating property \eqref{fmjsaturation}, let us denote the unknown coefficients as
\begin{subequations}
\begin{align}
&f^{(l=2)}_{2,2}(y_\alpha=1;x_1)=1+4x_1+a_2x_1^2+c_{2,3}x_1^3+b_0x_1^4,\\
&f^{(l=2)}_{3,2}(y_\alpha=1;x_1)=1+4x_1+a_2x_1^2+a_3x_1^3+c_{3,4}x_1^4+b_1x_1^5+b_0x_1^6,\\
&f^{(l=2)}_{4,2}(y_\alpha=1;x_1)=1+4x_1+a_2x_1^2+a_3x_1^3+a_4x_1^4+c_{4,5}x_1^5+b_2x_1^6+b_1x_1^7+b_0x_1^8,\\
&f^{(l=2)}_{5,2}(y_\alpha=1;x_1)=1+4x_1+a_2x_1^2+a_3x_1^3+a_4x_1^4+a_5x_1^5+c_{5,6}x_1^6+b_3x_1^7+b_2x_1^8+b_1x_1^9\nonumber \\
&\quad +b_0x_1^{10},\\
&\quad \vdots.\nonumber
\end{align}
\end{subequations}
Then, by subtracting the right-hand side of \eqref{singlesum} with the summation over the wrapping number truncated for $m\le 3$ from the left hand side, we obtain 
\begin{subequations}
\begin{align}
N&=1:\cr
0&= x_1^2\Bigl[(9-a_2)x_1+(20-a_2-c_{2,3})x_1^2+(29-a_2-b_0-c_{2,3})x_1^3\nonumber \\
&\qquad\quad +(25-a_2+a_3-b_0-2 c_{2,3})x_1^4+\cdots\Bigr],\\
N&=2:\cr
0&= x_1^4\Bigl[(9-a_2)x_1+(20-a_2-c_{2,3})x_1^2+(38-2 a_2-b_0-c_{2,3})x_1^3\nonumber \\
&\qquad\quad +(40-a_2-b_0-2 c_{2,3})x_1^4+\cdots\Bigr],\\
N&=3:\cr
0&= x_1^6\Bigl[(9-a_2)x_1+(20-a_2-c_{2,3})x_1^2+(38-2 a_2-b_0-c_{2,3})x_1^3\nonumber \\
&\qquad\quad +(49-2 a_2-b_0-2 c_{2,3})x_1^4+\cdots\Bigr].
\end{align}
\end{subequations}
From the saturated coefficients of $x_1^{2N+\nu}$ with $\nu\le 3$, we determine $a_2,c_{2,3},b_0$ as $a_2=9$, $c_{2,3}=11$, $b_0=9$.
After substituting these values, we find that the saturation of the coefficients continue to higher order as 
\begin{subequations}
\begin{align}
N&=1:\cr
0&=x_1^3\Bigl[(-15+a_3)x_1^3+(-36+a_3+c_{3,4})x_1^4+(-67+2 a_3+b_1+c_{3,4})x_1^5\nonumber \\
&\qquad\quad +(-88+2 a_3+b_1+2 c_{3,4})x_1^6+(-115+3 a_3-a_4+2 b_1+3 c_{3,4})x_1^7+\cdots\Bigr],\cr\\
N&=2:\cr
0&=x_1^6\Bigl[(-15+a_3)x_1^3+(-36+a_3+c_{3,4})x_1^4+(-67+2 a_3+b_1+c_{3,4})x_1^5\nonumber \\
&\qquad\quad+(-103+3 a_3+b_1+2 c_{3,4})x_1^6+(140+3 a_3+2 b_1+3 c_{3,4})x_1^7+\cdots\Bigr],\\
N&=3:\cr
0&=x_1^9\Bigl[(-15+a_3)x_1^3+(-36+a_3+c_{3,4})x_1^4+(-67+2 a_3+b_1+c_{3,4})x_1^5\nonumber \\
&\qquad\quad +(-103+3 a_3+b_1+2 c_{3,4})x_1^6+(-155+4 a_3+2 b_1+3 c_{3,4})x_1^7+\cdots\Bigr].\cr
\end{align}
\end{subequations}
From the saturated coefficients of $x_1^{3N+\nu}$ with $\nu\le 6$, we determine $a_3,c_{3,4},b_1$ as $a_3=15$, $c_{3,4}=21$, $b_1=16$.

\begin{table}
\begin{center}
\begin{align*}
\begin{array}{|c|c|c|c||c|c|}
\hline
l&m&j^{(A)}_\text{max}&j_\text{max}^{(A)}(y_\alpha=1)&j^{(B)}_\text{max}&j_\text{max}^{(B)}(y_\alpha=1)\\ \hline \hline
   2&1&           7&                \text{-}&                                      7&\text{-}\\ \cline{2-6}
 &2&           6&                       9&                                        6&9\\ \cline{2-6}
 &3&           4&                       7&                                        6&7\\ \cline{2-6}
 &4&           3&                       5&                                        6&7\\ \cline{2-6}
 &5&           3&                       4&                                        6&7\\ \cline{2-6}
 &6&           3&                       4&                                        6&7\\ \cline{2-6}
 &7&           3&                       4&                                        5&7\\ \cline{2-6}
 &8&           2&                       4&                                        5&7\\ \hline \hline
3&1&           3&                \text{-}&                                        3&\text{-}\\ \cline{2-6}
 &2&           3&                       6&                                        3&6\\ \cline{2-6}
 &3&           3&                       5&                                        3&6\\ \cline{2-6}
 &4\le m\le 8&           3&                       4&                                        3&6\\ \hline \hline
4&1&           3&                \text{-}&                                        3&\text{-}\\ \cline{2-6}
 &2&           3&                       3&                                        3&3\\ \cline{2-6}
 &3\le m\le 8&           2&                       3&                                        3&3\\ \hline
\end{array}
\end{align*}
\caption{
List of $j_\text{max}^{(A)}$ (resp.~$j_\text{max}^{(B)}$) for which we have determined $\hat{{\cal F}}_m^{(l)}(y_\alpha;x_1;x_2)$ to the order ${\cal O}(x_2^{j_\text{max}^{(A)}})$ (resp.~${\cal O}(x_2^{j_\text{max}^{(B)}})$) without (resp.~with) assuming inverse giant graviton expansion. 
In order to determine the giant graviton indices, we have used the exact expressions of ${\cal I}^{U(N)\text{ADHM-}[2]}_H$ for $N\le 14$ (resp.~$N\le 10$) with (resp.~without) setting $y_\alpha=1$, ${\cal I}^{U(N)\text{ADHM-}[3]}_H$ for  $N\le 7$ (resp.~$N\le 5$) with (resp.~without) setting $y_\alpha=1$ and ${\cal I}^{U(N)\text{ADHM-}[4]}_H$ for $N\le 4$ (resp.~$N\le 3$) with (resp.~without) setting $y_\alpha=1$.
}
\label{listofjmax}
\end{center}
\end{table}

Note that after the change of variables ${\cal F}_m^{(l)}(y_\alpha;x_1;x_2)=\hat{{\cal F}}_m^{(l)}(y_\alpha;x_1^{-1};x_1x_2)$, ${\cal F}_m^{(l)}(y_\alpha;x_1;x_2)$ is expected to be the supersymmetric indices of the M5-brane giant gravitons in the twisted limit. 
As we comment in section \ref{sec_inversegg}, we can determine more coefficients by assuming that these ${\cal F}_m^{(l)}(y_\alpha;x_1;x_2)$ facilitate the inverse giant graviton expansion.
In Table \ref{listofjmax} we summarize the orders of $x_2$ to which we have determined the giant graviton indices $\hat{{\cal F}}_m^{(l)}(y_\alpha;x_1;x_2)$ with or without assuming the inverse giant graviton expansion.
See Appendix \ref{app:listofFlm} for the final results, which are written in terms of $F_m^{(l)}(y_\alpha;x_1;x_2)$ defined by
\begin{align}\label{Fml}
{ F_m^{(l)}(y_\alpha;x_1;x_2)=\hat{{\cal F}}_m^{(l)}(y_\alpha;x_1^{-1};x_1x_2)\prod_{n=1}^m(x_1^n;x_1x_2)_\infty.}
\end{align}
In the following sections, except in section \ref{sec_inversegg} and section \ref{sec_inversegggenerall}, we utilize all of these data to analyze various structures of the M5-brane indices.
In section \ref{sec_inversegg} and section \ref{sec_inversegggenerall}, we rely only on the data obtained without using the inverse giant graviton expansion.

\section{Single sum expansion for $l=2$}
\label{sec_l=2}
Let us first consider the single sum expansion of the Higgs index for $l=2$
\begin{align}
\frac{
{\cal I}^{U(N)\text{ADHM-[}2]}_H(x,y;\mathfrak{t})
}{
{\cal I}^{U(\infty)\text{ADHM-[}2]}_H(x,y;\mathfrak{t})
}
=
1+\sum_{m=1}^\infty x_1^{mN} \hat{{\cal F}}_m^{(l=2)}(y;x_1;x_2). 
\end{align}
Here we have denoted $y_1=y_2^{-1}$ as $y$.
The term $x_1^{mN}$ in the expansion would correspond to 
the ground state configuration characterized by the supersymmetric surface wrapped by the M5-brane giant graviton. 
The coefficients $\hat{{\cal F}}_m^{(l=2)}(y;x_1;x_2)$ would be contributed from the M5-brane giant gravitons of wrapping number $m$. 

Remarkably, the coefficients $\hat{{\cal F}}_m^{(l=2)}(y;x_1;x_2)$ can be viewed as 
the giant graviton indices, the indices of the theory realized on the M5-brane giant gravitons. 
It follows that they are related to the supersymmetric indices of the world-volume theory of the M5-branes by a simple variable change of the fugacities $x_1$ and $x_2$
\begin{align}
{\cal F}_m^{(l=2)}(y;x_1;x_2)= \hat{{\cal F}}_m^{(l=2)}(y;x_1^{-1};x_1x_2).
\label{variablechange}
\end{align}
We shall refer to this ${\cal F}_m^{(l=2)}(y;x_1;x_2)$ as the M5-brane indices. 

\subsection{A single M5-brane}
For $m=1$, we find that 
\begin{align}
\label{GG_l2m1}
\hat{{\cal F}}_1^{(l=2)}(y;x_1;x_2)=\frac{1}{(x_1^{-1};x_2)_\infty}\chi_{\widehat{\mathfrak{su}}(2)_1}(y;x_2),
\end{align}
where
\begin{align}
\label{ch_su2k1}
\chi_{\widehat{\mathfrak{su}}(2)_1}(y;q)
&=\frac{1}{(q;q)_{\infty}}\sum_{m\in \mathbb{Z}}q^{m^2} y^{2m}
\end{align}
is the character of the vacuum module of the affine Kac-Moody algebra $\widehat{\mathfrak{su}}(2)_1$. 
The expression (\ref{ch_su2k1}) gives rise to the refinement of the result (\ref{hatcalFm1y1}). 
The expansion coefficient (\ref{GG_l2m1}) can be viewed as the index 
contributed from fluctuation modes of a single M5-brane giant wrapped around a supersymmetric $5$-cycle in $S^7/\mathbb{Z}_2$. 
The M5-brane index is
\begin{align}
\mathcal{F}_{1}^{(l=2)}(y;x_1;x_2)&=
\frac{1}{(x_1;x_1x_2)_{\infty}}
\chi_{\widehat{\mathfrak{su}}(2)_1}(y;x_1x_2).
\end{align}
The factor $(x_1;x_1x_2)_{\infty}^{-1}$ is the $1/4$-BPS index 
that is obtained as the twisted limit \cite{Hayashi:2024aaf} of the supersymmetric index of the 6d (2,0) theory for a free tensor multiplet \cite{Bhattacharya:2008zy}. 
It can be also found from the giant graviton index (\ref{hatcalFm1y1}) for $l=1$ upon a change of variables $x_1\rightarrow x_1^{-1}$, $x_2\rightarrow x_1x_2$. 
The other factor is the vacuum character (\ref{ch_su2k1}) of the affine Kac-Moody algebra $\widehat{\mathfrak{su}}(2)_1$. 

The appearance of the affine Kac-Moody algebra would be associated with 
the configuration of the M5-brane supported on the two-centered Taub-NUT space $TN_2$. 
Reducing the configuration on the $S^1$ fibration of the $TN_2$, 
one finds a D4-brane intersecting with two D6-branes along the two-dimensional space in Type IIA string theory. 
This intersection locus is called the I-brane \cite{Green:1996dd,Bachas:1997kn,Itzhaki:2005tu}. 
The massless spectrum on the I-brane consists of the chiral fermions that arise from the RR sector of the $4$-$6$ open strings stretching between the D4- and D6-branes. 
The chiral fermions are coupled to the $U(1)$ and $U(2)$ gauge fields, 
which live on the world-volumes of the D4-brane and D6-branes respectively. 
In general, the complex fermions transform as $(+,\overline{\square})$ and $(-,\square)$ under the $U(1)\times U(2)$ gauge group 
so that the they can realize the affine Kac-Moody algebra $\widehat{\mathfrak{u}}(2)_1$ at level $1$. 
However, in our case, the $U(2)$ gauge field on the D6-brane is not dynamical as the Taub-NUT geometry in M-theory is fixed, 
whereas the $U(1)$ gauge field on the D4-brane is dynamical. 
This implies that the effective theory is given by the coset
\begin{align}
\frac{\widehat{\mathfrak{u}}(2)_1}{\widehat{\mathfrak{u}}(1)_{-2}}&=\widehat{\mathfrak{su}}(2)_{1}. 
\end{align}
In fact, the vacuum character (\ref{ch_su2k1}) is equal to the graded character 
\begin{align}
(q;q)_{\infty}\oint \frac{ds}{2\pi is}
(q^{\frac12}s^{\pm}y^{\pm};q)_{\infty}
(q^{\frac12}s^{\pm}y^{\mp};q)_{\infty}
\end{align}
of the boundary chiral algebra of the two free fermions in two-dimensions 
coupled to the $U(1)$ Chern-Simons theory with level $-2$. 

\subsection{Multiple M5-branes}
\subsubsection{$m=2$}
Let us consider the case with $m=2$. 
In this case the M5-brane index is expected to encodes the spectrum of the theory of a stack of two M5-branes. 
We write 
\begin{align}
{\cal F}_2^{(l=2)}(y;x_1;x_2)&=\frac{1}{(x_1;x_1x_2)_\infty(x_1^2;x_1x_2)_\infty}F_2^{(l=2)}(y;x_1;x_2). 
\label{calFvsF}
\end{align}
Again the factor $(x_1;x_1x_2)_\infty^{-1}(x_1^2;x_1x_2)_\infty^{-1}$ that also appears for $l=1$ 
is the twisted limit \cite{Hayashi:2024aaf} of the 6d $(2,0)$ theory describing a stack of two M5-branes. 
The factor $F_2^{(l=2)}(y;x_1;x_2)$ is expected to be contributed from the spectrum of excitation modes for the I-brane. 
We find that the small $x_2$ expansion of $F_2^{(l=2)}(y;x_1;x_2)$ is given by
\begin{align}
\label{Fl2_m2_general}
&F_2^{(l=2)}(y;x_1;x_2)=
1
+\Bigl[
y^{-2}
+1
+y^2
+(
y^{-2}
+1
+y^2)
x_1
\Bigr]    x_1x_2\nonumber \\
&\quad +\Bigl[
y^{-4}
+2y^{-2}
+3
+2 y^2
+y^4
+x_1 (
2y^{-2}
+3
+2 y^2)
+x_1^2
\Bigr]    (x_1x_2)^2\nonumber \\
&\quad
+\Bigl[
y^{-4}
+4y^{-2}
+5
+4 y^2
+y^4
+(
y^{-4}
+5y^{-2}
+7
+5 y^2
+y^4)x_1
+(
y^{-2}
+2
+y^2)x_1^2 
\Bigr]    (x_1x_2)^3\nonumber \\
&\quad
+\Bigl[
3y^{-4}
+7y^{-2}
+10
+7 y^2
+3 y^4
+(
3y^{-4}
+10y^{-2}
+14
+10 y^2
+3 y^4)x_1 \nonumber \\
&\quad\quad+(
y^{-4}
+3y^{-2}
+6
+3 y^2
+y^4)x_1^2 
\Bigr]    (x_1x_2)^4\nonumber \\
&\quad
+\Bigl[
y^{-6}
+5y^{-4}
+13y^{-2}
+ 16
+13 y^2
+5 y^4
+y^6
\nonumber\\
&\quad 
+(
y^{-6}
+7y^{-4}
+20y^{-2}
+27
+20 y^2
+7 y^4
+y^6)x_1\nonumber \\
&\quad\quad +(
2y^{-4}
+8y^{-2}
+12
+8 y^2
+2 y^4)x_1^2 
+(
y^{-2}
+1
+y^2)x_1^3 
\Bigr]    (x_1x_2)^5\nonumber \\
&\quad +\Bigl[
2y^{-6}+10y^{-4}+21y^{-2}+28+21 y^2+10 y^4+2 y^6\nonumber \\
&\quad\quad +(2y^{-6}+14y^{-4}+36y^{-2}+48+36 y^2+14 y^4+2 y^6)x_1\nonumber \\
&\quad\quad +(6y^{-4}+17y^{-2}+26+17 y^2+6 y^4)x_1^2 
+(2y^{-2}+3+2 y^2)x_1^3 
\Bigr](x_1x_2)^6
+{\cal O}(x_2^7). 
\end{align}

On the I-brane we have the four complex fermions transforming as the bifundamental representation 
$(\square,\overline{\square})$ and $(\overline{\square},\square)$ under the $U(2)\times U(2)$ gauge group. 
Since the fermions carry opposite charges under the two $U(1)$ factors, the diagonal $U(1)$ is decoupled. 
Hence they effectively couple to $SU(2)\times SU(2)\times U(1)$. 
These chiral fermions give rise to a realization of the affine Kac-Moody algebra $\widehat{\mathfrak{u}}(4)_1$ with level $1$. 
There exists the following conformal embedding 
\begin{align}
\label{aff_emb_u(4)_1}
\widehat{\mathfrak{su}}(2)_2
\oplus 
\widehat{\mathfrak{su}}(2)_2
\oplus 
\widehat{\mathfrak{u}}(1)_{4}
\subset \widehat{\mathfrak{u}}(4)_1. 
\end{align}
The embedding (\ref{aff_emb_u(4)_1}) can be viewed as a splitting of the bilinear current formed by the free fermions 
and it is the affine analog of the gauge symmetry $SU(2)\times SU(2)\times U(1)$. 
When the two D4-branes are separated, one of the $SU(2)$ gauge factors is broken to $U(1)$ 
so that the fermions can generically couple to the $U(1)\times SU(2)\times U(1)$. 
Accordingly, we have the following splitting \cite{Itzhaki:2005tu}
\begin{align}
\widehat{\mathfrak{u}}(1)_{2}
\oplus 
(\widehat{\mathfrak{su}}(2)_1)^{2}
\oplus 
\widehat{\mathfrak{u}}(1)_{4}
\subset \widehat{\mathfrak{u}}(4)_1, 
\end{align}
where we have used the following equivalence of the cosets \cite{DiFrancesco:1997nk}
\begin{align}
\frac{\widehat{\mathfrak{su}}(2)_2}{\widehat{\mathfrak{u}}(1)_2}
&=
\frac{(\widehat{\mathfrak{su}}(2)_1)^{2}}{\widehat{\mathfrak{su}}(2)_2}. 
\end{align}
From the perspective of the effective theory on the separated D4-branes, 
the free fermions are dynamically coupled to the $U(1)$ gauge field. 
Accordingly, the effective theory can be described by the coset \cite{Tan:2008wp}
\begin{align}
\label{coset_l2_m2}
\frac{\widehat{\mathfrak{u}}(1)_{2}
\oplus 
\widehat{\mathfrak{u}}(1)_{4}
\oplus 
(\widehat{\mathfrak{su}}(2)_1)^{2}}
{\widehat{\mathfrak{u}}(1)_{2}
\oplus  
\widehat{\mathfrak{u}}(1)_{4}}.
\end{align}
Notice that the $U(1)/U(1)$ coset models appearing in the coset (\ref{coset_l2_m2}) are the topological field theories \cite{Witten:1991mm,Spiegelglas:1992jg}. 
They merely describe global degrees of freedom with no propagating fields 
and the physical states are restricted to be zero eigenstates of the $L_0$ \cite{Spiegelglas:1992jg}. 
So the actual contributions to the index will be expressed in terms of the product of two characters of the affine Kac-Moody algebra $\widehat{\mathfrak{su}}(2)_1$. 
In fact, we observe that the expansion (\ref{Fl2_m2_general}) contains the expected contributions. 
We see that it is natural to turn off the fugacity $x_1$ 
to extract the degeneracy of the sates as the expansion coefficients of $F_2^{(l=2)}$ with respect to $x_2$. 
In fact, when $x_1$ is set to unity, it is simply given by
\begin{align}
F_2^{(l=2)}(y;1;x_2)=\chi_{\widehat{\mathfrak{su}}(2)_{1}}(y;x_2)^2, 
\end{align}
where $\chi_{\widehat{\mathfrak{su}}(2)_{1}}(y;q)$ is the vacuum character (\ref{ch_su2k1}) of the $\widehat{\mathfrak{su}}(2)_1$. 
This would imply that the function $F_2^{(l=2)}(y;1;x_2)$ encodes the spectrum of excitations of a stack of two separated M5-branes 
which is effectively described by the coset (\ref{coset_l2_m2}).\footnote{A product of two $\mathfrak{su}(2)_1$ characters also appears in the elliptic genus of the M-string SCFT of rank two on $T^2 \times \mathbb{C}^2/\mathbb{Z}_2$ \cite{DelZotto:2023rct}.}

Next we focus on the terms with $(x_1x_2)^j$, $j=0,1,2,\cdots$ in the expansion (\ref{Fl2_m2_general}), 
which can be extracted by taking the fugacity limit $x_1\rightarrow 0,x_2\rightarrow\infty$ with $x_1x_2=q$ fixed. 
In this limit we find that 
the full index $\mathcal{F}_2^{(l=2)}(y;x_1;x_2)$ reduces to 
\begin{align}
\label{Fl2_m2}
&\lim_{\substack{x_1\rightarrow 0,x_2\rightarrow\infty\\ (x_1x_2=q\text{: fixed})}} \mathcal{F}_2^{(l=2)}(y;x_1;x_2)\nonumber \\
&=
1
+( y^{-2} +1 +y^2)    q
+( y^{-4} +2y^{-2} +3 +2 y^2 +y^4)    q^2\nonumber \\
&\quad +( y^{-4} +4y^{-2} +5 +4 y^2 +y^4)    q^3
+( 3y^{-4} +7y^{-2} +10 +7 y^2 +3 y^4)    q^4\nonumber \\
&\quad +( y^{-6} +5y^{-4} +13y^{-2} + 16 +13 y^2 +5 y^4 +y^6)    q^5\nonumber \\
&\quad +( 2y^{-6}+10y^{-4}+21y^{-2}+28+21 y^2+10 y^4+2 y^6)q^6
+{\cal O}(q^7). 
\end{align}
We observe that (\ref{Fl2_m2}) matches with the character 
\begin{align}
\label{ch_su2k2}
\chi_{\widehat{\mathfrak{su}}(2)_2}(y;q)
&=\frac{1}{(q;q)_{\infty}}
\sum_{m\in \mathbb{Z}}\frac{q^{2m^2}}{(q^{1+2m}y^2;q)_{\infty}(q^{1-2m}y^{-2};q)_{\infty}}
\end{align}
of the vacuum module of the affine Kac-Moody algebra $\widehat{\mathfrak{su}}(2)_2$ at level $2$. Namely, we find
\begin{equation}
 \lim_{\substack{x_1\rightarrow 0,x_2\rightarrow\infty\\ (x_1x_2=q\text{: fixed})}} \mathcal{F}_2^{(l=2)}(y;x_1;x_2) = \chi_{\widehat{\mathfrak{su}}(2)_2}(y;q).
\end{equation}
When the fugacity $y$ is turned off, we obtain the power series in $q$
\begin{align}
&
\lim_{\substack{x_1\rightarrow 0,x_2\rightarrow\infty\\ (x_1x_2=q\text{: fixed})}} \mathcal{F}_2^{(l=2)}({1;}x_1;x_2)
\nonumber\\
&=
1 +3  q +9  q^{2} +15 q^{3} +30 q^{4} +54 q^{5} +94 q^{6} +153 q^{7} +252 q^{8}+391 q^{9} +{\cal O}(q^{10}). 
\end{align}
This agrees with the unflavored vacuum character of the $\widehat{\mathfrak{su}}(2)_2$. 

Note that the vacuum character (\ref{ch_su2k2}) is equivalent to  the graded character 
\begin{align}
\frac12 (q;q)_{\infty}^2 \oint \prod_{i=1}^2 \frac{ds_i}{2\pi is_i}
(s_1^{\pm}s_2^{\mp};q)_{\infty}
\prod_{i=1}^2
(q^{\frac12}s_i^{\pm} y;q)_{\infty}
(q^{\frac12}s_i^{\pm} y^{-1};q)_{\infty} 
\end{align}
of the chiral algebra arising from four free fermions 
coupled to the 3d $U(2)_{-2}$ Chern-Simons theory with level $-2$ according to the level-rank duality \cite{MR1152377}. 

We note that the appearance of the affine Kac-Moody algebra $\widehat{\mathfrak{su}}(2)_2$ 
is analogous to the proposal \cite{Belavin:2011pp,Belavin:2012aa} that 
two M5-branes on $\mathbb{C}^2/\mathbb{Z}_2$ with $\Omega$-deformation leads to a system endowed with the $\widehat{\mathfrak{su}}(2)_2$ algebra (also see \cite{MR1302318,MR1441880} for the related mathematical facts). It would be nice to clarify the relation to the limits we have taken here. 

\subsubsection{General $m$}
Similarly, let us write the M5-brane indices for $m>2$ as
\begin{align}
{\cal F}_m^{(l=2)}(y;x_1,x_2)
&=
\prod_{n=1}^m
\frac{1}{(x_1^n;x_1x_2)_\infty} F_m^{(l=2)}(y;x_1;x_2). 
\label{Fmdefinitionl2}
\end{align}
Again the factor $\prod_{n=1}^m (x_1^n;x_1x_2)_\infty^{-1}$ is the $1/4$-BPS index of the $m$ M5-branes 
which is obtained from the twisted limit \cite{Hayashi:2024aaf}. 
When the fugacity $x_2$ is turned off, it becomes the unrefined index, 
that is equivalent to the vacuum character of the of the W-algebra $\mathcal{W}(\mathfrak{gl}(m))$ \cite{Kim:2012ava,Beem:2014kka}. 
The function $F_m^{(l=2)}(y;x_1;x_2)$ is expected to capture the spectrum of the I-brane configuration 
with $m$ D4-branes intersecting with two D6-branes. 
The small $x_2$ expansion of $F_m^{(l=2)}(y;x_1;x_2)$ are listed in Appendix \ref{app:listofFlm_l2withy}.

For general $m$ there exist $2m$ complex fermions on the I-brane. 
They transform in the bifundamental representation $(\square,\overline{\square})$ and $(\overline{\square},\square)$ of the $U(m)\times U(2)$ gauge group. 
As one of the $U(1)$ is decoupled, they effectively couple to $SU(m)\times SU(2)\times U(1)$. 
They can realize the affine Kac-Moody algebra $\widehat{\mathfrak{u}}(2m)_{1}$, 
for which there exists the conformal embedding
\begin{align}
\widehat{\mathfrak{su}}(m)_2
\oplus 
\widehat{\mathfrak{su}}(2)_m
\oplus 
\widehat{\mathfrak{u}}(1)_{2m}
\subset \widehat{\mathfrak{u}}(2m)_1. 
\end{align}
When the $m$ D4-branes are non-coincident, the $SU(m)$ gauge group is broken to $U(1)^{m-1}$ 
and the fermions may couple to the $U(1)^{m-1}\times SU(2)\times U(1)$. 
The resulting splitting of the current algebra is \cite{Itzhaki:2005tu}
\begin{align}
(\widehat{\mathfrak{u}}(1)_{2})^{m-1}
\oplus 
(\widehat{\mathfrak{su}}(2)_{1})^{m}
\oplus  
\widehat{\mathfrak{u}}(1)_{2m}
\subset \widehat{\mathfrak{u}}(2m)_1, 
\end{align}
where we have used the equivalent coset realization \cite{DiFrancesco:1997nk}
\begin{align}
\frac{\widehat{\mathfrak{su}}(m)_2}{\widehat{\mathfrak{u}}(1)_2^{m-1}}
&=
\frac{\widehat{\mathfrak{su}}(2)_1^{m}}{\widehat{\mathfrak{su}}(2)_m}. 
\end{align}
While the $SU(2)$ gauge field on the two D6-branes is not dynamical, 
the $U(1)$ gauge fields on the D4-branes are dynamical. 
This leads to the coset \cite{Tan:2008wp}
\begin{align}
\label{coset_l2_m}
\frac{(\widehat{\mathfrak{u}}(1)_{2})^{m-1}
\oplus  
\widehat{\mathfrak{u}}(1)_{2m}
\oplus  
\widehat{(\mathfrak{su}}(2)_{1})^{m}}
{(\widehat{\mathfrak{u}}(1)_{2})^{m-1}
\oplus  
\widehat{\mathfrak{u}}(1)_{2m}}. 
\end{align}
Again the $U(1)/U(1)$ coset models are the topological field theories \cite{Witten:1991mm,Spiegelglas:1992jg}. 
Hence the non-trivial contributions to the index is given by the product of $m$ characters of the affine Kac-Moody algebra $\widehat{\mathfrak{su}}(2)_1$. 
In fact, we observe that if we set $x_1=1$, then
\begin{align}
F_m^{(l=2)}(y;1,x_2)=\chi_{\widehat{\mathfrak{su}}(2)_1}(y;x_2)^m
\label{x11property}, 
\end{align}
where $\chi_{\widehat{\mathfrak{su}}(2)_1}(y;q)$ is the vacuum character (\ref{ch_su2k1}) of the $\widehat{\mathfrak{su}}(2)_1$. 
This implies that the index $F_m^{(l=2)}$ encodes the spectrum of the configuration of the I-brane with $m$ separated M5-branes described by the coset (\ref{coset_l2_m}). 

Furthermore, we observe that the full M5-brane index $\mathcal{F}_m^{(l=2)}(y;x_1;x_2)$ has a nice special fugacity limit. 
When we keep only the monomials of the form $(x_1x_2)^j$, with $j=0,1,2,\cdots$ in ${\cal F}_m^{(l=2)}(y;x_1;x_2)$, 
which can be obtained by taking the limit $x_1\rightarrow 0,x_2\rightarrow\infty$ with $x_1x_2=q$ fixed, 
we find 
\begin{align}
\lim_{\substack{x_1\rightarrow 0,x_2\rightarrow\infty\\ (x_1x_2=q\text{: fixed})}}
\mathcal{F}_m^{(l=2)}(y;x_1;x_2)=\chi_{\widehat{\mathfrak{su}}(2)_m}(y;q)
\label{x1^0(x1x2)^jproperty}. 
\end{align}
Here $\chi_{\widehat{\mathfrak{su}}(2)_k}(y;q)$ is the vacuum character of the affine Kac-Mooday $\widehat{\mathfrak{su}}(2)_k$ at level $k$. 
According to the Kac-Weyl formula, it is given by 
\begin{align}
\chi_{\widehat{\mathfrak{su}}(2)_k}(y;q)
&=q^{\frac{k}{8k+16}}\frac{\Theta_{1}^{(k+2)}(y;q)-\Theta_{-1}^{(k+2)}(y;q)}
{\Theta_{1}^{(2)}(y;q)-\Theta_{-1}^{(2)}(y;q)}, 
\label{chisu2k}
\end{align}
where 
\begin{align}
\Theta_l^k(y;q)&=\sum_{m\in \mathbb{Z}}
q^{k\left(m+\frac{l}{2k}\right)^2}y^{-2k(m+\frac{l}{2k})}. 
\end{align}
\begin{table}
\begin{center}
{\fontsize{9pt}{1pt}\selectfont
\begin{align*}
\begin{array}{|c|l|}
\hline
&\\[3pt]
m&\lim_{\substack{x_1\rightarrow 0,x_2\rightarrow\infty \\(x_1x_2=1\text{: fixed})}}{\cal F}^{(l=2)}_m(1;x_1;x_2)\\ \hline
&\\[3pt]
2&1 +3  q +9  q^{2} +15 q^{3} +30 q^{4} +54 q^{5} +94 q^{6} +153 q^{7} +252 q^{8} +391 q^{9} +{\cal O}(q^{10})\\ \hline
&\\[3pt]
3&1 +3  q +9  q^2 +22 q^3 +42 q^4 +81 q^5 +151 q^6 +264 q^7 +{\cal O}(q^8)\\ \hline
&\\[3pt]
4&1 +3  q +9  q^2 +22  q^3 +51  q^4 +97  q^5 +188  q^6 +343  q^7+{\cal O}(q^8)\\ \hline
&\\[3pt]
5&1 +3  q +9  q^2 +22  q^3 +51  q^4 +108 q^5 +208  q^6 +390  q^7+{\cal O}(q^8)\\ \hline
&\\[3pt]
6&1 +3  q +9  q^2 +22  q^3 +51  q^4 +108 q^5 +221  q^6 +414  q^7+{\cal O}(q^8)\\ \hline
&\\[3pt]
m\ge 7&1 +3  q +9  q^2 +22  q^3 +51  q^4 +108 q^5 +221  q^6 +429  q^7+{\cal O}(q^8)\\ \hline
\end{array}
\end{align*}
\begin{align*}
\begin{array}{|c|l|}
\hline
&\\[3pt]
m&\lim_{\substack{x_1\rightarrow 0,x_2\rightarrow\infty \\(x_1x_2=1\text{: fixed})}}{\cal F}^{(l=2)}_m(y;x_1;x_2)\\ \hline
&\\[3pt]
2&1 +( y^{-2} +1 +y^2)    q +( y^{-4} +2y^{-2} +3 +2 y^2 +y^4)    q^2 +( y^{-4} +4y^{-2} +5 +4 y^2 +y^4)    q^3\\
&\\[3pt]
& +( 3y^{-4} +7y^{-2} +10 +7 y^2 +3 y^4)    q^4 +( y^{-6} +5y^{-4} +13y^{-2} + 16 +13 y^2 +5 y^4 +y^6)    q^5\\
&\\[3pt]
& +( 2y^{-6}+10y^{-4}+21y^{-2}+28+21 y^2+10 y^4+2 y^6)q^6 +{\cal O}(q^7)\\ \hline
&\\[3pt]
3&1 +( y^{-2} +1 +y^2)    q +( y^{-4} +2y^{-2} +3 +2 y^2 +y^4)    q^2 +( y^{-6} +2y^{-4} +5y^{-2} +6 +5 y^2 +2 y^4 +y^6)    q^3\\
&\\[3pt]
& +( y^{-6} +5y^{-4} +9y^{-2} +12 +9 y^2 +5 y^4 +y^6)    q^4 + ( 3y^{-6} +9y^{-4} +18y^{-2} +21 +18y^2 +9y^4 +3y^6) q^5\\
&\\[3pt]
&+( y^{-8} +6y^{-6} +18y^{-4} +31y^{-2} +39 +31 y^2 +18 y^4 +6 y^6 +y^8)q^6 +{\cal O}(q^7)\\ \hline
&\\[3pt]
4&1 +(y^{-2} +1 +y^2)                                                                      q +(y^{-4} +2y^{-2} +3 +2 y^2 +y^4)                                           q^2 +(y^{-6} +2y^{-4} +5y^{-2} +6 +5 y^2 +2 y^4 +y^6)                       q^3\\
&\\[3pt]
&+(y^{-8} +2y^{-6} +6y^{-4} +10y^{-2} +13 +10y^2 +6y^4 +2y^6 +y^8)   q^4\\
&\\[3pt]
& +(y^{-8} +5y^{-6} +11y^{-4} +20y^{-2} +23 +20y^2 +11y^4 +5y^6 +y^8) q^5\\
&\\[3pt]
&+(3y^{-8} +10y^{-6} +23y^{-4} +36y^{-2} +44 +36 y^2 +23 y^4 +10 y^6 +3 y^8)q^6 +{\cal O}(q^7)\\ \hline
&\\[3pt]
5&1 +(y^{-2} +1 +y^2)                     q +(y^{-4} +2y^{-2} +3 +2 y^2 +y^4)                     q^2 +(y^{-6} +2y^{-4} +5y^{-2} +6 +5y^2 +2y^4 +y^6)                     q^3\\
&\\[3pt]
&+(y^{-8} +2y^{-6} +6y^{-4} +10y^{-2} +13 +10y^2 +6y^4 +2y^6 +y^8)                     q^4\\
&\\[3pt]
&+(y^{-10} +2y^{-8} +6y^{-6} +12y^{-4} +21y^{-2} +24 +21y^2 +12y^4 +6y^6 +2y^8 +y^{10})q^5\\
&\\[3pt]
&+(y^{-10} +5y^{-8} +12y^{-6} +25y^{-4} +38y^{-2} +46 +38 y^2 +25 y^4 +12 y^6 +5 y^8 +y^{10})q^6
+{\cal O}(q^7)\\ \hline
&\\[3pt]
6&1 +(y^{-2} +1 +y^2)                     q +(y^{-4} +2y^{-2} +3 +2 y^2 +y^4)                     q^2 +(y^{-6} +2y^{-4} +5y^{-2} +6 +5y^2 +2y^4 +y^6)                     q^3\\
&\\[3pt]
&+(y^{-8} +2y^{-6} +6y^{-4} +10y^{-2} +13 +10y^2 +6y^4 +2y^6 +y^8)                     q^4\\
&\\[3pt]
&+(y^{-10} +2y^{-8} +6y^{-6} +12y^{-4} +21y^{-2} +24 +21y^2 +12y^4 +6y^6 +2y^8 +y^{10})q^5\\
&\\[3pt]
&+( y^{-12} +2y^{-10} +6y^{-8} +13y^{-6} +26y^{-4} +39y^{-2} +47 +39 y^2 +26 y^4 +13 y^6 +6 y^8 +2 y^{10} +y^{12})q^6\\
&\\[3pt]
&+{\cal O}(q^7)\\ \hline
&\\[3pt]
m\ge 7&1 +(y^{-2} +1 +y^2)                     q +(y^{-4} +2y^{-2} +3 +2 y^2 +y^4)                     q^2 +(y^{-6} +2y^{-4} +5y^{-2} +6 +5y^2 +2y^4 +y^6)                     q^3\\
&\\[3pt]
&+(y^{-8} +2y^{-6} +6y^{-4} +10y^{-2} +13 +10y^2 +6y^4 +2y^6 +y^8)                     q^4\\
&\\[3pt]
&+(y^{-10} +2y^{-8} +6y^{-6} +12y^{-4} +21y^{-2} +24 +21y^2 +12y^4 +6y^6 +2y^8 +y^{10})q^5
+{\cal O}(q^6)\\ \hline
\end{array}
\end{align*}
}
\caption{
Coefficients of the monomials $q^j$ $=$ $(x_1x_2)^j$ in ${\cal F}^{(l=2)}_m(y;x_1;x_2)$.
To the order $q^m$, the coefficients of $q^j$ in ${\cal F}^{(l=2)}_{m'}(y;x_1;x_2)$ with $m'\ge m$ coincides with those in ${\cal F}^{(l=2)}_m(1;x_1;x_2)$.
}
\label{x1^0(x1x2)^j_l2_y1andwithy}
\end{center}
\end{table}
For example, for $m\le 8$, we find from \eqref{listofFlm_l2withy} and \eqref{listofFlm_l2y1higherorderonly} the coefficients of the monomials $(x_1x_2)^j$ as listed in Table \ref{x1^0(x1x2)^j_l2_y1andwithy}, which indeed agree with \eqref{chisu2k}.

The level-rank duality \cite{MR1152377} also implies that 
the expression (\ref{x1^0(x1x2)^jproperty}) is evaluated as the graded character 
of the $2m$ free fermions coupled to the $U(m)_{-2}$ Chern-Simons theory with level $-2$ 
\begin{align}
\frac{1}{m!}(q;q)_{\infty}^m 
\oint \prod_{i=1}^m \frac{ds_i}{2\pi is_i}
\prod_{i<j}(s_i^{\pm}s_j^{\mp};q)_{\infty}
\prod_{i=1}^m (q^{\frac12}s_i^{\pm}y;q)_{\infty}(q^{\frac12}s_i^{\pm}y^{-1};q)_{\infty}. 
\end{align}

The formula (\ref{x1^0(x1x2)^jproperty}) for the special fugacity limit of the M5-brane index shows 
that there is an action of the affine Kac-Moody algebra $\widehat{\mathfrak{su}}(2)_m$. 
A similar proposal was made in \cite{Nishioka:2011jk} that the affine Kac-Moody algebra $\widehat{\mathfrak{su}}(2)_m$ 
appears from $m$ M5-branes on $\mathbb{C}^2/\mathbb{Z}_2$ with $\Omega$-deformation. 
It would be an interesting future work to figure out the structure of the full M5-brane index $\mathcal{F}_m^{(l=2)}(y;x_1,x_2)$. 

\subsection{Inverse giant graviton expansion}
\label{sec_inversegg}
Let us investigate the inverse giant graviton expansion which was also observed for the Coulomb indices as well as for various other M2-brane setups \cite{Gaiotto:2021xce,Hayashi:2024aaf}.
In order to study the inverse giant graviton expansion, one needs to take the large $m$ limit of the M5-brane index (\ref{Fmdefinitionl2}). 
The large $m$ limit of the factor $\prod_{n}(x_1^n;x_1x_2)_{\infty}$ is identified with 
the twisted limit \cite{Hayashi:2024aaf} of the index \cite{Bhattacharya:2008zy} for the KK excitations on the supergravity background $AdS_7\times S^4$. 
On the other hand, as discussed above, the M5-brane index $\mathcal{F}_m^{(l=2)}(y;x_1;x_2)$ becomes the vacuum character of the $\widehat{\mathfrak{su}}(2)_m$ 
in the special fugacity limit where $x_1\rightarrow 0$ while keeping $x_1x_2=q$ finite. 
In the large $m$ limit, the vacuum character (\ref{chisu2k}) becomes 
\begin{align}
\frac{1}{(q;q)_{\infty}(y^2;q)_{\infty}(y^{-2};q)_{\infty}}. 
\end{align}
Making use of this observation, we find that the large $m$ limit of the whole expression agrees with 
\begin{align}
&
{\cal F}_\infty^{(l=2)}(y;x_1,x_2)
\nonumber\\
&=\prod_{n=1}^\infty\frac{1}{(x_1^n;x_1x_2)_\infty}
\frac{1}{(x_1^{n}x_2;x_1x_2)_{\infty} (x_1^{n}x_2 y^2;x_1x_2)_{\infty} (x_1^{n}x_2 y^{-2};x_1x_2)_{\infty}}. 
\end{align}
We have confirmed this formula to the order ${\cal O}(\mathfrak{t}^5)$ in the small $\mathfrak{t}$ expansion with $x$ fixed for $x_1=x^{-1}\mathfrak{t}$ and $x_2=x\mathfrak{t}$, by using the data of the full M5-brane indices including ${\cal F}_4^{(l=2)}(y_\alpha;x_1;x_2)$ to the order ${\cal O}(x_2^3)$.
For $y=1$, the formula has been confirmed to the order ${\cal O}(\mathfrak{t}^8)$ by using the data including ${\cal F}_7^{(l=2)}(1;x_1;x_2)$ to the order ${\cal O}(x_2^4)$.\footnote{
As summarized in Table \ref{listofjmax}, these data were obtained without assuming inverse giant graviton expansion.
}

Now let us consider the finite $m$ corrections of ${\cal F}_m^{(l=2)}(y;x_1;x_2)$ in the single sum expansion.
Then we observe
\begin{align}
\frac{{\cal F}_m^{(l=2)}(y;x_1;x_2)}{{\cal F}_\infty^{(l=2)}(y;x_1;x_2)}=1+\sum_{m'=1}^\infty x_1^{mm'}\hat{{\cal G}}^{(l=2)}_{m'}(y;x_1;x_2).
\end{align}
Substituting the above-mentioned data ${\cal F}^{(l=2)}_m(y;x_1;x_2)$ into the left-hand side and finding the saturation of the coefficients of $x_1^m$, $x_1^{2m}$, and so forth as $m$ increases, we can determine the first few coefficients $\hat{{\cal G}}_{m'}^{(l=2)}(y;x_1;x_2)$ as 
\begin{subequations}
\begin{align}
&\hat{\cal G}_1^{(l=2)}(y;x_1;x_2)\nonumber \\
&=\frac{1}{1-x_1^{-1}}\Bigl[
1+
(y^{-2}+1+y^2+x_1)x_2\nonumber \\
&\quad +\Bigl(y^{-4}+y^{-2}+1+y^2+y^4+(y^{-2}+1+y^2)x_1+x_1^2\Bigr)x_2^2
+{\cal O}(x_2^3)
\Bigr],\\
&\hat{\cal G}_2^{(l=2)}(y;x_1;x_2)\nonumber \\
&=\frac{1}{(1-x_1^{-1})(1-x_1^{-2})}\Bigl[
1+
\Bigl((y^{-2}+1+y^2+x_1)x_1^{-1}+y^{-2}+2+y^2+x_1\Bigr)x_2\nonumber \\
&\quad +\Bigl(
(y^{-4}+y^{-2}+1+y^2+y^4)x_1^{-2}
+(y^{-4}+2y^{-2}+2+2y^2+y^4)x_1^{-1}\nonumber \\
&\quad\quad 
+y^{-4}+3y^{-2}+4+3y^2+y^4
+(2y^{-2}+3+2y^2)x_1
+2x_1^2
\Bigr)x_2^2
+{\cal O}(x_2^3)
\Bigr],\\
&\hat{\cal G}_3^{(l=2)}(y;x_1;x_2)\nonumber \\
&=\frac{1}{(1-x_1^{-1})(1-x_1^{-2})(1-x_1^{-3})}\Bigl[
1 +
\Bigl((y^{-2}+1+y^2+x_1)x_1^{-2}
+(y^{-2}+2+y^2)x_1^{-1}\nonumber \\
&\quad 
+y^{-2}+2+y^2
+x_1\Bigr)x_2\nonumber \\
&\quad +\Bigl(
(y^{-4}+y^{-2}+1+y^2+y^4)x_1^{-4}
+(y^{-4}+2y^{-2}+2+2y^2+y^4)x_1^{-3}\nonumber \\
&\quad\quad
+(2y^{-4}+4y^{-2}+5+4y^2+2y^4)x_1^{-2}
+(y^{-4}+5y^{-2}+6+5y^2+y^4)x_1^{-1}\nonumber \\
&\quad\quad
+y^{-4}+4y^{-2}+7+4y^2+y^4
+(2y^{-2}+4+2y^2)x_1
+2x_1^2
\Bigr)x_2^2
+{\cal O}(x_2^3)
\Bigr],
\end{align}
\end{subequations}
and so on.
For $y=1$ we can also determine the higher order terms in $\hat{{\cal G}}_1^{(l=2)}(1;x_1;x_2)$ and $\hat{{\cal G}}_2^{(l=2)}(1;x_1;x_2)$ as
\begin{subequations}
\begin{align}
&\hat{{\cal G}}_1^{(l=2)}(1;x_1;x_2)\nonumber \\
&=
\frac{1}{1-x_1^{-1}}\Bigl[
1
+(3+x_1)x_2
+(5+3x_1+x_1^2)x_2^2
+(7+5x_1+3x_1^2+x_1^3)x_2^3\nonumber \\
&\quad +(9+7x_1+5x_1^2+3x_1^3+x_1^4)x_2^4
+{\cal O}(x_2^5)
\Bigr],\\
&\hat{{\cal G}}_2^{(l=2)}(1;x_1;x_2)\nonumber \\
&=\frac{1}{(1-x_1^{-1})(1-x_1^{-2})}\Bigl[
1
+(3x_1^{-1}+4+x_1)x_2
+(5x_1^{-2}+8x_1^{-1}+12+7x_1+2x_1^2)x_2^2\nonumber \\
&\quad +(7 x_1^{-3} + 12 x_1^{-2} + 20 x_1^{-1} + 29 + 23 x_1 + 11 x_1^2 + 2 x_1^3)x_2^3
+{\cal O}(x_2^4)\Bigr].
\end{align}
\end{subequations}
Interestingly, if we define ${\cal G}_{m'}^{(l=2)}(y;x_1;x_2)$ as
\begin{align}
{\cal G}_{m'}^{(l=2)}(y;x_1;x_2)=
\hat{\cal G}_{m'}^{(l=2)}(y;x_1^{-1};x_1x_2),
\end{align}
which is the same change of variables as \eqref{variablechange}, then these ${\cal G}_{m'}^{(l=2)}(y;x_1^{-1};x_1x_2)$ precisely agree with the small $x_2$ expansion of the Higgs indices ${\cal I}^{U(m')\text{ADHM-}[2]}_H(x,y;\mathfrak{t})$ with $x_1=x^{-1}\mathfrak{t}$ and $x_2=x\mathfrak{t}$.
This implies that the inverse giant graviton expansion works. 
It would be interesting to understand the inverse giant graviton from the gravity side through the expansion by analyzing the fluctuation modes on wrapped M2-branes, which was studied in e.g.~\cite{Gautason:2023igo,Beccaria:2023sph}.

Conversely, if we assume that the inverse giant graviton expansion works, we can determine the M5-brane indices ${\cal F}_m^{(l=2)}(y;x_1;x_2)$ directly from ${\cal I}^{U(N)\text{ADHM-}[2]}_H(x,y;\mathfrak{t})$ as
\begin{align}
{\cal F}_m^{(l=2)}(y;x_1;x_2)={\cal F}_\infty^{(l=2)}(y;x_1;x_2)\biggl(1+\sum_{n=1}^\infty x_1^{mn}\Bigl[{\cal I}^{U(n)\text{ADHM-}[2]}_H(x,y;\mathfrak{t})\Bigr]_{x_1\rightarrow x_1^{-1},x_2\rightarrow x_1x_2}\biggr).
\label{useinversegg}
\end{align}
Note that the coefficients of the single sum expansion of the Higgs indices ${\cal I}^{U(n)\text{ADHM-}[2]}_H(x,y;\mathfrak{t})$ has the following structure 
\begin{align}
\Bigl[{\cal I}^{U(n)\text{ADHM-}[2]}_H(x,y;\mathfrak{t})\Bigr]_{x_1\rightarrow x_1^{-1},x_2\rightarrow x_1x_2}=\sum_{j=0}^\infty x_2^j\times {\cal O}(x_1^{\frac{n(n+1)}{2}-nj+j}).
\end{align}
Therefore, in order to calculate the coefficient of $x_2^j$ in the M5-brane index ${\cal F}_m^{(l=2)}(y;x_1;x_2)$ to the order $x_1^\nu$ from the inverse expansion, 
it is sufficient to truncate the summation over $n$ on the right-hand side of \eqref{useinversegg} by the constraint
\begin{align}
mn+\frac{n(n+1)}{2}-nj+j\le \nu.
\label{boundofnbymjnu}
\end{align}
We confirmed that the giant graviton indices thus determined are indeed consistent with the coefficients $\hat{{\cal F}}^{(l=2)}_m(y;x_1;x_2)$ determined by the method explained in section \ref{app:howtodeterminehatcalFm} without assuming the inverse expansion.

\section{Single sum expansions for general $l$}
\label{sec_general_l}
Let us consider the single sum expansions of the Higgs indices for higher $l$. 
The gravity dual geometry contains the M5-brane giant gravitons wrapped around a supersymmetric $5$-cycle in $S^7/\mathbb{Z}_l$. 
The single sum expansions take the form 
\begin{align}
\frac{
{\cal I}^{U(N)\text{ADHM-[}l]}_H(x,y_\alpha;\mathfrak{t})
}{
{\cal I}^{U(\infty)\text{ADHM-[}l]}_H(x,y_\alpha;\mathfrak{t})
}
=
1+\sum_{m=1}^\infty x_1^{mN} \hat{{\cal F}}_m^{(l)}(y_\alpha;x_1;x_2).
\end{align}
The term $x_1^{mN}$ appears as a contribution from the ground state configuration associated with the supersymmetric surface wrapped by the M5-brane giant graviton. 
The expansion coefficients $\hat{{\cal F}}_m^{(l)}(y_\alpha;x_1;x_2)$ are identified with the index for the effective theory on the M5-brane giant gravitons of wrapping number $m$. 
As we see below, we find that our observations in the previous section for $\hat{{\cal F}}^{(l=2)}_m(y_\alpha;x_1;x_2)$ naturally extend to the cases with general $l$. 


Again, the index of the low-energy theory on a stack of $m$ M5-branes can be obtained from the giant graviton indices $\hat{{\cal F}}_m^{(l)}(y_\alpha;x_1;x_2)$ 
by a variable change of the fugacities $x_1$ and $x_2$
\begin{align}
{\cal F}_m^{(l)}(y_\alpha;x_1;x_2)
&=\hat{{\cal F}}_m^{(l)}(y_\alpha;x_1^{-1};x_1x_2). 
\end{align}

\subsection{A single M5-brane}
First let us look at the flavored index $\hat{{\cal F}}_1^{(l)}(y_\alpha;x_1;x_2)$ for a single M5-brane giant graviton. 
For $l=3$ we find that it can be expanded as 
\begin{align}
\label{calFv}
\hat{{\cal F}}_1^{(3)}(y_\alpha;x_1;x_2)
&=\frac{1}{(x_1^{-1};x_2)_\infty}\Biggl[
1
+\biggl(
2
+\frac{y_1}{y_2}
+\frac{y_2}{y_1}
+\frac{y_1}{y_3}
+\frac{y_2}{y_3}
+\frac{y_3}{y_1}
+\frac{y_3}{y_2}
\biggr)
x_2\nonumber \\
&\quad +\biggl(
5
+2\Bigl(
\frac{y_1}{y_2}
+\frac{y_2}{y_1}
+\frac{y_1}{y_3}
+\frac{y_2}{y_3}
+\frac{y_3}{y_1}
+\frac{y_3}{y_2}
\Bigr)
\biggr)
x_2^2\nonumber \\
&\quad +\biggl(
10
+5\Bigl(
 \frac{y_1}{y_2}
+\frac{y_2}{y_1}
+\frac{y_1}{y_3}
+\frac{y_2}{y_3}
+\frac{y_3}{y_1}
+\frac{y_3}{y_2}
\Bigr)\nonumber \\
&\quad\quad +\frac{y_1 y_2}{y_3^2}
+\frac{y_1^2}{y_2 y_3}
+\frac{y_2^2}{y_1 y_3}
+\frac{y_1 y_3}{y_2^2}
+\frac{y_2 y_3}{y_1^2}
+\frac{y_3^2}{y_1 y_2}
\biggr)
x_2^3
+{\cal O}(x_2^4)
\Biggr], 
\end{align}
with $y_1y_2y_3=1$. 
For $l=4$ we get
\begin{align}
\hat{{\cal F}}_1^{(4)}(y_\alpha;x_1;x_2)
&=\frac{1}{(x_1^{-1};x_2)_\infty}
\Biggl[
1+\left(3+\sum_{\alpha\neq \beta}^4 \frac{y_{\alpha}}{y_{\beta}}\right)x_2
\nonumber\\
&+\left(9+3\sum_{\alpha\neq \beta}^4 \frac{y_{\alpha}}{y_{\beta}} +
\sum_{\alpha<\beta}
\sum_{\substack{\gamma<\delta\\ (\gamma,\delta\neq \alpha,\beta)}}
\frac{y_\alpha y_\beta}{y_\gamma y_\delta}
\right)x_2^2+\mathcal{O}(x_2^3)
\Biggr], 
\end{align}
with $y_1y_2y_3y_4=1$. 

The factor $(x_1^{-1};x_2)_{\infty}^{-1}$ is the $1/4$-BPS index obtained from the twisted limit \cite{Hayashi:2024aaf} 
of the supersymmetric index \cite{Bhattacharya:2008zy} of the theory of a single M5-brane. 
We find that once the $1/4$-BPS index has been factored out 
the giant graviton index is given by the character of the vacuum module of the affine Kac-Moody algebra of type $A_{l-1}$ type at level $1$ 
\begin{align}
\label{GG_lm1}
\hat{{\cal F}}_1^{(l)}(y_\alpha;x_1;x_2)=\frac{\chi_{\widehat{\mathfrak{su}}(l)_1}(y_\alpha;x_2)}{(x_1^{-1};x_2)_\infty}, 
\end{align}
where 
\begin{align}
\label{ch_sulk1}
\chi_{\widehat{\mathfrak{su}}(l)_1}(y_\alpha;q)
&=\frac{1}{(q;q)_{\infty}^{l-1}}
\sum_{
\begin{smallmatrix}
m_1,\cdots,m_{l}\in \mathbb{Z}\\
m_1+\cdots+m_{l}=0\\
\end{smallmatrix}}
\frac{q^{\frac12\sum_{\alpha=1}^{l}m_{\alpha}^2}\prod_{\alpha=1}^{l} y_{\alpha}^{m_\alpha}}
{\prod_{\alpha<\beta}(q^{1\pm m_{\alpha}\mp m_{\beta}}y_{\alpha}^{\pm}y_{\beta}^{\mp};q)_{\infty}}, 
\end{align}
with $\prod_{\alpha}y_{\alpha}=1$. 
The emergent expression (\ref{ch_sulk1}) is the vacuum character of the affine Kac-Moody algebra $\widehat{\mathfrak{su}}(l)_{1}$ of level $1$. 
The formula (\ref{GG_lm1}) can be regarded as the promoted version of (\ref{hatcalFm1y1}) with the flavored fugacities $y_{\alpha}$ turned on.   

Recall that the configuration of the M5-brane wrapped on the $l$-centered Taub-NUT space $TN_{l}$ 
reduces on the $S^1$ fibration of the $TN_l$ to the configuration of the I-brane \cite{Green:1996dd,Bachas:1997kn,Itzhaki:2005tu} 
which contains the D4-brane intersecting with $l$ D6-branes in Type IIA string theory. 
The $4$-$6$ open strings suspended between the D4- and D6-branes give rise to the massless spectrum as two-dimensional free fermions on the I-brane 
transforming as $(+,\overline{\square})$ and $(-,\square)$ under the $U(1)\times U(l)$ gauge group, 
where $U(1)$ and $U(l)$ are the gauge group on the D4- and that on the D6-brane respectively. 
Such free fermions give a realization of the affine Kac-Moody algebra $\widehat{\mathfrak{u}}(l)_{1}$ at level $1$, 
however, from the perspective of the effective theory on the D4-brane, one needs to dynamically couple the $U(1)$ gauge field on the D4-branes. 
This means that we have the coset
\begin{align}
\label{coset_l_m1}
\frac{\widehat{\mathfrak{u}}(l)_1}{\widehat{\mathfrak{u}}(1)_{-l}}
&=\widehat{\mathfrak{su}}(l)_{1}. 
\end{align}
Note that the Chern-Simons term of level $-l$ of the $U(1)$ gauge field is required from the cancellation of the gauge anomaly.  
In fact, we can also reproduce the vacuum character (\ref{ch_sulk1}) from the coset (\ref{coset_l_m1}) 
by considering the $l$ fermions with the $U(1)$ gauge charge $+1$ 
\begin{align}
(q;q)_{\infty}\oint \frac{ds}{2\pi is}
\prod_{\alpha=1}^{l}
(q^{\frac12}s^{\pm}y_{\alpha}^{\pm};q)_{\infty}, 
\end{align}
with $\prod_{\alpha}y_{\alpha}=1$. 

\subsection{Multiple M5-branes}
For $m>1$ the giant graviton indices should capture the spectrum of a stack of the multiple M5-branes. 
Let us assume that the M5-brane index ${\cal F}_m^{(l)}(y_\alpha;x_1;x_2)$ is factorized as
\begin{align}
\label{M5ind_factorize}
{\cal F}_m^{(l)}(y_\alpha;x_1;x_2)
&=
\frac{1}{\prod_{n=1}^m(x_1^n;x_1x_2)_\infty}F_m^{(l)}(y_\alpha;x_1;x_2). 
\end{align}
Again the factor $\prod_{n=1}^m (x_1^n;x_1x_2)_\infty^{-1}$ is the twisted limit \cite{Hayashi:2024aaf} 
of the supersymmetric index of the 6d $(2,0)$ theory for $m$ M5-branes. 
It reduces to the unrefined index as the vacuum character of the of the W-algebra $\mathcal{W}(\mathfrak{gl}(m))$ \cite{Kim:2012ava,Beem:2014kka} when $x_2$ is set to unity. 
The other factor $F_m^{(l)}(y;x_1;x_2)$ will be contributed from the excitation modes associated with the I-brane configuration 
involving $m$ D4-branes intersecting with $l$ D6-branes. 
The small $x_2$ expansions of $F_m^{(l=3)}(y;x_1;x_2)$ and $F_m^{(l=4)}(y;x_1;x_2)$ are listed in Appendix \ref{app:listofFlm_l3withy} and \ref{app:listofFlm_l4withy}.

From the $4$-$6$ open strings suspended between the $m$ D4-branes and $l$ D6-branes one finds $lm$ two-dimensional complex free fermions 
transforming as the bifundamental representation $(\square,\overline{\square})$ and $(\overline{\square},\square)$ under the $U(m)\times U(l)$ gauge group. 
They carry charges $(+,-)$ under the two $U(1)$ factors, whereas the diagonal $U(1)$ is decoupled. 
Hence they effectively couple to $SU(m)\times SU(l)\times U(1)$ gauge group. 
The system of $lm$ complex free fermions can realize the affine Kac-Moody algebra $\widehat{\mathfrak{u}}(lm)_1$ at level $1$, 
for which we have the conformal embedding
\begin{align}
\label{emb_m_l}
\widehat{\mathfrak{su}}(m)_l
\oplus  
\widehat{\mathfrak{su}}(l)_m
\oplus  
\widehat{\mathfrak{u}}(1)_{lm}
\subset 
\widehat{\mathfrak{u}}(lm)_{1}, 
\end{align}
which preserve the central charge 
\begin{align}
\frac{l(m^2-1)}{(l+m)}
+\frac{m(l^2-1)}{(m+l)}
+1
&=lm. 
\end{align}
When the D4-branes are separated in the transverse directions, the $SU(m)$ gauge group is broken down to $U(1)^{m-1}$. 
In this case, the free fermions may couple to $U(1)^{m-1}\times SU(l)\times U(1)$ gauge group 
so that the current algebra built from bilinear currents of the free fermions split into \cite{Itzhaki:2005tu}
\begin{align}
\label{emb_separate_m_l}
(\widehat{\mathfrak{u}}(1)_l)^{m-1}
\oplus  
(\widehat{\mathfrak{su}}(l)_1)^{m}
\oplus  
\widehat{\mathfrak{u}}(1)_{lm}
\subset 
\widehat{\mathfrak{u}}(lm)_{1}, 
\end{align}
where the coset factors have been replaced according to the relation \cite{DiFrancesco:1997nk}
\begin{align}
\frac{\widehat{\mathfrak{su}}(m)_l}{(\widehat{\mathfrak{u}}(1)_l)^{m-1}}
&=\frac{(\widehat{\mathfrak{su}}(l)_1)^{m}}{\widehat{\mathfrak{su}}(l)_{m}}. 
\end{align}
Note that the affine embedding (\ref{emb_separate_m_l}) also keeps the central charge
\begin{align}
(m-1)+m\frac{l^2-1}{1+l}+1&=lm. 
\end{align}
As discussed for $l=2$, the free fermions are required to couple dynamical gauge fields. 
While they dynamically couple to $U(1)^{m-1}\times U(1)$ gauge fields living on a stack of $m$ M5-branes, 
the $SU(l)$ gauge group that is associated with the fixed multi-centered Taub-NUT geometry in M-theory should not be dynamical. 
Alternatively, from the view point of the effective theory on a stack of $m$ D4-branes in the I-brane setup, 
$SU(l)$ gauge field on the $l$ non-compact D6-branes should not be dynamical.
So the effective theory on a stack of $m$ M5-branes will be describe by the coset 
\begin{align}
\label{coset_l_m}
\frac{(\widehat{\mathfrak{u}}(1)_{l})^{m-1}
\oplus  
\widehat{\mathfrak{u}}(1)_{lm}
\oplus  
(\widehat{\mathfrak{su}}(l)_{1})^{m}}
{(\widehat{\mathfrak{u}}(1)_{l})^{m-1}
\oplus  
\widehat{\mathfrak{u}}(1)_{lm}}. 
\end{align}
The $U(1)/U(1)$ coset models are the topological field theories \cite{Witten:1991mm,Spiegelglas:1992jg}. 
The physical states accounted by the $U(1)/U(1)$ character does not contain any descendant states \cite{Spiegelglas:1992jg}. 
Therefore the actual contributions to the index will be given by the product of $m$ characters of the affine Kac-Moody algebra $\widehat{\mathfrak{su}}(l)_1$. 
Indeed, we observe that if we set $x_1=1$, then 
\begin{align}
\label{x11property_genetal}
F_m^{(l)}(y_{\alpha};1;x_2)=\chi_{\widehat{\mathfrak{su}}(l)_1}(y_{\alpha};x_2)^m. 
\end{align}
\begin{table}
\begin{center}
{\fontsize{9pt}{1pt}\selectfont
\begin{align*}
\begin{array}{|c|l|}
\hline
&\\[3pt]
m&\lim_{\substack{x_1\rightarrow 0,x_2\rightarrow\infty \\(x_1x_2=1\text{: fixed})}}{\cal F}^{(l=3)}_m(x_1;x_2)\\ \hline
&\\[3pt]
2&1 +8   q +44  q^2 +128 q^3 +376 q^4 +944 q^5 +2236q^6 +{\cal O}(q^7)\\ \hline
&\\[3pt]
3&1 +8   q +44  q^2 +192 q^3 +601 q^4 +1772q^5 +4668 q^6 +{\cal O}(q^7)\\ \hline
&\\[3pt]
4&1 +8    q +44   q^2 +192  q^3 +726  q^4 +2248 q^5 +6496 q^6 +{\cal O}(q^7)\\ \hline
&\\[3pt]
5&1 +8   q +44  q^2 +192 q^3 +726 q^4 +2464q^5 +7361q^6 +{\cal O}(q^7)\\ \hline
&\\[3pt]
m\ge 6&1 +8   q +44  q^2 +192 q^3 +726 q^4 +2464q^5 +7704q^6 +{\cal O}(q^7)\\ \hline
\end{array}
\end{align*}
\begin{align*}
\begin{array}{|c|l|}
\hline
&\\[3pt]
m&\lim_{\substack{x_1\rightarrow 0,x_2\rightarrow\infty \\(x_1x_2=1\text{: fixed})}}{\cal F}^{(l=3)}_m(y_\alpha;x_1;x_2)\\ \hline
&\\[3pt]
2&1 +(2 + A^{(3)}_1)q +(8 + 4A^{(3)}_1 + A^{(3)}_2 + B^{(3)}_2)q^2 +(20 + 12A^{(3)}_1 + 2A^{(3)}_2 + 4B^{(3)}_2)q^3 +{\cal O}(q^4)\\ \hline
&\\[3pt]
m\ge 3&1 +(2 + A^{(3)}_1)q +(8 + 4A^{(3)}_1 + A^{(3)}_2 + B^{(3)}_2)q^2 +(24 + 15A^{(3)}_1 + 4A^{(3)}_2 + A^{(3)}_3 + 6B^{(3)}_2 + B^{(3)}_3)q^3\\
&\\[3pt]
&+{\cal O}(q^4)\\ \hline
\end{array}
\end{align*}
\begin{align*}
\begin{array}{|c|l|}
\hline
&\\[3pt]
m&\lim_{\substack{x_1\rightarrow 0,x_2\rightarrow\infty \\(x_1x_2=1\text{: fixed})}}{\cal F}^{(l=4)}_m(y_\alpha;x_1;x_2)\\ \hline
&\\[3pt]
2&1 +(3+A^{(4)}_1)q +(15+6A^{(4)}_1+A^{(4)}_2+2B^{(4)}_2+C^{(4)}_2)q^2\\
&\\[3pt]
& +(50+25A^{(4)}_1+3A^{(4)}_2+11B^{(4)}_2+6C^{(4)}_2+B^{(4)}_3)q^3 +{\cal O}(q^4)\\ \hline
&\\[3pt]
m\ge 3&1 +(3+A^{(4)}_1)q +(15+6A^{(4)}_1+A^{(4)}_2+2B^{(4)}_2+C^{(4)}_2)q^2\\
&\\[3pt]
&+(60+31A^{(4)}_1+6A^{(4)}_2+A^{(4)}_3+14B^{(4)}_2+9C^{(4)}_2+2B^{(4)}_3+C^{(4)}_3+D^{(4)}_3)q^3 +{\cal O}(q^4)\\ \hline
\end{array}
\end{align*}
}
\caption{
Coefficients of the monomials $q^j$ $=$ $(x_1x_2)^j$ in ${\cal F}^{(l=3)}_m(y_\alpha;x_1;x_2)$ and ${\cal F}^{(l=4)}_m(y_\alpha;x_1;x_2)$.
To the order $q^m$, the coefficients of $q^j$ in ${\cal F}^{(l)}_{m'}(y_\alpha;x_1;x_2)$ ($l=3,4$) with $m'\ge m$ coincides with those in ${\cal F}^{(l)}_m(y_\alpha;x_1;x_2)$.
}
\label{x1^0(x1x2)^j_l3l4_y1andwithy}
\end{center}
\end{table}
The formula (\ref{x11property_genetal}) would imply that 
the index of the M5-brane giant gravitons of wrapping number $m$ contains 
the excitation modes accounted by the $1/4$-BPS M5-brane index 
and the contributions from the spectrum of the I-brane with $m$ separated M5-branes described by the coset (\ref{coset_l_m}). 

Another interesting property is that 
the full M5-brane index $\mathcal{F}_m^{(l)}(y_{\alpha};x_1;x_2)$ admits a different specialization of the fugacities 
for which it reduces to the character of the affine Kac-Moody algebra. 
Keeping $q$ $=$ $x_1 x_2$ fixed and taking $x_1$ $\rightarrow$ $0$, $x_2$ $\rightarrow$ $\infty$, 
we find
the M5-brane indices, which are given by \eqref{M5ind_factorize} with \eqref{listofFlm_l3withy}, \eqref{listofFlm_l3y1higherorderonly} and \eqref{listofFlm_l4withy}, reduce as displayed in Table \ref{x1^0(x1x2)^j_l3l4_y1andwithy}. Here the results are written in terms of $A_i^{(l)}$, $B_i^{(l)}$ and $C_i^{(l)}$ defined by \eqref{ABCl3} and \eqref{ABCl4}.
In this limit, we find that
\begin{align}
\label{lim_F_m^l}
\lim_{\substack{x_1\rightarrow 0,x_2\rightarrow\infty\\ (x_1x_2=q\text{: fixed})}}
\mathcal{F}_m^{(l)}(y_\alpha;x_1;x_2)=\chi_{\widehat{\mathfrak{su}}(l)_m}(y_\alpha;q). 
\end{align}
Here 
\begin{align}
\label{ch_sulk}
\chi_{\widehat{\mathfrak{su}}(l)_m}(y_\alpha;q)
&=\frac{1}{(q;q)_{\infty}^{l-1}}
\sum_{
\begin{smallmatrix}
m_1,\cdots,m_{l}\in \mathbb{Z}\\
m_1+\cdots+m_{l}=0\\
\end{smallmatrix}}
\frac{q^{\frac{m}{2}\sum_{\alpha=1}^{l}m_{\alpha}^2}\prod_{\alpha=1}^{l} y_{\alpha}^{m m_{\alpha}}}
{\prod_{\alpha<\beta}(q^{1\pm m_{\alpha}\mp m_{\beta}}y_{\alpha}^{\pm}y_{\beta}^{\mp};q)_{\infty}}, 
\end{align}
where $\prod_{\alpha}y_{\alpha}=1$, is the vacuum character of the affine Kac-Moody algebra $\widehat{\mathfrak{su}}(l)_m$ of level $m$. 

Alternatively, the expression can be also obtained from the 
graded character of the $lm$ free fermions coupled to the $U(m)_{-l}$ Chern-Simons theory with level $-l$ 
by taking into account the level-rank duality \cite{MR1152377}
\begin{align}
\frac{1}{m!}(q;q)_{\infty}^m 
\oint \prod_{i=1}^m \frac{ds_i}{2\pi is_i}
\prod_{i<j}(s_i^{\pm}s_j^{\mp};q)_{\infty}
\prod_{i=1}^m
\prod_{\alpha=1}^l (q^{\frac12}s_i^{\pm}y_{\alpha}^{\pm};q)_{\infty}. 
\end{align}

The fact that the full index $\mathcal{F}_m^{(l)}(y_{\alpha};x_1;x_2)$ for the theory of $m$ M5-branes on $\mathbb{C}^2/\mathbb{Z}_l$ admits the special fugacity limit 
(\ref{lim_F_m^l}) in which it reduces to the vacuum character of the affine Kac-Moody algebra $\widehat{\mathfrak{su}}(l)_m$ 
is analogous to the proposal in \cite{Nishioka:2011jk} that the affine Kac-Moody algebra $\widehat{\mathfrak{su}}(l)_m$ 
appears from $m$ M5-branes on $\mathbb{C}^2/\mathbb{Z}_l$ with $\Omega$-deformation. 
Here we remark that the full M5-brane index $\mathcal{F}_m^{(l)}(y;x_1,x_2)$ contains extra excitation modes 
which may carry the non-trivial charges coupled to the fugacity $x_1$. 
While the factorized form (\ref{M5ind_factorize}) reveals the non-trivial Kac-Moody algebraic structure as in (\ref{x11property_genetal}), 
it may not be convenient to see the exact form of the full index. 
In fact, it is suggested \cite{Nishioka:2011jk} that 
there is a further action of the para-$\mathcal{W}$ algebra \cite{Christe:1988vc,Bowcock:1988vs,MR1062148,Ahn:1990gn,Bernard:1990ti,Nemeschansky:1991pr,LeClair:1992xi} 
on the world-volume theory of $m$ M5-branes on $\mathbb{C}^2/\mathbb{Z}_l$ with $\Omega$-deformation. 
It would be an interesting future work to shed light on the exact closed-form expression for the full M5-brane index. 

\subsection{Inverse giant graviton expansion}
\label{sec_inversegggenerall}
Next let us look at the inverse giant graviton expansion, 
i.e.~the giant graviton expansion of the index ${\cal F}_m^{(l)}(y_\alpha;x_1;x_2)$ for a stack of $m$ M5-branes. 
Given that the index is factorized as (\ref{M5ind_factorize}), 
the $1/4$-BPS index simply becomes $\prod_{n=1}^{\infty}(x_1^n;x_1x_2)_{\infty}^{-1}$ in the large $m$ limit, 
which is equal to the twisted limit \cite{Hayashi:2024aaf} of the index \cite{Bhattacharya:2008zy} for the KK modes on the $AdS_7\times S^4$. 
Also, as seen from Table \ref{x1^0(x1x2)^j_l3l4_y1andwithy}, ${\cal F}_m^{(l)}(y_{\alpha};x_1;x_2)$ admits the large $m$ limit. 
In the special fugacity limit where we send $x_1$ to $0$ by keeping $x_1x_2=q$ finite, 
it is shown to become
\begin{align}
\frac{(q ;q)_{\infty}}{\prod_{\alpha,\beta=1}^l (q\frac{y_{\alpha}}{y_{\beta}};q)_{\infty}}. 
\end{align}
More generally, we find that the large $m$ limit of the M5-brane index ${\cal F}_m^{(l)}(y_\alpha;x_1;x_2)$ is given by
\begin{align}
{\cal F}_\infty^{(l)}(y_\alpha;x_1;x_2)=\prod_{n=1}^\infty \frac{1}{(x_1^n;x_1x_2)_\infty}
\frac{(x_1^nx_2;x_1x_2)_{\infty}}{\prod_{\alpha,\beta=1}^l(x_1^n x_2 \frac{y_{\alpha}}{y_{\beta}};x_1x_2)_{\infty}}.
\end{align}
We have confirmed for $l=3$ to the order ${\cal O}(\mathfrak{t}^6)$ and for $l=4$ to the order ${\cal O}(\mathfrak{t}^4)$ in the small $\mathfrak{t}$ expansion with $x$ fixed for $x_1=x^{-1}\mathfrak{t}$ and $x_2=x\mathfrak{t}$, by using the data of the full M5-brane indices including ${\cal F}_5^{(l=3)}(y_\alpha;x_1;x_2)$ to the order ${\cal O}(x_2^3)$ and ${\cal F}_3^{(l=4)}(y_\alpha;x_1;x_2)$ to the order ${\cal O}(x_2^2)$.
For $y_\alpha=1$, the formula has been confirmed for $l=3$ to the order ${\cal O}(\mathfrak{t}^8)$ and for $l=4$ to the order ${\cal O}(\mathfrak{t}^6)$ by using the data including ${\cal F}_7^{(l=3)}(1;x_1;x_2)$ to the order ${\cal O}(x_2^4)$ and ${\cal F}_5^{(l=4)}(1;x_1;x_2)$ to the order ${\cal O}(x_2^3)$.\footnote{
As summarized in Table \ref{listofjmax}, these data were obtained without assuming inverse giant graviton expansion.
}
The single particle index is
\begin{align}
f_{\infty}^{(l)}(y_{\alpha};x_1;x_2)
&=\frac{x_1+x_1 x_2 \chi_{\textrm{adj}}^{SU(l)}(y_\alpha)}{(1-x_1)(1-x_1x_2)}, 
\end{align}
where $\chi_{\textrm{adj}}^{SU(l)}(y_\alpha)$ is the adjoint character of $SU(l)$. 

As we have done for $l=2$ in section \ref{sec_inversegg}, let us consider the finite $m$ corrections of ${\cal F}_m^{(l)}(y;x_1;x_2)$ in the single sum expansion
\begin{align}
\frac{{\cal F}_m^{(l)}(y;x_1;x_2)}{{\cal F}_\infty^{(l)}(y;x_1;x_2)}=1+\sum_{m'=1}^\infty x_1^{mm'}\hat{{\cal G}}^{(l)}_{m'}(y;x_1;x_2),
\end{align}
where $\hat{{\cal G}}^{(l)}_{m'}(y;x_1;x_2)$ are determined from the data ${\cal F}^{(l)}_m(y;x_1;x_2)$ as
\begin{subequations}
\begin{align}
&\hat{{\cal G}}_1^{(l=3)}(y_\alpha;x_1;x_2)\nonumber \\
&=\frac{1}{1-x_1^{-1}}\Bigl[
1 + (2 + A^{(3)}_1 + x_1)x_2 + \Bigl(3 + 2A^{(3)}_1 + A^{(3)}_2 + B^{(3)}_2 + (2 + A^{(3)}_1)x_1 + x_1^2\Bigr)x_2^2\nonumber \\
&\quad +{\cal O}(x_2^3)\Bigr],\\
&\hat{{\cal G}}_2^{(l=3)}(y_\alpha;x_1;x_2)\nonumber \\
&=\frac{1}{(1-x_1^{-1})(1-x_1^{-2})}\Bigl[
1
+ \Bigl((2 + A^{(3)}_1) x_1^{-1} + 3 + A^{(3)}_1 + x_1\Bigr)x_2\nonumber \\
&\quad + \Bigl((3 + 2 A^{(3)}_1 + A^{(3)}_2 + B^{(3)}_2) x_1^{-2} + (7 + 5 A^{(3)}_1 + A^{(3)}_2 + 2 B^{(3)}_2)  x_1^{-1}\nonumber \\
&\quad\quad + 10 + 5 A^{(3)}_1 + A^{(3)}_2 + B^{(3)}_2 + (5 + 2 A^{(3)}_1) x_1 + 2 x_1^2\Bigr)x_2^2
+{\cal O}(x_2^3)\Bigr],\\
&\hat{{\cal G}}_3^{(l=3)}(y_\alpha;x_1;x_2)\nonumber \\
&=\frac{1}{(1-x_1^{-1})(1-x_1^{-2})(1-x_1^{-3})}\Bigl[
1
+ \Bigl((2 + A^{(3)}_1) x_1^{-2} + (3 + A^{(3)}_1) x_1^{-1} + 3 + A^{(3)}_1 + x_1\Bigr)x_2\nonumber \\
&\quad
+ \Bigl((3 + 2 A^{(3)}_1 + A^{(3)}_2 + B^{(3)}_2) x_1^{-4} + (7 + 5 A^{(3)}_1 + A^{(3)}_2 + 2 B^{(3)}_2)  x_1^{-3}\nonumber \\
&\quad\quad + (15 + 9 A^{(3)}_1 + 2 A^{(3)}_2 + 3 B^{(3)}_2)  x_1^{-2} + (16 + 9 A^{(3)}_1 + A^{(3)}_2 + 2 B^{(3)}_2) x_1^{-1}\nonumber \\
&\quad\quad + 14 + 6 A^{(3)}_1 + A^{(3)}_2 + B^{(3)}_2 + (6 + 2 A^{(3)}_1) x_1 + 2 x_1^2\Bigr)x_2^2 
+{\cal O}(x_2^3)\Bigr],\\
&\hat{{\cal G}}_1^{(l=4)}(y_\alpha;x_1;x_2)\nonumber \\
&=\frac{1}{1-x_1^{-1}}\Bigl[
1
+ (3 + A^{(4)}_1 + x_1)x_2\nonumber \\
&\quad
+ \Bigl(6 + 3A^{(4)}_1 + A^{(4)}_2 + B^{(4)}_2 + C^{(4)}_2 + (3 + A^{(4)}_1)x_1 + x_1^2\Bigr)x_2^2
+{\cal O}(x_2^3)\Bigr],\\
&\hat{{\cal G}}_2^{(l=4)}(y_\alpha;x_1;x_2)\nonumber \\
&=\frac{1}{(1-x_1^{-1})(1-x_1^{-2})}\Bigl[
1
+ \Bigl((3 + A^{(4)}_1)x_1^{-1} + 4 + A^{(4)}_1 + x_1\Bigr)x_2\nonumber \\
&\quad + \Bigl((6 + 3A^{(4)}_1 + A^{(4)}_2 + B^{(4)}_2 + C^{(4)}_2) x_1^{-2}
+ (15 + 8A^{(4)}_1 + A^{(4)}_2 + 3B^{(4)}_2 + 2C^{(4)}_2)x_1^{-1}\nonumber \\
&\quad\quad + 18 + 7A^{(4)}_1 + A^{(4)}_2 + 2B^{(4)}_2 + C^{(4)}_2 + (7 + 2A^{(4)}_1)x_1 + (2)x_1^2\Bigr)x_2^2
+{\cal O}(x_2^3)\Bigr],\\
&\hat{{\cal G}}_3^{(l=4)}(y_\alpha;x_1;x_2)\nonumber \\
&=\frac{1}{(1-x_1^{-1})(1-x_1^{-2})(1-x_1^{-3})}\Bigl[
1
+ \Bigl((3 + A^{(4)}_1)x_1^{-2} + (4 + A^{(4)}_1)x_1^{-1} + 4 + A^{(4)}_1 + x_1\Bigr)x_2\nonumber \\
&\quad + \Bigl((6 + 3A^{(4)}_1 + A^{(4)}_2 + B^{(4)}_2 + C^{(4)}_2) x_1^{-4} + (15 + 8A^{(4)}_1 + A^{(4)}_2 + 3B^{(4)}_2 + 2C^{(4)}_2) x_1^{-3}\nonumber \\
&\quad\quad + (30 + 14A^{(4)}_1 + 2A^{(4)}_2 + 5B^{(4)}_2 + 3C^{(4)}_2) x_1^{-2}
+ (30 + 13A^{(4)}_1 + A^{(4)}_2 + 4B^{(4)}_2 + 2C^{(4)}_2)x_1^{-1}\nonumber \\
&\quad\quad + 23 + 8A^{(4)}_1 + A^{(4)}_2 + 2B^{(4)}_2 + C^{(4)}_2 + (8 + 2A^{(4)}_1)x_1 + 2x_1^2\Bigr)x_2^2
+{\cal O}(x_2^3)\Bigr],
\end{align}
\end{subequations}
and so on.
Here
$A^{(l)}_i$, $B^{(l)}_i$ and $C^{(4)}_2$ are defined as \eqref{ABCl3} and \eqref{ABCl4}.
For $y_\alpha=1$ we can also determine the higher order coefficients as
\begin{subequations}
\begin{align}
&\hat{{\cal G}}_1^{(l=3)}(1;x_1;x_2)\nonumber \\
&=\frac{1}{1-x_1^{-1}}\Bigl[
1
+ (8 + x_1) x_2
+ (27 + 8 x_1 + x_1^2)  x_2^2
+ (64 + 27 x_1 + 8 x_1^2 + x_1^3)  x_2^3\nonumber \\
&\quad
+ (125 + 64 x_1 + 27 x_1^2 + 8 x_1^3 + x_1^4) x_2^4
+{\cal O}(x_2^5)\Bigr],\\
&\hat{{\cal G}}_2^{(l=3)}(1;x_1;x_2)\nonumber \\
&=\frac{1}{(1-x_1^{-1})(1-x_1^{-2})}\Bigl[
1
+ (8 x_1^{-1} + 9 + x_1) x_2
+ (27x_1^{-2} + 55 x_1^{-1} + 52 + 17 x_1 + 2 x_1^2)  x_2^2\nonumber \\
&\quad
+ (64x_1^{-3} + 161x_1^{-2} + 243 x_1^{-1} + 237 + 115 x_1 + 26 x_1^2 + 2 x_1^3) x_2^3
+{\cal O}(x_2^4)\Bigr],\\
%
%
%
&\hat{{\cal G}}_1^{(l=4)}(1;x_1;x_2)\nonumber \\
&=\frac{1}{1-x_1^{-1}}\Bigl[
1 + (15 + x_1) x_2 + (84 + 15 x_1 + x_1^2)  x_2^2 + (300 + 84 x_1 + 15 x_1^2 + x_1^3) x_2^3
+{\cal O}(x_2^4)\Bigr],\\
&\hat{{\cal G}}_2^{(l=4)}(1;x_1;x_2)\nonumber \\
&=\frac{1}{(1-x_1^{-1})(1-x_1^{-2})}\Bigl[
1
+ (15x_1^{-1} + 16 + x_1) x_2
+ (84x_1^{-2} + 189x_1^{-1} + 150 + 31 x_1 + 2 x_1^2) x_2^2\nonumber \\
&\quad
+ (300x_1^{-3} + 896x_1^{-2} + 1344x_1^{-1} + 1057 + 354 x_1 + 47 x_1^2 + 2 x_1^3)x_2^3
+{\cal O}(x_2^4)\Bigr].
%
\end{align}
\end{subequations}
After the change of variables ${\cal G}_{m'}^{(l)}(y_\alpha;x_1;x_2)=\hat{\cal G}_{m'}^{(l)}(y_\alpha;x_1^{-1};x_1x_2)$, we find that ${\cal G}_{m'}^{(l)}(y_\alpha;x_1;x_2)$ precisely coincide with the small $x_2$ expansion of the Higgs indices ${\cal I}^{U(m')\text{ADHM-}[l]}_H(x,y_\alpha;\mathfrak{t})$ with $x_1=x^{-1}\mathfrak{t}$ and $x_2=x\mathfrak{t}$.
This strongly suggests that the inverse giant graviton expansion works for general $l$ flavors as well.

As we mentioned in section \ref{sec_inversegg}, once we assume that the inverse giant graviton expansion holds, it provides new constraints to determine the M5-brane indices.
We observe that the leading powers in $x_1$ of each coefficient of the small $x_2$ expansion of the Higgs indices ${\cal I}^{U(n)\text{ADHM-}[l]}_H(x,y_\alpha;\mathfrak{t})$ for general $l$ are also given by \eqref{boundofnbymjnu}.
Therefore by following the same argument as in section \ref{sec_inversegg}, we can calculate the coefficient of $x_2^j$ in ${\cal F}_m^{(l)}(y_\alpha;x_1;x_2)$ to the order $x_1^\nu$ directly from the Higgs indices by
\begin{align}
&{\cal F}_m^{(l)}(y_\alpha;x_1;x_2)\nonumber \\
&={\cal F}_\infty^{(l)}(y_\alpha;x_1;x_2)\biggl(1+\sum_{\substack{n\ge 1\\
(mn+\frac{n(n+1)}{2}-nj+j\le \nu)}} x_1^{mn}\Bigl[{\cal I}^{U(n)\text{ADHM-}[l]}_H(x,y_\alpha;\mathfrak{t})\Bigr]_{x_1\rightarrow x_1^{-1},x_2\rightarrow x_1x_2}\biggr)\nonumber \\
&\quad +\cdots.
\end{align}
We confirmed that the giant graviton indices thus determined are indeed consistent with the coefficients $\hat{{\cal F}}^{(l)}_m(y_\alpha;x_1;x_2)$ determined by the method explained in section \ref{app:howtodeterminehatcalFm} without assuming the inverse expansion.


\acknowledgments

The authors would like to thank Masahide Manabe and Douglas J.~Smith for useful discussions and comments.
The work of H.H.~is supported in part by JSPS KAKENHI Grand Number JP23K03396.
The work of T.N.~is supported by the Startup Funding no.~2302-SRFP-2024-0012 of Shanghai Institute for Mathematics and Interdisciplinary Sciences.
The work of T.O.~was supported by the Startup Funding no.~4007012317 of the Southeast University.

\appendix

\section{List of $F^{(l)}_m(y_\alpha;x_1;x_2)$}
\label{app:listofFlm}

In this appendix we list the giant graviton indices $\hat{{\cal F}}_m^{(l)}(y_\alpha;x_1,x_2)$ obtained by the analysis explained in section \ref{app:howtodeterminehatcalFm} and also by using the inverse giant graviton expansion explained in section \ref{sec_inversegg}.
The results are displayed in terms of the function $F^{(l)}_m(y_\alpha;x_1;x_2)$ defined as \eqref{Fml}, namely,
\begin{align}
F_m^{(l)}(y_\alpha;x_1;x_2)=\hat{{\cal F}}_m^{(l)}(y_\alpha;x_1^{-1},x_1x_2)\prod_{n=1}^m(x_1^n;x_1x_2)_\infty.
\end{align}

\subsection{$l=2$}
\label{app:listofFlm_l2withy}

We find that $F^{(l=2)}_m(y;x_1;x_2)$ are given as
{\fontsize{9pt}{1pt}\selectfont
\begin{subequations}
\label{listofFlm_l2withy}
\begin{align}
&F_1^{(l=2)}(y;x_1;x_2)=
1
+(y^{-2}+1+y^2)x_1x_2
+(y^{-2}+2+y^2)(x_1x_2)^2
+(2y^{-2}+3+2 y^2)(x_1x_2)^3\nonumber \\
&\quad
+(y^{-4}+3y^{-2}+5+3 y^2+y^4)(x_1x_2)^4
+(y^{-4}+5y^{-2}+7+5 y^2+y^4)(x_1x_2)^5\nonumber \\
&\quad
+(2y^{-4}+7y^{-2}+11+7 y^2+2 y^4)(x_1x_2)^6
+(3y^{-4}+11y^{-2}+15+11 y^2+3 y^4)(x_1x_2)^7
+{\cal O}(x_2^8),\\
&F_2^{(l=2)}(y;x_1;x_2)=
1
+\Bigl[
y^{-2}
+1
+y^2
+(
y^{-2}
+1
+y^2)
x_1
\Bigr]    x_1x_2\nonumber \\
&\quad +\Bigl[
y^{-4}
+2y^{-2}
+3
+2 y^2
+y^4
+x_1 (
2y^{-2}
+3
+2 y^2)
+x_1^2
\Bigr]    (x_1x_2)^2\nonumber \\
&\quad
+\Bigl[
y^{-4}
+4y^{-2}
+5
+4 y^2
+y^4
+(
y^{-4}
+5y^{-2}
+7
+5 y^2
+y^4)x_1
+(
y^{-2}
+2
+y^2)x_1^2 
\Bigr]    (x_1x_2)^3\nonumber \\
&\quad
+\Bigl[
3y^{-4}
+7y^{-2}
+10
+7 y^2
+3 y^4
+(
3y^{-4}
+10y^{-2}
+14
+10 y^2
+3 y^4)x_1 \nonumber \\
&\quad\quad+(
y^{-4}
+3y^{-2}
+6
+3 y^2
+y^4)x_1^2 
\Bigr]    (x_1x_2)^4\nonumber \\
&\quad
+\Bigl[
y^{-6}
+5y^{-4}
+13y^{-2}
+ 16
+13 y^2
+5 y^4
+y^6
+(
y^{-6}
+7y^{-4}
+20y^{-2}
+27
+20 y^2
+7 y^4
+y^6)x_1\nonumber \\
&\quad\quad +(
2y^{-4}
+8y^{-2}
+12
+8 y^2
+2 y^4)x_1^2 
+(
y^{-2}
+1
+y^2)x_1^3 
\Bigr]    (x_1x_2)^5\nonumber \\
&\quad +\Bigl[
2y^{-6}+10y^{-4}+21y^{-2}+28+21 y^2+10 y^4+2 y^6
+(2y^{-6}+14y^{-4}+36y^{-2}+48+36 y^2+14 y^4+2 y^6)x_1\nonumber \\
&\quad\quad +(6y^{-4}+17y^{-2}+26+17 y^2+6 y^4)x_1^2 
+(2y^{-2}+3+2 y^2)x_1^3 
\Bigr](x_1x_2)^6
+{\cal O}(x_2^7),\\
&F_3^{(l=2)}(y;x_1;x_2)=
1
+\Bigl[
y^{-2}
+1
+y^2
+(
y^{-2}
+1
+y^2)x_1 
+(
y^{-2}
+1
+y^2)x_1^2 
\Bigr]    x_1x_2\nonumber \\
&\quad
+\Bigl[
y^{-4}
+2y^{-2}
+3
+2 y^2
+y^4
+(
y^{-4}
+3y^{-2}
+4
+3 y^2
+y^4)x_1 
+(
y^{-4}
+3y^{-2}
+5
+3 y^2
+y^4)x_1^2\nonumber \\
&\quad\quad
+(
y^{-2}
+2
+y^2)x_1^3 
+x_1^4
\Bigr]    (x_1x_2)^2\nonumber \\
&\quad
+\Bigl[
y^{-6}
+2y^{-4}
+5y^{-2}
+6
+5 y^2
+2 y^4
+y^6
+(
3y^{-4}
+8y^{-2}
+10
+8 y^2
+3 y^4)x_1 
\nonumber \\
&\quad\quad
+(
3y^{-4}
+10y^{-2}
+13
+10 y^2
+3 y^4)x_1^2
+(
y^{-4}
+5y^{-2}
+8
+5 y^2
+y^4)x_1^3 
+(
2y^{-2}
+3
+2 y^2)x_1^4 
\Bigr]    (x_1x_2)^3\nonumber \\
&\quad
+\Bigl[
y^{-6}
+5y^{-4}
+9y^{-2}
+12
+9 y^2
+5 y^4
+y^6
+(
y^{-6}
+8y^{-4}
+18y^{-2}
+23
+18 y^2
+8 y^4
+y^6)x_1 \nonumber \\
&\quad\quad
+(
y^{-6}
+10y^{-4}
+24y^{-2}
+33
+24 y^2
+10 y^4
+y^6)x_1^2
+(
5y^{-4}
+16y^{-2}
+23
+16 y^2
+5 y^4)x_1^3 \nonumber \\
&\quad\quad
+(
2y^{-4}
+7y^{-2}
+12
+7 y^2
+2 y^4)x_1^4 
+(
y^{-2}
+2
+y^2)x_1^5 
\Bigr]    (x_1x_2)^4\nonumber \\
&\quad +
\Bigl[
3y^{-6} +9y^{-4} +18y^{-2} +21 +18y^2 +9y^4 +3y^6
+( 4y^{-6} +18y^{-4} +38y^{-2} +47 +38y^2 +18y^4 +4y^6)x_1\nonumber \\
&\quad\quad +( 5y^{-6} +24y^{-4} +56y^{-2} +71 +56y^2 +24y^4 +5y^6)x_1^2\nonumber \\
&\quad\quad
+( 2y^{-6} +16y^{-4} +43y^{-2} +58 +43y^2 +16y^4 +2y^6)x_1^3\nonumber \\
&\quad\quad
+( y^{-6} +7y^{-4} +24y^{-2} +33 +24y^2 +7y^4 +y^6)x_1^4
+( y^{-4} +6y^{-2} +9 +6y^2 +y^4)x_1^5\nonumber \\
&\quad\quad
+( y^{-2} +1 +y^2)x_1^6 \Bigr]    (x_1x_2)^5\nonumber \\
&\quad
+\Bigl[
y^{-8} +6y^{-6} +18y^{-4} +31y^{-2} +39 +31 y^2 +18 y^4 +6 y^6 +y^8\nonumber \\
&\quad\quad
+( y^{-8} +10y^{-6} +38y^{-4} +74y^{-2} +91 +74 y^2 +38 y^4 +10 y^6 +y^8)    x_1   \nonumber \\
&\quad\quad +( y^{-8} +13y^{-6} +56y^{-4} +115y^{-2} +147 +115 y^2 +56 y^4 +13 y^6 +y^8) x_1^2 \nonumber \\
&\quad\quad +( 8y^{-6} +43y^{-4} +101y^{-2} +133 +101 y^2 +43 y^4 +8 y^6)                   x_1^3 \nonumber \\
&\quad\quad +( 3y^{-6} +24y^{-4} +62y^{-2} +86 +62 y^2 +24 y^4 +3 y^6)                     x_1^4
+( 6y^{-4} +21y^{-2} +30 +21 y^2 +6 y^4)                                          x_1^5 \nonumber \\
&\quad\quad +( y^{-4} +4y^{-2} +7 +4 y^2 +y^4)                                                x_1^6
\Bigr](x_1x_2)^6
+{\cal O}(x_2^7),\\
&F_4^{(l=2)}(y;x_1;x_2)=
1
+\Bigl[
y^{-2} +1 +y^2
+( y^{-2} +1 +y^2)x_1 
+( y^{-2} +1 +y^2)x_1^2 +( y^{-2} +1 +y^2)x_1^3 
\Bigr]    x_1x_2\nonumber \\
&\quad +\Bigl[
y^{-4} +2y^{-2} +3 +2 y^2 +y^4
+( y^{-4} +3y^{-2} +4 +3 y^2 +y^4)x_1 
+( 2y^{-4} +4y^{-2} +6 +4 y^2 +2 y^4)x_1^2 \nonumber \\
&\quad\quad
+( y^{-4} +4y^{-2} +6 +4 y^2 +y^4)x_1^3 
+( y^{-4} +2y^{-2} +4 +2 y^2 +y^4)x_1^4 
+( y^{-2} +2 +y^2)x_1^5
+x_1^6
\Bigr]    (x_1x_2)^2\nonumber \\
&\quad
+\Bigl[
y^{-6} +2y^{-4} +5y^{-2} +6 +5 y^2 +2 y^4 +y^6
+( y^{-6} +4y^{-4} +9y^{-2} +11 +9 y^2 +4 y^4 +y^6)x_1 \nonumber \\
&\quad\quad
+( y^{-6} +6y^{-4} +14y^{-2} +17 +14 y^2 +6 y^4 +y^6)x_1^2 
+( y^{-6} +6y^{-4} +16y^{-2} +21 +16 y^2 +6 y^4 +y^6)x_1^3 \nonumber \\
&\quad\quad
+( 4y^{-4} +12y^{-2} +16 +12 y^2 +4 y^4)x_1^4 
+( 2y^{-4} +8y^{-2} +11 +8 y^2 +2 y^4)x_1^5 
+( +3y^{-2} +5 +3 y^2)x_1^6\nonumber \\
&\quad\quad
+( y^{-2} +1 +y^2)x_1^7 
\Bigr]    (x_1x_2)^3\nonumber \\
&\quad
+\Bigl[
y^{-8} +2y^{-6} +6y^{-4} +10y^{-2} +13 +10y^2 +6y^4 +2y^6 +y^8\nonumber \\
&\quad\quad
+( 3y^{-6} +11y^{-4} +21y^{-2} +26 +21y^2 +11y^4 +3y^6)x_1
+( 4y^{-6} 19y^{-4} 36y^{-2} +46 +36y^2 +19y^4 +4y^6)x_1^2\nonumber \\
&\quad\quad
+( 4y^{-6} +21y^{-4} +46y^{-2} +59 +46y^2 +21y^4 +4y^6)x_1^3\nonumber \\
&\quad\quad
+( 2y^{-6} +18y^{-4} +41y^{-2} +56 +41y^2 +18y^4 +2y^6)x_1^4\nonumber \\
&\quad\quad
+( y^{-6} +11y^{-4} +31y^{-2} +42 +31y^2 +11y^4 +y^6)x_1^5
+( 5y^{-4} +16y^{-2} +25 +16y^2 +5y^4)x_1^6\nonumber \\
&\quad\quad
+( y^{-4} +6y^{-2} +9 +6y^2 +y^4)x_1^7
+( y^{-2} +3 +y^2)x_1^8
\Bigr](x_1x_2)^4\nonumber \\
&\quad
+\Bigl[
y^{-8} +5y^{-6} +11y^{-4} +20y^{-2} +23 +20y^2 +11y^4 +5y^6 +y^8\nonumber \\
&\quad\quad
+( y^{-8} +9y^{-6} +26y^{-4} +47y^{-2} +56 +47y^2 +26y^4 +9y^6 +y^8)x_1\nonumber \\
&\quad\quad
+( y^{-8} +14y^{-6} +46y^{-4} +87y^{-2} +104 +87y^2 +46y^4 +14y^6 +y^8)x_1^2\nonumber \\
&\quad\quad
+( y^{-8} +16y^{-6} +59y^{-4} +119y^{-2} +147 +119y^2 +59y^4 +16y^6 +y^8)x_1^3\nonumber \\
&\quad\quad
+( 12y^{-6} +55y^{-4} +120y^{-2} +151 +120y^2 +55y^4 +12y^6)x_1^4\nonumber \\
&\quad\quad
+( 8y^{-6} +42y^{-4} +101y^{-2} +130 +101y^2 +42y^4 +8y^6)x_1^5\nonumber \\
&\quad\quad
+( 3y^{-6} +22y^{-4} +63y^{-2} +85 +63y^2 +22y^4 +3y^6)x_1^6\nonumber \\
&\quad\quad
+( y^{-6} +9y^{-4} +31y^{-2} +42 +31y^2 +9y^4 +y^6)x_1^7
+( 2y^{-4} +10y^{-2} +15 +10y^2 +2y^4)x_1^8\nonumber \\
&\quad\quad
+( 2y^{-2} +3 +2y^2)x_1^9
\Bigr]
(x_1x_2)^5\nonumber \\
&\quad
+
\Bigl[3y^{-8} +10y^{-6} +23y^{-4} +36y^{-2} +44 +36 y^2 +23 y^4 +10 y^6 +3 y^8\nonumber \\
&\quad\quad +( 4y^{-8} +21y^{-6} +56y^{-4} +95y^{-2} +113 +95 y^2 +56 y^4 +21 y^6 +4 y^8)     x_1    \nonumber \\
&\quad\quad +( 6y^{-8} +36y^{-6} +108y^{-4} +187y^{-2} +226 +187 y^2 +108 y^4 +36 y^6 +6 y^8) x_1^2  \nonumber \\
&\quad\quad +( 6y^{-8} +46y^{-6} +147y^{-4} +274y^{-2} +335 +274 y^2 +147 y^4 +46 y^6 +6 y^8) x_1^3  \nonumber \\
&\quad\quad +( 4y^{-8} +41y^{-6} +155y^{-4} +302y^{-2} +378 +302 y^2 +155 y^4 +41 y^6 +4 y^8) x_1^4  \nonumber \\
&\quad\quad +( 2y^{-8} +31y^{-6} +130y^{-4} +276y^{-2} +349 +276 y^2 +130 y^4 +31 y^6 +2 y^8) x_1^5  \nonumber \\
&\quad\quad +( y^{-8} +16y^{-6} +85y^{-4} +195y^{-2} +259 +195 y^2 +85 y^4 +16 y^6 +y^8)      x_1^6  \nonumber \\
&\quad\quad +( 6y^{-6} +42y^{-4} +110y^{-2} +147 +110 y^2 +42 y^4 +6 y^6)                       x_1^7  \nonumber \\
&\quad\quad +( y^{-6} +16y^{-4} +46y^{-2} +67 +46 y^2 +16 y^4 +y^6)                             x_1^8
+( 3y^{-4} +13y^{-2} +20 +13 y^2 +3 y^4)                                              x_1^9  \nonumber \\
&\quad\quad +( y^{-4} +2y^{-2} +4 +2 y^2 +y^4)                                                    x_1^{10}
\Bigr](x_1x_2)^6
+{\cal O}(x_2^7),\\
&F_5^{(l=2)}(y;x_1;x_2)=
1
+\Bigl[
y^{-2}
+1
+y^2
+(
y^{-2}
+1
+y^2)x_1 
+(
y^{-2}
+1
+y^2)x_1^2 
+(
y^{-2}
+1
+y^2)x_1^3 \nonumber \\
&\quad\quad +(
y^{-2}
+1
+y^2)x_1^4 
\Bigr]    x_1x_2\nonumber \\
&\quad
+\Bigl[
y^{-4}
+2y^{-2}
+3
+2 y^2
+y^4
+(
y^{-4}
+3y^{-2}
+4
+3 y^2
+y^4)x_1 \nonumber \\
&\quad\quad
+(
2y^{-4}
+4y^{-2}
+6
+4 y^2
+2 y^4)x_1^2 
+(
2y^{-4}
+5y^{-2}
+7
+5 y^2
+2 y^4)x_1^3 \nonumber \\
&\quad\quad +(
2y^{-4}
+5y^{-2}
+8
+5 y^2
+2 y^4)x_1^4
+(
y^{-4}
+3y^{-2}
+5
+3 y^2
+y^4)x_1^5 \nonumber \\
&\quad\quad +(
y^{-4}
+2y^{-2}
+4
+2 y^2
+y^4)x_1^6
+(
y^{-2}
+2
+y^2)x_1^7 
+x_1^8
\Bigr]    (x_1x_2)^2\nonumber \\
&\quad +\Bigl[
y^{-6} +2y^{-4} +5y^{-2} +6 +5y^2 +2y^4 +y^6
+( y^{-6} +4y^{-4} +9y^{-2} +11 +9y^2 +4y^4 +y^6)x_1\nonumber \\
&\quad\quad 
+( 2y^{-6} +7y^{-4} +15y^{-2} +18 +15y^2 +7y^4 +2y^6)x_1^2\nonumber \\
&\quad\quad
+( 2y^{-6} +9y^{-4} +20y^{-2} +25 +20y^2 +9y^4 +2y^6)x_1^3\nonumber \\
&\quad\quad
+( 2y^{-6} +10y^{-4} +24y^{-2} +30 +24y^2 +10y^4 +2y^6)x_1^4\nonumber \\
&\quad\quad
+( y^{-6} +8y^{-4} +20y^{-2} +26 +20y^2 +8y^4 +y^6)x_1^5
+( y^{-6} +6y^{-4} +17y^{-2} +22 +17y^2 +6y^4 +y^6)x_1^6\nonumber \\
&\quad\quad
+( 3y^{-4} +11y^{-2} +15 +11y^2 +3y^4)x_1^7
+( y^{-4} +6y^{-2} +8 +6y^2 +y^4)x_1^8
+( 2y^{-2} +3 +2y^2)x_1^9\nonumber \\
&\quad\quad
+( y^{-2} +1 +y^2)x_1^{10}
\Bigr](x_1x_2)^3\nonumber \\
&\quad +\Bigl[
y^{-8} +2y^{-6} +6y^{-4} +10y^{-2} +13 +10y^2 +6y^4 +2y^6 +y^8\nonumber \\
&\quad\quad
+( y^{-8} +4y^{-6} +12y^{-4} +22y^{-2} +27 +22y^2 +12y^4 +4y^6 +y^8)x_1\nonumber \\
&\quad\quad
+( y^{-8} +7y^{-6} +23y^{-4} +40y^{-2} +50 +40y^2 +23y^4 +7y^6 +y^8)x_1^2\nonumber \\
&\quad\quad
+( y^{-8} +9y^{-6} +32y^{-4} +60y^{-2} +74 +60y^2 +32y^4 +9y^6 +y^8)x_1^3\nonumber \\
&\quad\quad
+( y^{-8} +10y^{-6} +40y^{-4} +77y^{-2} +98 +77y^2 +40y^4 +10y^6 +y^8)x_1^4\nonumber \\
&\quad\quad
+( 8y^{-6} +37y^{-4} +77y^{-2} +98 +77y^2 +37y^4 +8y^6)x_1^5\nonumber \\
&\quad\quad
+( 6y^{-6} +33y^{-4} +71y^{-2} +93 +71y^2 +33y^4 +6y^6)x_1^6\nonumber \\
&\quad\quad
+( 3y^{-6} +22y^{-4} +55y^{-2} +73 +55y^2 +22y^4 +3y^6)x_1^7\nonumber \\
&\quad\quad
+( y^{-6} +13y^{-4} +35y^{-2} +50 +35y^2 +13y^4 +y^6)x_1^8
+(5y^{-4} +18y^{-2} +26 +18y^2 +5y^4)x_1^9\nonumber \\
&\quad\quad
+( 2y^{-4} +8y^{-2} +13 +8y^2 +2y^4)x_1^{10}
+( 2y^{-2} +4 +2y^2)x_1^{11}
+x_1^{12}
\Bigr](x_1x_2)^4\nonumber \\
&\quad +\Bigl[
y^{-10} +2y^{-8} +6y^{-6} +12y^{-4} +21y^{-2} +24 +21y^2 +12y^4 +6y^6 +2y^8 +y^{10}\nonumber \\
&\quad\quad
+( 3y^{-8} +12y^{-6} +29y^{-4} +50y^{-2} +59 +50y^2 +29y^4 +12y^6 +3y^8)x_1\nonumber \\
&\quad\quad
+( 4y^{-8} +23y^{-6} +58y^{-4} +100y^{-2} +117 +100y^2 +58y^4 +23y^6 +4y^8)x_1^2\nonumber \\
&\quad\quad
+( 5y^{-8} +32y^{-6} +90y^{-4} +159y^{-2} +189 +159y^2 +90y^4 +32y^6 +5y^8)x_1^3\nonumber \\
&\quad\quad
+( 5y^{-8} +40y^{-6} +119y^{-4} +220y^{-2} +264 +220y^2 +119y^4 +40y^6 +5y^8)x_1^4\nonumber \\
&\quad\quad
+( 3y^{-8} +37y^{-6} +126y^{-4} +242y^{-2} +296 +242y^2 +126y^4 +37y^6 +3y^8)x_1^5\nonumber \\
&\quad\quad
+( 2y^{-8} +33y^{-6} +121y^{-4} +245y^{-2} +301 +245y^2 +121y^4 +33y^6 +2y^8)x_1^6\nonumber \\
&\quad\quad
+( y^{-8} +22y^{-6} +96y^{-4} +209y^{-2} +264 +209y^2 +96y^4 +22y^6 +y^8)x_1^7\nonumber \\
&\quad\quad
+( 13y^{-6} +65y^{-4} +156y^{-2} +200 +156y^2 +65y^4 +13y^6)x_1^8\nonumber \\
&\quad\quad
+( 5y^{-6} +36y^{-4} +96y^{-2} +127 +96y^2 +36y^4 +5y^6)x_1^9\nonumber \\
&\quad\quad
+( 2y^{-6} +17y^{-4} +53y^{-2} +72 +53y^2 +17y^4 +2y^6)x_1^{10}
+( 5y^{-4} +21y^{-2} +31 +21y^2 +5y^4)x_1^{11}\nonumber \\
&\quad\quad
+( y^{-4} +7y^{-2} +10 +7y^2 +y^4)x_1^{12}
+( y^{-2} +2 +y^2)x_1^{13}
\Bigr](x_1x_2)^5\nonumber \\
&\quad +\Bigl[
y^{-10} +5y^{-8} +12y^{-6} +25y^{-4} +38y^{-2} +46 +38 y^2 +25 y^4 +12 y^6 +5 y^8 +y^{10}\nonumber \\
&\quad\quad +( y^{-10} +9y^{-8} +29y^{-6} +65y^{-4} +104y^{-2} +122 +104 y^2 +65 y^4 +29 y^6 +9 y^8 +y^{10})      x_1    \nonumber \\
&\quad\quad +( y^{-10} +15y^{-8} +58y^{-6} +139y^{-4} +221y^{-2} +261 +221 y^2 +139 y^4 +58 y^6 +15 y^8 +y^{10})  x_1^2  \nonumber \\
&\quad\quad +( y^{-10} +20y^{-8} +90y^{-6} +227y^{-4} +376y^{-2} +444 +376 y^2 +227 y^4 +90 y^6 +20 y^8 +y^{10})  x_1^3  \nonumber \\
&\quad\quad +( y^{-10} +24y^{-8} +119y^{-6} +322y^{-4} +548y^{-2} +656 +548 y^2 +322 y^4 +119 y^6 +24 y^8 +y^{10})x_1^4  \nonumber \\
&\quad\quad +( 20y^{-8} +126y^{-6} +369y^{-4} +656y^{-2} +790 +656 y^2 +369 y^4 +126 y^6 +20 y^8)                   x_1^5  \nonumber \\
&\quad\quad +( 17y^{-8} +121y^{-6} +384y^{-4} +705y^{-2} +862 +705 y^2 +384 y^4 +121 y^6 +17 y^8)                   x_1^6  \nonumber \\
&\quad\quad +( 11y^{-8} +96y^{-6} +337y^{-4} +657y^{-2} +813 +657 y^2 +337 y^4 +96 y^6 +11 y^8)                     x_1^7  \nonumber \\
&\quad\quad +( 6y^{-8} +65y^{-6} +261y^{-4} +535y^{-2} +678 +535 y^2 +261 y^4 +65 y^6 +6 y^8)                       x_1^8  \nonumber \\
&\quad\quad +( 2y^{-8} +36y^{-6} +169y^{-4} +375y^{-2} +483 +375 y^2 +169 y^4 +36 y^6 +2 y^8)                       x_1^9  \nonumber \\
&\quad\quad +( y^{-8} +17y^{-6} +98y^{-4} +232y^{-2} +309 +232 y^2 +98 y^4 +17 y^6 +y^8)                            x_1^{10}\nonumber \\
&\quad\quad +( 5y^{-6} +43y^{-4} +117y^{-2} +161 +117 y^2 +43 y^4 +5 y^6)                                             x_1^{11}\nonumber \\
&\quad\quad +( y^{-6} +16y^{-4} +48y^{-2} +71 +48 y^2 +16 y^4 +y^6)                                                   x_1^{12}
+( 4y^{-4} +15y^{-2}+23  +15 y^2 +4 y^4)                                                                    x_1^{13}\nonumber \\
&\quad\quad +( y^{-4} +3y^{-2} +6 +3 y^2 +y^4)                                                                          x_1^{14}
\Bigr](x_1x_2)^6
+{\cal O}(x_2^7),\\
&F_6^{(l=2)}(y;x_1;x_2)=
1
+\Bigl[
y^{-2} +1 +y^{2}
+( y^{-2} +1 +y^{2}) x_1    
+( y^{-2} +1 +y^{2}) x_1^{2}
+( y^{-2} +1 +y^{2}) x_1^{3}\nonumber \\
&\quad\quad
+( y^{-2} +1 +y^{2}) x_1^{4}
+( y^{-2} +1 +y^{2}) x_1^{5}
\Bigr]x_1x_2\nonumber \\
&\quad +\Bigl[
y^{-4} +2y^{-2} +3 +2y^{2} +y^{4}
+( y^{-4} +3y^{-2} +4 +3y^{2} +y^{4})      x_1     
+( 2y^{-4} +4y^{-2} +6 +4y^{2} +2y^{4})    x_1^{2}\nonumber \\
&\quad\quad
+( 2y^{-4} +5y^{-2} +7 +5y^{2} +2y^{4})    x_1^{3} 
+( 3y^{-4} +6y^{-2} +9 +6y^{2} +3y^{4})    x_1^{4} \nonumber \\
&\quad\quad
+( 2y^{-4} +6y^{-2} +9 +6y^{2} +2y^{4})    x_1^{5}
+( 2y^{-4} +4y^{-2} +7 +4y^{2} +2y^{4})    x_1^{6} 
+( y^{-4} +3y^{-2} +5 +3y^{2} +y^{4})      x_1^{7} \nonumber \\
&\quad\quad
+( y^{-4} +2y^{-2} +4 +2y^{2} +y^{4})      x_1^{8}
+( y^{-2} +2 +y^{2})                         x_1^{9} 
+                                              x_1^{10}
\Bigr](x_1x_2)^2\nonumber \\
&\quad +\Bigl[
y^{-6} +2y^{-4} +5y^{-2} +6 +5y^{2} +2y^{4} +y^{6}
+( y^{-6} +4y^{-4} +9y^{-2} +11 +9y^{2} +4y^{4} +y^{6})         x_1\nonumber \\
&\quad\quad
+( 2y^{-6} +7y^{-4} +15y^{-2} +18 +15y^{2} +7y^{4} +2y^{6})     x_1^{2} \nonumber \\
&\quad\quad
+( 3y^{-6} +10y^{-4} +21y^{-2} +26 +21y^{2} +10y^{4} +3y^{6})   x_1^{3}\nonumber \\
&\quad\quad
+( 3y^{-6} +13y^{-4} +28y^{-2} +34 +28y^{2} +13y^{4} +3y^{6})   x_1^{4} \nonumber \\
&\quad\quad
+( 3y^{-6} +14y^{-4} +32y^{-2} +40 +32y^{2} +14y^{4} +3y^{6})   x_1^{5}\nonumber \\
&\quad\quad
+( 3y^{-6} +13y^{-4} +30y^{-2} +38 +30y^{2} +13y^{4} +3y^{6})   x_1^{6} \nonumber \\
&\quad\quad
+( 2y^{-6} +11y^{-4} +27y^{-2} +34 +27y^{2} +11y^{4} +2y^{6})   x_1^{7}\nonumber \\
&\quad\quad
+( y^{-6} +8y^{-4} +22y^{-2} +28 +22y^{2} +8y^{4} +y^{6})       x_1^{8} 
+( y^{-6} +5y^{-4} +16y^{-2} +21 +16y^{2} +5y^{4} +y^{6})       x_1^{9}\nonumber \\
&\quad\quad
+( 2y^{-4} +9y^{-2} +12 +9y^{2} +2y^{4})                          x_1^{10}
+( y^{-4} +5y^{-2} +6 +5y^{2} +y^{4})                             x_1^{11}
+( 2y^{-2} +3 +2y^{2})                                              x_1^{12}\nonumber \\
&\quad\quad
+( y^{-2} +1 +y^{2})                                                x_1^{13}
\Bigr](x_1x_2)^3\nonumber \\
&\quad +\Bigl[
y^{-8} +2y^{-6} +6y^{-4} +10y^{-2} +13 +10y^{2} +6y^{4} +2y^{6} +y^{8}\nonumber \\
&\quad\quad
+( y^{-8} +4y^{-6} +12y^{-4} +22y^{-2} +27 +22y^{2} +12y^{4} +4y^{6} +y^{8})             x_1\nonumber \\
&\quad\quad
+(2y^{-8} +8y^{-6} +24y^{-4} +41y^{-2} +51 +41y^{2} +24y^{4} +8y^{6} +2y^{8})            x_1^{2}\nonumber \\
&\quad\quad
+( 2y^{-8} +12y^{-6} +36y^{-4} +64y^{-2} +78 +64y^{2} +36y^{4} +12y^{6} +2y^{8})         x_1^{3}\nonumber \\
&\quad\quad
+(3y^{-8} +16y^{-6} +52y^{-4} +92y^{-2} +114 +92y^{2} +52y^{4} +16y^{6} +3y^{8})         x_1^{4}\nonumber \\
&\quad\quad
+( 2y^{-8} +18y^{-6} +61y^{-4} +115y^{-2} +142 +115y^{2} +61y^{4} +18y^{6} +2y^{8})      x_1^{5}\nonumber \\
&\quad\quad
+( 2y^{-8} +18y^{-6} +66y^{-4} +124y^{-2} +156 +124y^{2} +66y^{4} +18y^{6} +2y^{8})      x_1^{6}\nonumber \\
&\quad\quad
+( y^{-8} +16y^{-6} +62y^{-4} +124y^{-2} +155 +124y^{2} +62y^{4} +16y^{6} +y^{8})        x_1^{7}\nonumber \\
&\quad\quad
+( y^{-8} +12y^{-6} +55y^{-4} +113y^{-2} +146 +113y^{2} +55y^{4} +12y^{6} +y^{8})        x_1^{8}\nonumber \\
&\quad\quad
+( 8y^{-6} +41y^{-4} +93y^{-2} +120 +93y^{2} +41y^{4} +8y^{6})                             x_1^{9} \nonumber \\
&\quad\quad
+( 4y^{-6} +28y^{-4} +66y^{-2} +90 +66y^{2} +28y^{4} +4y^{6})                              x_1^{10}\nonumber \\
&\quad\quad
+( 2y^{-6} +16y^{-4} +43y^{-2} +58 +43y^{2} +16y^{4} +2y^{6})                              x_1^{11}
+( 8y^{-4} +24y^{-2} +36 +24y^{2} +8y^{4})                                                   x_1^{12}\nonumber \\
&\quad\quad
+( 3y^{-4} +12y^{-2} +18 +12y^{2} +3y^{4})                                                   x_1^{13}
+( y^{-4} +4y^{-2} +8 +4y^{2} +y^{4})                                                        x_1^{14}
+( y^{-2} +2 +y^{2})                                                                           x_1^{15}\nonumber \\
&\quad\quad
+                                                                                                x_1^{16}
\Bigr](x_1x_2)^4\nonumber \\
&\quad +\Bigl[
y^{-10} +2y^{-8} +6y^{-6} +12y^{-4} +21y^{-2} +24 +21y^{2} +12y^{4} +6y^{6} +2y^{8} +y^{10}\nonumber \\
&\quad\quad
+( y^{-10} +4y^{-8} +13y^{-6} +30y^{-4} +51y^{-2} +60 +51y^{2} +30y^{4} +13y^{6} +4y^{8} +y^{10})x_1\nonumber \\
&\quad\quad
+( y^{-10} +7y^{-8} +27y^{-6} +62y^{-4} +104y^{-2} +121 +104y^{2} +62y^{4} +27y^{6} +7y^{8} +y^{10})x_1^{2}\nonumber \\
&\quad\quad
+( y^{-10} +10y^{-8} +43y^{-6} +104y^{-4} +174y^{-2} +204 +174y^{2} +104y^{4} +43y^{6} +10y^{8} +y^{10})x_1^{3}\nonumber \\
&\quad\quad
+( y^{-10} +13y^{-8} +62y^{-6} +156y^{-4} +266y^{-2} +312 +266y^{2} +156y^{4} +62y^{6} +13y^{8} +y^{10})x_1^{4}\nonumber \\
&\quad\quad
+( y^{-10} +14y^{-8} +76y^{-6} +203y^{-4} +355y^{-2} +422 +355y^{2} +203y^{4} +76y^{6} +14y^{8} +y^{10})x_1^{5}\nonumber \\
&\quad\quad
+( 13y^{-8} +83y^{-6} +234y^{-4} +418y^{-2} +499 +418y^{2} +234y^{4} +83y^{6} +13y^{8})x_1^{6}\nonumber \\
&\quad\quad
+( 11y^{-8} +82y^{-6} +246y^{-4} +452y^{-2} +544 +452y^{2} +246y^{4} +82y^{6} +11y^{8})x_1^{7}\nonumber \\
&\quad\quad
+( 8y^{-8} +72y^{-6} +234y^{-4} +448y^{-2} +545 +448y^{2} +234y^{4} +72y^{6} +8y^{8})x_1^{8}\nonumber \\
&\quad\quad
+( 5y^{-8} +57y^{-6} +202y^{-4} +405y^{-2} +499 +405y^{2} +202y^{4} +57y^{6} +5y^{8})x_1^{9}\nonumber \\
&\quad\quad
+( 2y^{-8} +38y^{-6} +154y^{-4} +327y^{-2} +410 +327y^{2} +154y^{4} +38y^{6} +2y^{8})x_1^{10}\nonumber \\
&\quad\quad
+( y^{-8} +24y^{-6} +108y^{-4} +243y^{-2} +308 +243y^{2} +108y^{4} +24y^{6} +y^{8})x_1^{11}\nonumber \\
&\quad\quad
+( 11y^{-6} +65y^{-4} +161y^{-2} +210 +161y^{2} +65y^{4} +11y^{6})x_1^{12}\nonumber \\
&\quad\quad
+( 5y^{-6} +35y^{-4} +97y^{-2} +129 +97y^{2} +35y^{4} +5y^{6})x_1^{13}\nonumber \\
&\quad\quad
+( y^{-6} +15y^{-4} +48y^{-2} +67 +48y^{2} +15y^{4} +y^{6})x_1^{14}\nonumber \\
&\quad\quad
+( 5y^{-4} +21y^{-2} +30 +21y^{2} +5y^{4})x_1^{15}
+( y^{-4} +7y^{-2} +11 +7y^{2} +y^{4})x_1^{16}\nonumber \\
&\quad\quad
+( 2y^{-2} +3 +2y^{2})x_1^{17}
\Bigr](x_1x_2)^5\nonumber \\
&\quad\Bigl[
y^{-12} +2y^{-10} +6y^{-8} +13y^{-6} +26y^{-4} +39y^{-2} +47 +39 y^2 +26 y^4 +13 y^6 +6 y^8 +2 y^{10} +y^{12}\nonumber \\
&\quad\quad +( 3y^{-10} +12y^{-8} +32y^{-6} +68y^{-4} +107y^{-2} +125 +107 y^2 +68 y^4 +32 y^6 +12 y^8 +3 y^{10})         x_1    \nonumber \\
&\quad\quad +( 4y^{-10} +24y^{-8} +70y^{-6} +152y^{-4} +234y^{-2} +274 +234 y^2 +152 y^4 +70 y^6 +24 y^8 +4 y^{10})       x_1^2  \nonumber \\
&\quad\quad +( 5y^{-10} +36y^{-8} +121y^{-6} +267y^{-4} +419y^{-2} +488 +419 y^2 +267 y^4 +121 y^6 +36 y^8 +5 y^{10})     x_1^3  \nonumber \\
&\quad\quad +( 6y^{-10} +52y^{-8} +184y^{-6} +426y^{-4} +674y^{-2} +789 +674 y^2 +426 y^4 +184 y^6 +52 y^8 +6 y^{10})     x_1^4  \nonumber \\
&\quad\quad +( 6y^{-10} +61y^{-8} +243y^{-6} +584y^{-4} +953y^{-2} +1119 +953 y^2 +584 y^4 +243 y^6 +61 y^8 +6 y^{10})    x_1^5  \nonumber \\
&\quad\quad +( 4y^{-10} +66y^{-8} +287y^{-6} +724y^{-4} +1198y^{-2} +1420 +1198 y^2 +724 y^4 +287 y^6 +66 y^8 +4 y^{10})  x_1^6  \nonumber \\
&\quad\quad +( 3y^{-10} +62y^{-8} +306y^{-6} +809y^{-4} +1379y^{-2} +1638 +1379 y^2 +809 y^4 +306 y^6 +62 y^8 +3 y^{10})  x_1^7  \nonumber \\
&\quad\quad +( 2y^{-10} +55y^{-8} +295y^{-6} +834y^{-4} +1457y^{-2} +1753 +1457 y^2 +834 y^4 +295 y^6 +55 y^8 +2 y^{10})  x_1^8  \nonumber \\
&\quad\quad +( y^{-10} +41y^{-8} +259y^{-6} +776y^{-4} +1412y^{-2} +1712 +1412 y^2 +776 y^4 +259 y^6 +41 y^8 +y^{10})     x_1^9  \nonumber \\
&\quad\quad +( 28y^{-8} +200y^{-6} +661y^{-4} +1243y^{-2} +1532 +1243 y^2 +661 y^4 +200 y^6 +28 y^8)                        x_1^{10}\nonumber \\
&\quad\quad +( 16y^{-8} +143y^{-6} +511y^{-4} +1008y^{-2} +1252 +1008 y^2 +511 y^4 +143 y^6 +16 y^8)                        x_1^{11}\nonumber \\
&\quad\quad +( 8y^{-8} +88y^{-6} +360y^{-4} +744y^{-2} +948 +744 y^2 +360 y^4 +88 y^6 +8 y^8)                               x_1^{12}\nonumber \\
&\quad\quad +( 3y^{-8} +48y^{-6} +225y^{-4} +501y^{-2} +646 +501 y^2 +225 y^4 +48 y^6 +3 y^8)                               x_1^{13}\nonumber \\
&\quad\quad +( y^{-8} +21y^{-6} +126y^{-4} +297y^{-2} +397 +297 y^2 +126 y^4 +21 y^6 +y^8)                                  x_1^{14}\nonumber \\
&\quad\quad +( 8y^{-6} +59y^{-4} +157y^{-2} +214 +157 y^2 +59 y^4 +8 y^6)                                                     x_1^{15}\nonumber \\
&\quad\quad +( 2y^{-6} +25y^{-4} +71y^{-2} +103 +71 y^2 +25 y^4 +2 y^6)                                                       x_1^{16}
+( 7y^{-4} +27y^{-2} +40 +27 y^2 +7 y^4)                                                                            x_1^{17}\nonumber \\
&\quad\quad +( 2y^{-4} +7y^{-2} +13 +7 y^2 +2 y^4)                                                                              x_1^{18}
+( y^{-2} +2 +y^2)                                                                                                    x_1^{19}
\Bigr](x_1x_2)^6+{\cal O}(x_2^7),\\
&F_7^{(l=2)}(y;x_1;x_2)=
1
+\Bigl[
y^{-2} +1 +y^{2}
+( y^{-2} +1 +y^{2})x_1    
+( y^{-2} +1 +y^{2})x_1^{2}
+( y^{-2} +1 +y^{2})x_1^{3}\nonumber \\
&\quad\quad
+( y^{-2} +1 +y^{2})x_1^{4}
+( y^{-2} +1 +y^{2})x_1^{5}
+( y^{-2} +1 +y^{2})x_1^{6}
\Bigr]x_1x_2\nonumber \\
&\quad+\Bigl[
y^{-4} +2y^{-2} +3 +2y^{2} +y^{4}
+( y^{-4} +3y^{-2} +4 +3y^{2} +y^{4})    x_1     
+( 2y^{-4} +4y^{-2} +6 +4y^{2} +2y^{4})  x_1^{2}\nonumber \\
&\quad\quad
+( 2y^{-4} +5y^{-2} +7 +5y^{2} +2y^{4})  x_1^{3} 
+( 3y^{-4} +6y^{-2} +9 +6y^{2} +3y^{4})  x_1^{4} \nonumber \\
&\quad\quad
+( 3y^{-4} +7y^{-2} +10 +7y^{2} +3y^{4}) x_1^{5}
+( 3y^{-4} +7y^{-2} +11 +7y^{2} +3y^{4}) x_1^{6} \nonumber \\
&\quad\quad
+( 2y^{-4} +5y^{-2} +8 +5y^{2} +2y^{4})  x_1^{7} 
+( 2y^{-4} +4y^{-2} +7 +4y^{2} +2y^{4})  x_1^{8}
+( y^{-4} +3y^{-2} +5 +3y^{2} +y^{4})    x_1^{9} \nonumber \\
&\quad\quad
+( y^{-4} +2y^{-2} +4 +2y^{2} +y^{4})    x_1^{10}
+( y^{-2} +2 +y^{2})                       x_1^{11}
+                                            x_1^{12}
\Bigr](x_1x_2)^2\nonumber \\
&\quad+\Bigl[
y^{-6} +2y^{-4} +5y^{-2} +6 +5y^{2} +2y^{4} +y^{6}
+( y^{-6} +4y^{-4} +9y^{-2} +11 +9y^{2} +4y^{4} +y^{6})         x_1\nonumber \\
&\quad\quad
+( 2y^{-6} +7y^{-4} +15y^{-2} +18 +15y^{2} +7y^{4} +2y^{6})     x_1^{2} \nonumber \\
&\quad\quad
+( 3y^{-6} +10y^{-4} +21y^{-2} +26 +21y^{2} +10y^{4} +3y^{6})   x_1^{3}\nonumber \\
&\quad\quad
+( 4y^{-6} +14y^{-4} +29y^{-2} +35 +29y^{2} +14y^{4} +4y^{6})   x_1^{4} \nonumber \\
&\quad\quad
+( 4y^{-6} +17y^{-4} +36y^{-2} +44 +36y^{2} +17y^{4} +4y^{6})   x_1^{5}\nonumber \\
&\quad\quad
+( 5y^{-6} +19y^{-4} +42y^{-2} +52 +42y^{2} +19y^{4} +5y^{6})   x_1^{6} \nonumber \\
&\quad\quad
+( 4y^{-6} +18y^{-4} +40y^{-2} +50 +40y^{2} +18y^{4} +4y^{6})   x_1^{7}\nonumber \\
&\quad\quad
+( 4y^{-6} +17y^{-4} +39y^{-2} +48 +39y^{2} +17y^{4} +4y^{6})   x_1^{8} \nonumber \\
&\quad\quad
+( 3y^{-6} +14y^{-4} +34y^{-2} +43 +34y^{2} +14y^{4} +3y^{6})   x_1^{9}\nonumber \\
&\quad\quad
+( 2y^{-6} +11y^{-4} +29y^{-2} +36 +29y^{2} +11y^{4} +2y^{6})   x_1^{10}\nonumber \\
&\quad\quad
+( y^{-6} +7y^{-4} +21y^{-2} +27 +21y^{2} +7y^{4} +y^{6})       x_1^{11}
+( y^{-6} +4y^{-4} +14y^{-2} +18 +14y^{2} +4y^{4} +y^{6})       x_1^{12}\nonumber \\
&\quad\quad
+( 2y^{-4} +8y^{-2} +10 +8y^{2} +2y^{4})                          x_1^{13}
+( y^{-4} +5y^{-2} +6 +5y^{2} +y^{4})                             x_1^{14}
+( 2y^{-2} +3 +2y^{2})                                              x_1^{15}\nonumber \\
&\quad\quad
+( y^{-2} +1 +y^{2})                                                x_1^{16}
\Bigr](x_1x_2)^3\nonumber \\
&\quad+\Bigl[
y^{-8} +2y^{-6} +6y^{-4} +10y^{-2} +13 +10y^{2} +6y^{4} +2y^{6} +y^{8}\nonumber \\
&\quad\quad
+( y^{-8} +4y^{-6} +12y^{-4} +22y^{-2} +27 +22y^{2} +12y^{4} +4y^{6} +y^{8})                  x_1\nonumber \\
&\quad\quad
+( 2y^{-8} +8y^{-6} +24y^{-4} +41y^{-2} +51 +41y^{2} +24y^{4} +8y^{6} +2y^{8})                x_1^{2}\nonumber \\
&\quad\quad
+( 3y^{-8} +13y^{-6} +37y^{-4} +65y^{-2} +79 +65y^{2} +37y^{4} +13y^{6} +3y^{8})              x_1^{3}\nonumber \\
&\quad\quad
+( 4y^{-8} +19y^{-6} +56y^{-4} +96y^{-2} +118 +96y^{2} +56y^{4} +19y^{6} +4y^{8})             x_1^{4}\nonumber \\
&\quad\quad
+( 4y^{-8} +24y^{-6} +73y^{-4} +130y^{-2} +158 +130y^{2} +73y^{4} +24y^{6} +4y^{8})           x_1^{5}\nonumber \\
&\quad\quad
+( 5y^{-8} +29y^{-6} +91y^{-4} +163y^{-2} +201 +163y^{2} +91y^{4} +29y^{6} +5y^{8})           x_1^{6}\nonumber \\
&\quad\quad
+( 4y^{-8} +30y^{-6} +97y^{-4} +179y^{-2} +220 +179y^{2} +97y^{4} +30y^{6} +4y^{8})           x_1^{7}\nonumber \\
&\quad\quad
+( 4y^{-8} +30y^{-6} +102y^{-4} +189y^{-2} +235 +189y^{2} +102y^{4} +30y^{6} +4y^{8})         x_1^{8}\nonumber \\
&\quad\quad
+( 3y^{-8} +27y^{-6} +96y^{-4} +186y^{-2} +231 +186y^{2} +96y^{4} +27y^{6} +3y^{8})           x_1^{9}\nonumber \\
&\quad\quad
+( 2y^{-8} +22y^{-6} +87y^{-4} +172y^{-2} +218 +172y^{2} +87y^{4} +22y^{6} +2y^{8})           x_1^{10}\nonumber \\
&\quad\quad
+( y^{-8} +16y^{-6} +69y^{-4} +146y^{-2} +186 +146y^{2} +69y^{4} +16y^{6} +y^{8})             x_1^{11}\nonumber \\
&\quad\quad
+( y^{-8} +11y^{-6} +53y^{-4} +114y^{-2} +150 +114y^{2} +53y^{4} +11y^{6} +y^{8})             x_1^{12}\nonumber \\
&\quad\quad
+( 6y^{-6} +35y^{-4} +82y^{-2} +108 +82y^{2} +35y^{4} +6y^{6})                                  x_1^{13}\nonumber \\
&\quad\quad
+( 3y^{-6} +23y^{-4} +56y^{-2} +77 +56y^{2} +23y^{4} +3y^{6})                                   x_1^{14}\nonumber \\
&\quad\quad
+( y^{-6} +12y^{-4} +34y^{-2} +48 +34y^{2} +12y^{4} +y^{6})                                     x_1^{15}
+( 6y^{-4} +18y^{-2} +28 +18y^{2} +6y^{4})                                                        x_1^{16}\nonumber \\
&\quad\quad
+( 2y^{-4} +8y^{-2} +13 +8y^{2} +2y^{4})                                                          x_1^{17}
+( y^{-4} +3y^{-2} +6 +3y^{2} +y^{4})                                                             x_1^{18}
+( y^{-2} +2 +y^{2})                                                                                x_1^{19}\nonumber \\
&\quad\quad
+                                                                                                     x_1^{20}
\Bigr](x_1x_2)^4\nonumber \\
&\quad+\Bigl[
y^{-10} +2y^{-8} +6y^{-6} +12y^{-4} +21y^{-2} +24 +21y^{2} +12y^{4} +6y^{6} +2y^{8} +y^{10}\nonumber \\
&\quad\quad
+( y^{-10} +4y^{-8} +13y^{-6} +30y^{-4} +51y^{-2} +60 +51y^{2} +30y^{4} +13y^{6} +4y^{8} +y^{10})               x_1\nonumber \\
&\quad\quad
+( 2y^{-10} +8y^{-8} +28y^{-6} +63y^{-4} +105y^{-2} +122 +105y^{2} +63y^{4} +28y^{6} +8y^{8} +2y^{10})          x_1^{2}\nonumber \\
&\quad\quad
+( 2y^{-10} +13y^{-8} +47y^{-6} +108y^{-4} +178y^{-2} +208 +178y^{2} +108y^{4} +47y^{6} +13y^{8} +2y^{10})      x_1^{3}\nonumber \\
&\quad\quad
+( 3y^{-10} +19y^{-8} +74y^{-6} +171y^{-4} +282y^{-2} +328 +282y^{2} +171y^{4} +74y^{6} +19y^{8} +3y^{10})      x_1^{4}\nonumber \\
&\quad\quad
+( 3y^{-10} +24y^{-8} +100y^{-6} +242y^{-4} +403y^{-2} +472 +403y^{2} +242y^{4} +100y^{6} +24y^{8} +3y^{10})    x_1^{5}\nonumber \\
&\quad\quad
+( 3y^{-10} +29y^{-8} +129y^{-6} +318y^{-4} +538y^{-2} +632 +538y^{2} +318y^{4} +129y^{6} +29y^{8} +3y^{10})    x_1^{6}\nonumber \\
&\quad\quad
+( 2y^{-10} +30y^{-8} +145y^{-6} +374y^{-4} +640y^{-2} +757 +640y^{2} +374y^{4} +145y^{6} +30y^{8} +2y^{10})    x_1^{7}\nonumber \\
&\quad\quad
+( 2y^{-10} +30y^{-8} +158y^{-6} +418y^{-4} +728y^{-2} +862 +728y^{2} +418y^{4} +158y^{6} +30y^{8} +2y^{10})    x_1^{8}\nonumber \\
&\quad\quad
+( y^{-10} +27y^{-8} +155y^{-6} +432y^{-4} +768y^{-2} +918 +768y^{2} +432y^{4} +155y^{6} +27y^{8} +y^{10})      x_1^{9}\nonumber \\
&\quad\quad
+( y^{-10} +22y^{-8} +144y^{-6} +421y^{-4} +770y^{-2} +925 +770y^{2} +421y^{4} +144y^{6} +22y^{8} +y^{10})      x_1^{10}\nonumber \\
&\quad\quad
+( 16y^{-8} +120y^{-6} +378y^{-4} +713y^{-2} +867 +713y^{2} +378y^{4} +120y^{6} +16y^{8})                         x_1^{11}\nonumber \\
&\quad\quad
+( 11y^{-8} +96y^{-6} +318y^{-4} +622y^{-2} +762 +622y^{2} +318y^{4} +96y^{6} +11y^{8})                           x_1^{12}\nonumber \\
&\quad\quad
+( 6y^{-8} +67y^{-6} +247y^{-4} +502y^{-2} +623 +502y^{2} +247y^{4} +67y^{6} +6y^{8})                             x_1^{13}\nonumber \\
&\quad\quad
+( 3y^{-8} +45y^{-6} +181y^{-4} +387y^{-2} +485 +387y^{2} +181y^{4} +45y^{6} +3y^{8})                             x_1^{14}\nonumber \\
&\quad\quad
+( y^{-8} +25y^{-6} +119y^{-4} +271y^{-2} +348 +271y^{2} +119y^{4} +25y^{6} +y^{8})                               x_1^{15}\nonumber \\
&\quad\quad
+( 13y^{-6} +72y^{-4} +178y^{-2} +231 +178y^{2} +72y^{4} +13y^{6})                                                  x_1^{16}\nonumber \\
&\quad\quad
+( 5y^{-6} +38y^{-4} +103y^{-2} +138 +103y^{2} +38y^{4} +5y^{6})                                                    x_1^{17}\nonumber \\
&\quad\quad
+( 2y^{-6} +18y^{-4} +56y^{-2} +76 +56y^{2} +18y^{4} +2y^{6})                                                       x_1^{18}
+( 7y^{-4} +26y^{-2} +37 +26y^{2} +7y^{4})                                                                            x_1^{19}\nonumber \\
&\quad\quad
+( 2y^{-4} +11y^{-2} +16 +11y^{2} +2y^{4})                                                                            x_1^{20}
+( 3y^{-2} +5 +3y^{2})                                                                                                  x_1^{21}
+( y^{-2} +1 +y^{2})                                                                                                    x_1^{22}
\Bigr](x_1x_2)^5
+{\cal O}(x_2^6),\\
&F_8^{(l=2)}(y;x_1;x_2)=
1
+\Bigl[
y^{-2} +1 +y^{2}
+( y^{-2} +1 +y^{2})x_1    
+( y^{-2} +1 +y^{2})x_1^{2}
+( y^{-2} +1 +y^{2})x_1^{3}\nonumber \\
&\quad\quad
+( y^{-2} +1 +y^{2})x_1^{4}
+( y^{-2} +1 +y^{2})x_1^{5}
+( y^{-2} +1 +y^{2})x_1^{6}
+( y^{-2} +1 +y^{2})x_1^{7}
\Bigr]x_1x_2\nonumber \\
&\quad+\Bigl[
y^{-4} +2y^{-2} +3 +2y^{2} +y^{4}
+( y^{-4} +3y^{-2} +4 +3y^{2} +y^{4})     x_1     
+( 2y^{-4} +4y^{-2} +6 +4y^{2} +2y^{4})   x_1^{2}\nonumber \\
&\quad\quad
+( 2y^{-4} +5y^{-2} +7 +5y^{2} +2y^{4})   x_1^{3} 
+( 3y^{-4} +6y^{-2} +9 +6y^{2} +3y^{4})   x_1^{4} \nonumber \\
&\quad\quad
+( 3y^{-4} +7y^{-2} +10 +7y^{2} +3y^{4})  x_1^{5}
+( 4y^{-4} +8y^{-2} +12 +8y^{2} +4y^{4})  x_1^{6} \nonumber \\
&\quad\quad
+( 3y^{-4} +8y^{-2} +12 +8y^{2} +3y^{4})  x_1^{7} 
+( 3y^{-4} +6y^{-2} +10 +6y^{2} +3y^{4})  x_1^{8}\nonumber \\
&\quad\quad
+( 2y^{-4} +5y^{-2} +8 +5y^{2} +2y^{4})   x_1^{9} 
+( 2y^{-4} +4y^{-2} +7 +4y^{2} +2y^{4})   x_1^{10}\nonumber \\
&\quad\quad
+( y^{-4} +3y^{-2} +5 +3y^{2} +y^{4})     x_1^{11}
+( y^{-4} +2y^{-2} +4 +2y^{2} +y^{4})     x_1^{12}
+( y^{-2} +2 +y^{2})                        x_1^{13}
+                                             x_1^{14}
\Bigr](x_1x_2)^2\nonumber \\
&\quad+\Bigl[
y^{-6} +2y^{-4} +5y^{-2} +6 +5y^{2} +2y^{4} +y^{6}
+( y^{-6} +4y^{-4} +9y^{-2} +11 +9y^{2} +4y^{4} +y^{6})       x_1\nonumber \\
&\quad\quad
+( 2y^{-6} +7y^{-4} +15y^{-2} +18 +15y^{2} +7y^{4} +2y^{6})   x_1^{2} \nonumber \\
&\quad\quad
+( 3y^{-6} +10y^{-4} +21y^{-2} +26 +21y^{2} +10y^{4} +3y^{6}) x_1^{3}\nonumber \\
&\quad\quad
+(4y^{-6} +14y^{-4} +29y^{-2} +35 +29y^{2} +14y^{4} +4y^{6})  x_1^{4} \nonumber \\
&\quad\quad
+( 5y^{-6} +18y^{-4} +37y^{-2} +45 +37y^{2} +18y^{4} +5y^{6}) x_1^{5}\nonumber \\
&\quad\quad
+( 6y^{-6} +22y^{-4} +46y^{-2} +56 +46y^{2} +22y^{4} +6y^{6}) x_1^{6} \nonumber \\
&\quad\quad
+( 6y^{-6} +24y^{-4} +52y^{-2} +64 +52y^{2} +24y^{4} +6y^{6}) x_1^{7}\nonumber \\
&\quad\quad
+( 6y^{-6} +24y^{-4} +52y^{-2} +64 +52y^{2} +24y^{4} +6y^{6}) x_1^{8} \nonumber \\
&\quad\quad
+( 6y^{-6} +23y^{-4} +51y^{-2} +63 +51y^{2} +23y^{4} +6y^{6}) x_1^{9}\nonumber \\
&\quad\quad
+( 5y^{-6} +21y^{-4} +48y^{-2} +59 +48y^{2} +21y^{4} +5y^{6}) x_1^{10}\nonumber \\
&\quad\quad
+( 4y^{-6} +18y^{-4} +43y^{-2} +53 +43y^{2} +18y^{4} +4y^{6}) x_1^{11}\nonumber \\
&\quad\quad
+( 3y^{-6} +14y^{-4} +36y^{-2} +45 +36y^{2} +14y^{4} +3y^{6}) x_1^{12}\nonumber \\
&\quad\quad
+( 2y^{-6} +10y^{-4} +28y^{-2} +35 +28y^{2} +10y^{4} +2y^{6}) x_1^{13}\nonumber \\
&\quad\quad
+( y^{-6} +6y^{-4} +19y^{-2} +24 +19y^{2} +6y^{4} +y^{6})     x_1^{14}
+( y^{-6} +4y^{-4} +13y^{-2} +16 +13y^{2} +4y^{4} +y^{6})     x_1^{15}\nonumber \\
&\quad\quad
+( 2y^{-4} +8y^{-2} +10 +8y^{2} +2y^{4})                        x_1^{16}
+( +y^{-4} +5y^{-2} +6 +5y^{2} +y^{4})                          x_1^{17}
+( 2y^{-2} +3 +2y^{2})                                            x_1^{18}\nonumber \\
&\quad\quad
+( y^{-2} +1 +y^{2})                                              x_1^{19}
\Bigr](x_1x_2)^3\nonumber \\
&\quad+\Bigl[
y^{-8} +2y^{-6} +6y^{-4} +10y^{-2} +13 +10y^{2} +6y^{4} +2y^{6} +y^{8}\nonumber \\
&\quad\quad
+( y^{-8} +4y^{-6} +12y^{-4} +22y^{-2} +27 +22y^{2} +12y^{4} +4y^{6} +y^{8})         x_1\nonumber \\
&\quad\quad
+( 2y^{-8} +8y^{-6} +24y^{-4} +41y^{-2} +51 +41y^{2} +24y^{4} +8y^{6} +2y^{8})       x_1^{2}\nonumber \\
&\quad\quad
+( 3y^{-8} +13y^{-6} +37y^{-4} +65y^{-2} +79 +65y^{2} +37y^{4} +13y^{6} +3y^{8})     x_1^{3}\nonumber \\
&\quad\quad
+( 5y^{-8} +20y^{-6} +57y^{-4} +97y^{-2} +119 +97y^{2} +57y^{4} +20y^{6} +5y^{8})    x_1^{4}\nonumber \\
&\quad\quad
+( 5y^{-8} +27y^{-6} +77y^{-4} +134y^{-2} +162 +134y^{2} +77y^{4} +27y^{6} +5y^{8})  x_1^{5}\nonumber \\
&\quad\quad
+( 7y^{-8} +35y^{-6} +103y^{-4} +178y^{-2} +217 +178y^{2} +103y^{4} +35y^{6} +7y^{8})x_1^{6}\nonumber \\
&\quad\quad
+( 7y^{-8} +41y^{-6} +122y^{-4} +218y^{-2} +265 +218y^{2} +122y^{4} +41y^{6} +7y^{8})x_1^{7}\nonumber \\
&\quad\quad
+( 8y^{-8} +45y^{-6} +138y^{-4} +245y^{-2} +301 +245y^{2} +138y^{4} +45y^{6} +8y^{8})x_1^{8}\nonumber \\
&\quad\quad
+( 7y^{-8} +47y^{-6} +145y^{-4} +264y^{-2} +322 +264y^{2} +145y^{4} +47y^{6} +7y^{8})x_1^{9}\nonumber \\
&\quad\quad
+( 7y^{-8} +46y^{-6} +149y^{-4} +272y^{-2} +336 +272y^{2} +149y^{4} +46y^{6} +7y^{8})x_1^{10}\nonumber \\
&\quad\quad
+( 5y^{-8} +42y^{-6} +141y^{-4} +267y^{-2} +329 +267y^{2} +141y^{4} +42y^{6} +5y^{8})x_1^{11}\nonumber \\
&\quad\quad
+( 5y^{-8} +36y^{-6} +130y^{-4} +249y^{-2} +313 +249y^{2} +130y^{4} +36y^{6} +5y^{8})x_1^{12}\nonumber \\
&\quad\quad
+( 3y^{-8} +29y^{-6} +109y^{-4} +219y^{-2} +275 +219y^{2} +109y^{4} +29y^{6} +3y^{8})x_1^{13}\nonumber \\
&\quad\quad
+( 2y^{-8} +21y^{-6} +88y^{-4} +179y^{-2} +231 +179y^{2} +88y^{4} +21y^{6} +2y^{8})  x_1^{14}\nonumber \\
&\quad\quad
+( y^{-8} +15y^{-6} +66y^{-4} +141y^{-2} +181 +141y^{2} +66y^{4} +15y^{6} +y^{8})    x_1^{15}\nonumber \\
&\quad\quad
+( y^{-8} +9y^{-6} +48y^{-4} +105y^{-2} +140 +105y^{2} +48y^{4} +9y^{6} +y^{8})      x_1^{16}\nonumber \\
&\quad\quad
+( 5y^{-6} +31y^{-4} +74y^{-2} +99 +74y^{2} +31y^{4} +5y^{6})                          x_1^{17}\nonumber \\
&\quad\quad
+( 2y^{-6} +19y^{-4} +47y^{-2} +67 +47y^{2} +19y^{4} +2y^{6})                          x_1^{18}\nonumber \\
&\quad\quad
+( y^{-6} +10y^{-4} +28y^{-2} +40 +28y^{2} +10y^{4} +y^{6})                            x_1^{19}
+( 5y^{-4} +14y^{-2} +23 +14y^{2} +5y^{4})                                               x_1^{20}\nonumber \\
&\quad\quad
+( 2y^{-4} +7y^{-2} +11 +7y^{2} +2y^{4})                                                 x_1^{21}
+( y^{-4} +3y^{-2} +6 +3y^{2} +y^{4})                                                    x_1^{22}
+(y^{-2} +2 +y^{2})                                                                        x_1^{23}\nonumber \\
&\quad\quad
+                                                                                            x_1^{24}
\Bigr](x_1x_2)^4\nonumber \\
&\quad+\Bigl[
y^{-10} +2y^{-8} +6y^{-6} +12y^{-4} +21y^{-2} +24 +21y^{2} +12y^{4} +6y^{6} +2y^{8} +y^{10}\nonumber \\
&\quad\quad
+( y^{-10} +4y^{-8} +13y^{-6} +30y^{-4} +51y^{-2} +60 +51y^{2} +30y^{4} +13y^{6} +4y^{8} +y^{10})               x_1\nonumber \\
&\quad\quad
+( 2y^{-10} +8y^{-8} +28y^{-6} +63y^{-4} +105y^{-2} +122 +105y^{2} +63y^{4} +28y^{6} +8y^{8} +2y^{10})          x_1^{2}\nonumber \\
&\quad\quad
+( 3y^{-10} +14y^{-8} +48y^{-6} +109y^{-4} +179y^{-2} +209 +179y^{2} +109y^{4} +48y^{6} +14y^{8} +3y^{10})      x_1^{3}\nonumber \\
&\quad\quad
+( 4y^{-10} +22y^{-8} +78y^{-6} +175y^{-4} +286y^{-2} +332 +286y^{2} +175y^{4} +78y^{6} +22y^{8} +4y^{10})      x_1^{4}\nonumber \\
&\quad\quad
+( 5y^{-10} +30y^{-8} +112y^{-6} +257y^{-4} +419y^{-2} +488 +419y^{2} +257y^{4} +112y^{6} +30y^{8} +5y^{10})    x_1^{5}\nonumber \\
&\quad\quad
+( 6y^{-10} +40y^{-8} +154y^{-6} +358y^{-4} +587y^{-2} +683 +587y^{2} +358y^{4} +154y^{6} +40y^{8} +6y^{10})    x_1^{6}\nonumber \\
&\quad\quad
+( 6y^{-10} +48y^{-8} +193y^{-6} +460y^{-4} +762y^{-2} +892 +762y^{2} +460y^{4} +193y^{6} +48y^{8} +6y^{10})    x_1^{7}\nonumber \\
&\quad\quad
+( 6y^{-10} +54y^{-8} +228y^{-6} +553y^{-4} +923y^{-2} +1082 +923y^{2} +553y^{4} +228y^{6} +54y^{8} +6y^{10})   x_1^{8}\nonumber \\
&\quad\quad
+( 6y^{-10} +58y^{-8} +255y^{-6} +632y^{-4} +1064y^{-2} +1251 +1064y^{2} +632y^{4} +255y^{6} +58y^{8} +6y^{10}) x_1^{9}\nonumber \\
&\quad\quad
+( 5y^{-10} +58y^{-8} +270y^{-6} +687y^{-4} +1172y^{-2} +1383 +1172y^{2} +687y^{4} +270y^{6} +58y^{8} +5y^{10}) x_1^{10}\nonumber \\
&\quad\quad
+( 4y^{-10} +54y^{-8} +270y^{-6} +711y^{-4} +1232y^{-2} +1461 +1232y^{2} +711y^{4} +270y^{6} +54y^{8} +4y^{10}) x_1^{11}\nonumber \\
&\quad\quad
+( 3y^{-10} +48y^{-8} +256y^{-6} +700y^{-4} +1237y^{-2} +1475 +1237y^{2} +700y^{4} +256y^{6} +48y^{8} +3y^{10}) x_1^{12}\nonumber \\
&\quad\quad
+( 2y^{-10} +40y^{-8} +230y^{-6} +654y^{-4} +1181y^{-2} +1418 +1181y^{2} +654y^{4} +230y^{6} +40y^{8} +2y^{10}) x_1^{13}\nonumber \\
&\quad\quad
+( y^{-10} +30y^{-8} +193y^{-6} +577y^{-4} +1068y^{-2} +1292 +1068y^{2} +577y^{4} +193y^{6} +30y^{8} +y^{10})   x_1^{14}\nonumber \\
&\quad\quad
+( y^{-10} +22y^{-8} +156y^{-6} +489y^{-4} +927y^{-2} +1128 +927y^{2} +489y^{4} +156y^{6} +22y^{8} +y^{10})     x_1^{15}\nonumber \\
&\quad\quad
+( 14y^{-8} +116y^{-6} +391y^{-4} +766y^{-2} +942 +766y^{2} +391y^{4} +116y^{6} +14y^{8})                         x_1^{16}\nonumber \\
&\quad\quad
+( 8y^{-8} +82y^{-6} +297y^{-4} +605y^{-2} +751 +605y^{2} +297y^{4} +82y^{6} +8y^{8})                             x_1^{17}\nonumber \\
&\quad\quad
+( 4y^{-8} +52y^{-6} +210y^{-4} +448y^{-2} +564 +448y^{2} +210y^{4} +52y^{6} +4y^{8})                             x_1^{18}\nonumber \\
&\quad\quad
+( 2y^{-8} +32y^{-6} +140y^{-4} +315y^{-2} +401 +315y^{2} +140y^{4} +32y^{6} +2y^{8})                             x_1^{19}\nonumber \\
&\quad\quad
+( 16y^{-6} +85y^{-4} +204y^{-2} +265 +204y^{2} +85y^{4} +16y^{6})                                                  x_1^{20}\nonumber \\
&\quad\quad
+( 8y^{-6} +49y^{-4} +127y^{-2} +166 +127y^{2} +49y^{4} +8y^{6})                                                    x_1^{21}\nonumber \\
&\quad\quad
+( 3y^{-6} +25y^{-4} +72y^{-2} +97 +72y^{2} +25y^{4} +3y^{6})                                                       x_1^{22}\nonumber \\
&\quad\quad
+( y^{-6} +11y^{-4} +38y^{-2} +52 +38y^{2} +11y^{4} +y^{6})                                                         x_1^{23}
+( 4y^{-4} +17y^{-2} +24 +17y^{2} +4y^{4})                                                                            x_1^{24}\nonumber \\
&\quad\quad
+( y^{-4} +7y^{-2} +10 +7y^{2} +y^{4})                                                                                x_1^{25}
+( 2y^{-2} +3 +2y^{2})                                                                                                  x_1^{26}
+( y^{-2} +1 +y^{2})                                                                                                    x_1^{27}
\Bigr](x_1x_2)^5
+{\cal O}(x_2^6),
\end{align}
\end{subequations}
}
where we have denoted $y_1=y_2^{-1}$ as $y$.

By setting $y=1$ we also determined the higher order terms in $x_2$ as
{\fontsize{9pt}{1pt}\selectfont
\begin{subequations}
\label{listofFlm_l2y1higherorderonly}
\begin{align}
&F_2^{(l=2)}(1;x_1;x_2)=
\cdots
+(153+275 x_1+144 x_1^{2}+22 x_1^{3})    (x_1x_2)^{7}\nonumber \\
&\quad +(252+468 x_1+278 x_1^{2}+50 x_1^{3}+x_1^{4})    (x_1x_2)^{8}
+(391+786 x_1+503 x_1^{2}+112 x_1^{3}+4 x_1^{4})    (x_1x_2)^{9}
+{\cal O}(x_2^{10}),\\
&F_3^{(l=2)}(1;x_1;x_2)=
\cdots
+(264+643 x_1+1042 x_1^2+967 x_1^3+646 x_1^4+251 x_1^5+62 x_1^6+4 x_1^7)    (x_1x_2)^7
+{\cal O}(x_2^8),\\
&F_4^{(l=2)}(1;x_1;x_2)=
\cdots
+(343+930 x_1+1881 x_1^{2}+2851 x_1^{3}+3280 x_1^{4}+3147 x_1^{5}+2389 x_1^{6}+1472 x_1^{7}+707 x_1^{8}\nonumber \\
&\quad\quad +253 x_1^{9}+60 x_1^{10} +7 x_1^{11}) (x_1x_2)^7
+{\cal O}(x_2^8),\\
&F_5^{(l=2)}(1;x_1;x_2)=
\cdots
+(390+1102 x_1+2438 x_1^{2}+4259 x_1^{3}+6404 x_1^{4}+7961 x_1^{5}+8907 x_1^{6}+8667 x_1^{7}+7472 x_1^{8}\nonumber \\
&\quad\quad +5605 x_1^{9}+3764 x_1^{10} +2126 x_1^{11}+1033 x_1^{12}+407 x_1^{13}+129 x_1^{14}+23 x_1^{15}+3 x_1^{16})(x_1x_2)^7
+{\cal O}(x_2^8),\\
&F_6^{(l=2)}(1;x_1;x_2)=
\cdots
+(414+1193 x_1+2737 x_1^{2}+5084 x_1^{3}+8422 x_1^{4}+12273 x_1^{5}+16006 x_1^{6}+19095 x_1^{7}\nonumber \\
&\quad\quad +20930 x_1^{8}+21119 x_1^{9} +19510 x_1^{10}+16689 x_1^{11}+13089 x_1^{12}+9454 x_1^{13}+6160 x_1^{14}+3644 x_1^{15}+1918 x_1^{16}\nonumber \\
&\quad\quad +888 x_1^{17}+341 x_1^{18} +108 x_1^{19}+23 x_1^{20}+3 x_1^{21})(x_1x_2)^7
+{\cal O}(x_2^8),\\
&F_7^{(l=2)}(1;x_1;x_2)=
\cdots
+(221+582 x_1+1286 x_1^{2}+2321 x_1^{3}+3849 x_1^{4}+5729 x_1^{5}+7948 x_1^{6}+9952 x_1^{7}+11816 x_1^{8}\nonumber \\
&\quad\quad +13069 x_1^{9}+13679 x_1^{10}+13358 x_1^{11}+12329 x_1^{12}+10630 x_1^{13}+8712 x_1^{14}+6629 x_1^{15}+4735 x_1^{16}+3109 x_1^{17}\nonumber \\
&\quad\quad +1907 x_1^{18}+1056 x_1^{19}+536 x_1^{20}+229 x_1^{21}+87 x_1^{22}+24 x_1^{23}+5 x_1^{24})(x_1x_2)^6\nonumber \\
&\quad +(429+1230 x_1+2876 x_1^{2}+5484 x_1^{3}+9493 x_1^{4}+14772 x_1^{5}+21346 x_1^{6}+28078 x_1^{7}+34851 x_1^{8}+40438 x_1^{9}\nonumber \\
&\quad\quad +44379 x_1^{10}+45713 x_1^{11}+44583 x_1^{12}+40902 x_1^{13}+35666 x_1^{14}+29195 x_1^{15}+22556 x_1^{16}+16272 x_1^{17}\nonumber \\
&\quad\quad +11041 x_1^{18} +6924 x_1^{19}+4038 x_1^{20}+2116 x_1^{21}+1010 x_1^{22}+416 x_1^{23}+151 x_1^{24}+39 x_1^{25}+7 x_1^{26})(x_1x_2)^7\nonumber \\
&\quad +{\cal O}(x_2^8),\\
&F_8^{(l=2)}(1;x_1;x_2)=
\cdots
+(221+582 x_1+1299 x_1^{2}+2365 x_1^{3}+3999 x_1^{4}+6129 x_1^{5}+8941 x_1^{6}+12073 x_1^{7}+15358 x_1^{8}\nonumber \\
&\quad\quad +18495 x_1^{9}+21306 x_1^{10}+23354 x_1^{11}+24545 x_1^{12}+24568 x_1^{13}+23503 x_1^{14}+21524 x_1^{15}+18899 x_1^{16}+15834 x_1^{17}\nonumber \\
&\quad\quad +12655 x_1^{18}+9613 x_1^{19}+6920 x_1^{20}+4723 x_1^{21}+3053 x_1^{22}+1831 x_1^{23}+1020 x_1^{24}+517 x_1^{25}+238 x_1^{26}+96 x_1^{27}\nonumber \\
&\quad\quad +34 x_1^{28}+8 x_1^{29}+x_1^{30})(x_1x_2)^6\nonumber \\
&\quad +(429+1245 x_1+2928 x_1^{2}+5651 x_1^{3}+9969 x_1^{4}+15996 x_1^{5}+24200 x_1^{6}+34103 x_1^{7}+45220 x_1^{8}+56902 x_1^{9}\nonumber \\
&\quad\quad +68303 x_1^{10}+78286 x_1^{11}+85914 x_1^{12}+90257 x_1^{13}+90732 x_1^{14}+87640 x_1^{15}+81207 x_1^{16}+72238 x_1^{17}\nonumber \\
&\quad\quad +61484 x_1^{18}+50163 x_1^{19}+39001 x_1^{20}+29024 x_1^{21}+20530 x_1^{22}+13783 x_1^{23}+8692 x_1^{24}+5167 x_1^{25}+2837 x_1^{26}\nonumber \\
&\quad\quad +1452 x_1^{27}+668 x_1^{28}+275 x_1^{29}+90 x_1^{30}+26 x_1^{31}+4 x_1^{32})(x_1x_2)^7
+{\cal O}(x_2^8),
\end{align}
\end{subequations}
}
where the lower order terms $\cdots$ are obtained from \eqref{listofFlm_l2withy} by setting $y=1$.

\subsection{$l=3$}
\label{app:listofFlm_l3withy}

We find that $F^{(l=3)}_m(y_\alpha;x_1;x_2)$ are given as
{\fontsize{9pt}{1pt}\selectfont
\begin{subequations}
\label{listofFlm_l3withy}
\begin{align}
&F_1^{(l=3)}(y_\alpha;x_1;x_2)=
1
+(2 + A^{(3)}_1)x_1x_2
+(5 + 2 A^{(3)}_1)(x_1x_2)^2
+(10 + 5 A^{(3)}_1 + B^{(3)}_2)(x_1x_2)^3
+{\cal O}(x_2^4),\\
&F_2^{(l=3)}(y_\alpha;x_1;x_2)=
1
+\Bigl[2 + A^{(3)}_1 + (2 + A^{(3)}_1)x_1\Bigr]x_1x_2\nonumber \\
&\quad +\Bigl[8 + 4A^{(3)}_1 + A^{(3)}_2 + B^{(3)}_2 + (9 + 5A^{(3)}_1 + B^{(3)}_2)x_1 + (3 + A^{(3)}_1)x_1^2\Bigr](x_1x_2)^2\nonumber \\
&\quad +\Bigl[20 + 12A^{(3)}_1 + 2A^{(3)}_2 + 4B^{(3)}_2 + (31 + 18A^{(3)}_1 + 2A^{(3)}_2 + 5B^{(3)}_2)x_1 + (12 + 6A^{(3)}_1 + B^{(3)}_2)x_1^2 + x_1^3\Bigr](x_1x_2)^3\nonumber \\
&\quad +{\cal O}(x_2^4),\\
&F_3^{(l=3)}(y_\alpha;x_1;x_2)=
1
+\Bigl[2 + A^{(3)}_1 + (2 + A^{(3)}_1)x_1 + (2 + A^{(3)}_1)x_1^2\Bigr]x_1x_2\nonumber \\
&\quad +\Bigl[8 + 4A^{(3)}_1 + A^{(3)}_2 + B^{(3)}_2 + (12 + 7A^{(3)}_1 + A^{(3)}_2 + 2B^{(3)}_2)x_1 + (15 + 8A^{(3)}_1 + A^{(3)}_2 + 2B^{(3)}_2)x_1^2\nonumber \\
&\quad\quad + (7 + 4A^{(3)}_1 + B^{(3)}_2)x_1^3 + (3 + A^{(3)}_1)x_1^4\Bigr](x_1x_2)^2\nonumber \\
&\quad +\Bigl[24 + 15A^{(3)}_1 + 4A^{(3)}_2 + A^{(3)}_3 + 6B^{(3)}_2 + B^{(3)}_3 + (48 + 31A^{(3)}_1 + 7A^{(3)}_2 + 12B^{(3)}_2 + B^{(3)}_3) x_1\nonumber \\
&\quad\quad + (67 + 42A^{(3)}_1 + 8A^{(3)}_2 + 15B^{(3)}_2 + B^{(3)}_3) x_1^2 + (49 + 29A^{(3)}_1 + 4A^{(3)}_2 + 9B^{(3)}_2) x_1^3\nonumber \\
&\quad\quad + (24 + 13A^{(3)}_1 + A^{(3)}_2 + 3B^{(3)}_2) x_1^4 + (5 + 2A^{(3)}_1)x_1^5 + x_1^6\Bigr] (x_1x_2)^3 +{\cal O}(x_2^4),\\
&F_4^{(l=3)}(y_\alpha;x_1;x_2)=
1
+\Bigl[
2+A^{(3)}_1
+(2+A^{(3)}_1)x_1  
+(2+A^{(3)}_1)x_1^2
+(2+A^{(3)}_1)x_1^3
\Bigr]x_1x_2\nonumber \\
&\quad +\Bigl[
8+4A^{(3)}_1+A^{(3)}_2+B^{(3)}_2
+(12+7A^{(3)}_1+A^{(3)}_2+2B^{(3)}_2)    x_1  
+(18+10A^{(3)}_1+2A^{(3)}_2+3B^{(3)}_2)  x_1^2\nonumber \\
&\quad\quad
+(19+11A^{(3)}_1+A^{(3)}_2+3B^{(3)}_2)   x_1^3
+(13+7A^{(3)}_1+A^{(3)}_2+2B^{(3)}_2)    x_1^4
+(7+4A^{(3)}_1+B^{(3)}_2)          x_1^5\nonumber \\
&\quad\quad +(3+A^{(3)}_1)               x_1^6
\Bigr](x_1x_2)^2\nonumber \\
&\quad +\Bigl[
24+15A^{(3)}_1+4A^{(3)}_2+A^{(3)}_3+6B^{(3)}_2+B^{(3)}_3
+(52+34A^{(3)}_1+9A^{(3)}_2+A^{(3)}_3+14B^{(3)}_2+2B^{(3)}_3)   x_1  \nonumber \\
&\quad\quad +(88+58A^{(3)}_1+15A^{(3)}_2+A^{(3)}_3+24B^{(3)}_2+3B^{(3)}_3)  x_1^2\nonumber \\
&\quad\quad +(115+74A^{(3)}_1+17A^{(3)}_2+A^{(3)}_3+29B^{(3)}_2+3B^{(3)}_3) x_1^3\nonumber \\
&\quad\quad +(103+66A^{(3)}_1+14A^{(3)}_2+25B^{(3)}_2+2B^{(3)}_3)     x_1^4
+(79+49A^{(3)}_1+9A^{(3)}_2+17B^{(3)}_2+B^{(3)}_3)        x_1^5\nonumber \\
&\quad\quad
+(44+25A^{(3)}_1+3A^{(3)}_2+7B^{(3)}_2)             x_1^6
+(17+9A^{(3)}_1+A^{(3)}_2+2B^{(3)}_2)               x_1^7
+(5+2A^{(3)}_1)                         x_1^8
+x_1^9
\Bigr]
(x_1x_2)^3\nonumber \\
&\quad +{\cal O}(x_2^4),\\
&F_5^{(l=3)}(y_\alpha;x_1;x_2)=
1
+\Bigl[2+A^{(3)}_1
+(2+A^{(3)}_1)x_1  
+(2+A^{(3)}_1)x_1^2
+(2+A^{(3)}_1)x_1^3
+(2+A^{(3)}_1)x_1^4
\Bigr]x_1x_2\nonumber \\
&\quad
+\Bigl[8+4A^{(3)}_1+A^{(3)}_2+B^{(3)}_2
+(12+7A^{(3)}_1+A^{(3)}_2+2B^{(3)}_2)      x_1  
+(18+10A^{(3)}_1+2A^{(3)}_2+3B^{(3)}_2)    x_1^2\nonumber \\
&\quad\quad
+(22+13A^{(3)}_1+2A^{(3)}_2+4B^{(3)}_2)    x_1^3
+(25+14A^{(3)}_1+2A^{(3)}_2+4B^{(3)}_2)    x_1^4
+(17+10A^{(3)}_1+A^{(3)}_2+3B^{(3)}_2)     x_1^5\nonumber \\
&\quad\quad
+(13+7A^{(3)}_1+A^{(3)}_2+2B^{(3)}_2)      x_1^6
+(7+4A^{(3)}_1+B^{(3)}_2)            x_1^7
+(3+A^{(3)}_1)                 x_1^8
\Bigr](x_1x_2)^2\nonumber \\
&\quad
+\Bigl[24+15A^{(3)}_1+4A^{(3)}_2+A^{(3)}_3+6B^{(3)}_2+B^{(3)}_3
+(52+34A^{(3)}_1+9A^{(3)}_2+A^{(3)}_3+14B^{(3)}_2+2B^{(3)}_3)      x_1   \nonumber \\
&\quad\quad +(92+61A^{(3)}_1+17A^{(3)}_2+2A^{(3)}_3+26B^{(3)}_2+4B^{(3)}_3)    x_1^2 \nonumber \\
&\quad\quad +(136+90A^{(3)}_1+24A^{(3)}_2+2A^{(3)}_3+38B^{(3)}_2+5B^{(3)}_3)   x_1^3 \nonumber \\
&\quad\quad +(173+114A^{(3)}_1+29A^{(3)}_2+2A^{(3)}_3+47B^{(3)}_2+6B^{(3)}_3)  x_1^4 \nonumber \\
&\quad\quad +(169+111A^{(3)}_1+27A^{(3)}_2+A^{(3)}_3+45B^{(3)}_2+5B^{(3)}_3)   x_1^5 \nonumber \\
&\quad\quad
+(153+99A^{(3)}_1+23A^{(3)}_2+A^{(3)}_3+39B^{(3)}_2+4B^{(3)}_3)    x_1^6
+(117+74A^{(3)}_1+15A^{(3)}_2+27B^{(3)}_2+2B^{(3)}_3)        x_1^7 \nonumber \\
&\quad\quad
+(74+45A^{(3)}_1+8A^{(3)}_2+15B^{(3)}_2+B^{(3)}_3)           x_1^8 
+(37+21A^{(3)}_1+3A^{(3)}_2+6B^{(3)}_2)                x_1^9 \nonumber \\
&\quad\quad +(17+9A^{(3)}_1+A^{(3)}_2+2B^{(3)}_2)                  x_1^{10}
+(5+2A^{(3)}_1)                            x_1^{11}
+x_1^{12}
\Bigr](x_1x_2)^3
+{\cal O}(x_2^4),\\
&F_6^{(l=3)}(y_\alpha;x_1;x_2)=
1
+\Bigl[2+A^{(3)}_1
+(2+A^{(3)}_1)x_1  
+(2+A^{(3)}_1)x_1^2
+(2+A^{(3)}_1)x_1^3
+(2+A^{(3)}_1)x_1^4\nonumber \\
&\quad\quad
+(2+A^{(3)}_1)x_1^5
\Bigr]x_1x_2\nonumber \\
&\quad
+\Bigl[8+4A^{(3)}_1+A^{(3)}_2+B^{(3)}_2
+(12+7A^{(3)}_1+A^{(3)}_2+2B^{(3)}_2)   x_1   
+(18+10A^{(3)}_1+2A^{(3)}_2+3B^{(3)}_2) x_1^2 \nonumber \\
&\quad\quad
+(22+13A^{(3)}_1+2A^{(3)}_2+4B^{(3)}_2) x_1^3
+(28+16A^{(3)}_1+3A^{(3)}_2+5B^{(3)}_2) x_1^4 
+(29+17A^{(3)}_1+2A^{(3)}_2+5B^{(3)}_2) x_1^5 \nonumber \\
&\quad\quad
+(23+13A^{(3)}_1+2A^{(3)}_2+4B^{(3)}_2) x_1^6
+(17+10A^{(3)}_1+A^{(3)}_2+3B^{(3)}_2)  x_1^7 
+(13+7A^{(3)}_1+A^{(3)}_2+2B^{(3)}_2)   x_1^8 \nonumber \\
&\quad\quad
+(7+4A^{(3)}_1+B^{(3)}_2)         x_1^9 
+(3+A^{(3)}_1)              x_1^{10}
\Bigr](x_1x_2)^2\nonumber \\
&\quad
+\Bigl[24+15A^{(3)}_1+4A^{(3)}_2+A^{(3)}_3+6B^{(3)}_2+B^{(3)}_3
+(52+34A^{(3)}_1+9A^{(3)}_2+A^{(3)}_3+14B^{(3)}_2+2B^{(3)}_3)      x_1   \nonumber \\
&\quad\quad
+(92+61A^{(3)}_1+17A^{(3)}_2+2A^{(3)}_3+26B^{(3)}_2+4B^{(3)}_3)    x_1^2 \nonumber \\
&\quad\quad
+(140+93A^{(3)}_1+26A^{(3)}_2+3A^{(3)}_3+40B^{(3)}_2+6B^{(3)}_3)   x_1^3 \nonumber \\
&\quad\quad
+(194+130A^{(3)}_1+36A^{(3)}_2+3A^{(3)}_3+56B^{(3)}_2+8B^{(3)}_3)  x_1^4 \nonumber \\
&\quad\quad
+(239+159A^{(3)}_1+42A^{(3)}_2+3A^{(3)}_3+67B^{(3)}_2+9B^{(3)}_3)  x_1^5 \nonumber \\
&\quad\quad
+(247+164A^{(3)}_1+43A^{(3)}_2+3A^{(3)}_3+69B^{(3)}_2+9B^{(3)}_3)  x_1^6 \nonumber \\
&\quad\quad
+(237+157A^{(3)}_1+40A^{(3)}_2+2A^{(3)}_3+65B^{(3)}_2+8B^{(3)}_3)  x_1^7 \nonumber \\
&\quad\quad
+(209+137A^{(3)}_1+33A^{(3)}_2+A^{(3)}_3+55B^{(3)}_2+6B^{(3)}_3)   x_1^8 \nonumber \\
&\quad\quad
+(167+107A^{(3)}_1+24A^{(3)}_2+A^{(3)}_3+41B^{(3)}_2+4B^{(3)}_3)   x_1^9 \nonumber \\
&\quad\quad
+(112+70A^{(3)}_1+14A^{(3)}_2+25B^{(3)}_2+2B^{(3)}_3)        x_1^{10}
+(67+41A^{(3)}_1+8A^{(3)}_2+14B^{(3)}_2+B^{(3)}_3)           x_1^{11}\nonumber \\
&\quad\quad
+(37+21A^{(3)}_1+3A^{(3)}_2+6B^{(3)}_2)                x_1^{12}
+(17+9A^{(3)}_1+A^{(3)}_2+2B^{(3)}_2)                  x_1^{13}
+(5+2A^{(3)}_1)                            x_1^{14}
+x_1^{15}\Bigr](x_1x_2)^3\nonumber \\
&\quad
+{\cal O}(x_2^4),\\
&F_7^{(l=3)}(y_\alpha;x_1;x_2)=
1
+\Bigl[2+A^{(3)}_1
+(2+A^{(3)}_1)x_1  
+(2+A^{(3)}_1)x_1^2
+(2+A^{(3)}_1)x_1^3
+(2+A^{(3)}_1)x_1^4
+(2+A^{(3)}_1)x_1^5\nonumber \\
&\quad\quad
+(2+A^{(3)}_1)x_1^6
\Bigr]x_1x_2\nonumber \\
&\quad
+\Bigl[8+4A^{(3)}_1+A^{(3)}_2+B^{(3)}_2
+(12+7A^{(3)}_1+A^{(3)}_2+2B^{(3)}_2)   x_1   
+(18+10A^{(3)}_1+2A^{(3)}_2+3B^{(3)}_2) x_1^2 \nonumber \\
&\quad\quad
+(22+13A^{(3)}_1+2A^{(3)}_2+4B^{(3)}_2) x_1^3
+(28+16A^{(3)}_1+3A^{(3)}_2+5B^{(3)}_2) x_1^4 
+(32+19A^{(3)}_1+3A^{(3)}_2+6B^{(3)}_2) x_1^5 \nonumber \\
&\quad\quad
+(35+20A^{(3)}_1+3A^{(3)}_2+6B^{(3)}_2) x_1^6
+(27+16A^{(3)}_1+2A^{(3)}_2+5B^{(3)}_2) x_1^7 
+(23+13A^{(3)}_1+2A^{(3)}_2+4B^{(3)}_2) x_1^8 \nonumber \\
&\quad\quad
+(17+10A^{(3)}_1+A^{(3)}_2+3B^{(3)}_2)  x_1^9
+(13+7A^{(3)}_1+A^{(3)}_2+2B^{(3)}_2)   x_1^{10}
+(7+4A^{(3)}_1+B^{(3)}_2)         x_1^{11}\nonumber \\
&\quad\quad 
+(3+A^{(3)}_1)              x_1^{12}
\Bigr](x_1x_2)^2\nonumber \\
&\quad
+\Bigl[24+15A^{(3)}_1+4A^{(3)}_2+A^{(3)}_3+6B^{(3)}_2+B^{(3)}_3
+(52+34A^{(3)}_1+9A^{(3)}_2+A^{(3)}_3+14B^{(3)}_2+2B^{(3)}_3)       x_1   \nonumber \\
&\quad\quad
+(92+61A^{(3)}_1+17A^{(3)}_2+2A^{(3)}_3+26B^{(3)}_2+4B^{(3)}_3)     x_1^2 \nonumber \\
&\quad\quad
+(140+93A^{(3)}_1+26A^{(3)}_2+3A^{(3)}_3+40B^{(3)}_2+6B^{(3)}_3)    x_1^3 \nonumber \\
&\quad\quad
+(198+133A^{(3)}_1+38A^{(3)}_2+4A^{(3)}_3+58B^{(3)}_2+9B^{(3)}_3)   x_1^4  \nonumber \\
&\quad\quad
+(260+175A^{(3)}_1+49A^{(3)}_2+4A^{(3)}_3+76B^{(3)}_2+11B^{(3)}_3)  x_1^5 \nonumber \\
&\quad\quad
+(317+212A^{(3)}_1+58A^{(3)}_2+5A^{(3)}_3+91B^{(3)}_2+13B^{(3)}_3)  x_1^6  \nonumber \\
&\quad\quad
+(331+222A^{(3)}_1+60A^{(3)}_2+4A^{(3)}_3+95B^{(3)}_2+13B^{(3)}_3)  x_1^7 \nonumber \\
&\quad\quad
+(333+223A^{(3)}_1+60A^{(3)}_2+4A^{(3)}_3+95B^{(3)}_2+13B^{(3)}_3)  x_1^8 \nonumber \\
&\quad\quad
+(313+208A^{(3)}_1+54A^{(3)}_2+3A^{(3)}_3+87B^{(3)}_2+11B^{(3)}_3)  x_1^9 \nonumber \\
&\quad\quad
+(277+183A^{(3)}_1+46A^{(3)}_2+2A^{(3)}_3+75B^{(3)}_2+9B^{(3)}_3)   x_1^{10}\nonumber \\
&\quad\quad
+(223+145A^{(3)}_1+34A^{(3)}_2+A^{(3)}_3+57B^{(3)}_2+6B^{(3)}_3)    x_1^{11}\nonumber \\
&\quad\quad
+(162+103A^{(3)}_1+23A^{(3)}_2+A^{(3)}_3+39B^{(3)}_2+4B^{(3)}_3)    x_1^{12}
+(105+66A^{(3)}_1+14A^{(3)}_2+24B^{(3)}_2+2B^{(3)}_3)         x_1^{13}\nonumber \\
&\quad\quad
+(67+41A^{(3)}_1+8A^{(3)}_2+14B^{(3)}_2+B^{(3)}_3)            x_1^{14}
+(37+21A^{(3)}_1+3A^{(3)}_2+6B^{(3)}_2)                 x_1^{15}\nonumber \\
&\quad\quad
+(17+9A^{(3)}_1+A^{(3)}_2+2B^{(3)}_2)                   x_1^{16}
+(5+2A^{(3)}_1)                             x_1^{17}
+x_1^{18}\Bigr](x_1x_2)^3
+{\cal O}(x_2^4),\\
&F_8^{(l=3)}(y_\alpha;x_1;x_2)=
1
+\Bigl[2+A^{(3)}_1
+(2+A^{(3)}_1)x_1  
+(2+A^{(3)}_1)x_1^2
+(2+A^{(3)}_1)x_1^3
+(2+A^{(3)}_1)x_1^4
+(2+A^{(3)}_1)x_1^5\nonumber \\
&\quad\quad +(2+A^{(3)}_1)x_1^6
+(2+A^{(3)}_1)x_1^7
\Bigr]x_1x_2\nonumber \\
&\quad
+\Bigl[8+4A^{(3)}_1+A^{(3)}_2+B^{(3)}_2
+(12+7A^{(3)}_1+A^{(3)}_2+2B^{(3)}_2)    x_1   
+(18+10A^{(3)}_1+2A^{(3)}_2+3B^{(3)}_2)  x_1^2 \nonumber \\
&\quad\quad
+(22+13A^{(3)}_1+2A^{(3)}_2+4B^{(3)}_2)  x_1^3 
+(28+16A^{(3)}_1+3A^{(3)}_2+5B^{(3)}_2)  x_1^4 
+(32+19A^{(3)}_1+3A^{(3)}_2+6B^{(3)}_2)  x_1^5 \nonumber \\
&\quad\quad
+(38+22A^{(3)}_1+4A^{(3)}_2+7B^{(3)}_2)  x_1^6 
+(39+23A^{(3)}_1+3A^{(3)}_2+7B^{(3)}_2)  x_1^7 
+(33+19A^{(3)}_1+3A^{(3)}_2+6B^{(3)}_2)  x_1^8 \nonumber \\
&\quad\quad
+(27+16A^{(3)}_1+2A^{(3)}_2+5B^{(3)}_2)  x_1^9 
+(23+13A^{(3)}_1+2A^{(3)}_2+4B^{(3)}_2)  x_1^{10}
+(17+10A^{(3)}_1+A^{(3)}_2+3B^{(3)}_2)   x_1^{11}\nonumber \\
&\quad\quad
+(13+7A^{(3)}_1+A^{(3)}_2+2B^{(3)}_2)    x_1^{12}
+(7+4A^{(3)}_1+B^{(3)}_2)          x_1^{13}
+(3+A^{(3)}_1)               x_1^{14}
\Bigr](x_1x_2)^2\nonumber \\
&\quad
+\Bigl[24+15A^{(3)}_1+4A^{(3)}_2+A^{(3)}_3+6B^{(3)}_2+B^{(3)}_3
+(52+34A^{(3)}_1+9A^{(3)}_2+A^{(3)}_3+14B^{(3)}_2+2B^{(3)}_3)       x_1   \nonumber \\
&\quad\quad
+(92+61A^{(3)}_1+17A^{(3)}_2+2A^{(3)}_3+26B^{(3)}_2+4B^{(3)}_3)     x_1^2 \nonumber \\
&\quad\quad
+(140+93A^{(3)}_1+26A^{(3)}_2+3A^{(3)}_3+40B^{(3)}_2+6B^{(3)}_3)    x_1^3 \nonumber \\
&\quad\quad
+(198+133A^{(3)}_1+38A^{(3)}_2+4A^{(3)}_3+58B^{(3)}_2+9B^{(3)}_3)   x_1^4 \nonumber \\
&\quad\quad
+(264+178A^{(3)}_1+51A^{(3)}_2+5A^{(3)}_3+78B^{(3)}_2+12B^{(3)}_3)  x_1^5 \nonumber \\
&\quad\quad
+(338+228A^{(3)}_1+65A^{(3)}_2+6A^{(3)}_3+100B^{(3)}_2+15B^{(3)}_3) x_1^6 \nonumber \\
&\quad\quad
+(401+270A^{(3)}_1+75A^{(3)}_2+6A^{(3)}_3+117B^{(3)}_2+17B^{(3)}_3) x_1^7 \nonumber \\
&\quad\quad
+(427+288A^{(3)}_1+80A^{(3)}_2+6A^{(3)}_3+125B^{(3)}_2+18B^{(3)}_3) x_1^8 \nonumber \\
&\quad\quad
+(437+294A^{(3)}_1+81A^{(3)}_2+6A^{(3)}_3+127B^{(3)}_2+18B^{(3)}_3) x_1^9 \nonumber \\
&\quad\quad
+(427+287A^{(3)}_1+78A^{(3)}_2+5A^{(3)}_3+123B^{(3)}_2+17B^{(3)}_3) x_1^{10}\nonumber \\
&\quad\quad
+(399+267A^{(3)}_1+71A^{(3)}_2+4A^{(3)}_3+113B^{(3)}_2+15B^{(3)}_3) x_1^{11}\nonumber \\
&\quad\quad
+(353+234A^{(3)}_1+60A^{(3)}_2+3A^{(3)}_3+97B^{(3)}_2+12B^{(3)}_3)  x_1^{12}\nonumber \\
&\quad\quad
+(291+191A^{(3)}_1+47A^{(3)}_2+2A^{(3)}_3+77B^{(3)}_2+9B^{(3)}_3)   x_1^{13}\nonumber \\
&\quad\quad
+(218+141A^{(3)}_1+33A^{(3)}_2+A^{(3)}_3+55B^{(3)}_2+6B^{(3)}_3)    x_1^{14}\nonumber \\
&\quad\quad
+(155+99A^{(3)}_1+23A^{(3)}_2+A^{(3)}_3+38B^{(3)}_2+4B^{(3)}_3)     x_1^{15}\nonumber \\
&\quad\quad
+(105+66A^{(3)}_1+14A^{(3)}_2+24B^{(3)}_2+2B^{(3)}_3)         x_1^{16}
+(67+41A^{(3)}_1+8A^{(3)}_2+14B^{(3)}_2+B^{(3)}_3)            x_1^{17}\nonumber \\
&\quad\quad
+(37+21A^{(3)}_1+3A^{(3)}_2+6B^{(3)}_2)                 x_1^{18}
+(17+9A^{(3)}_1+A^{(3)}_2+2B^{(3)}_2)                   x_1^{19}
+(5+2A^{(3)}_1)                             x_1^{20}
+x_1^{21}\Bigr](x_1x_2)^3\nonumber \\
&\quad
+{\cal O}(x_2^4),
\end{align}
\end{subequations}
}
with
\begin{subequations}
\label{ABCl3}
\begin{align}
&A^{(3)}_1=\sum_{\alpha\neq\beta}^3\frac{y_\alpha}{y_\beta},\quad
A^{(3)}_2=\sum_{\alpha\neq\beta}^3\frac{y_\alpha^2}{y_\beta^2},\quad
A^{(3)}_3=\sum_{\alpha\neq\beta}^3\frac{y_\alpha^3}{y_\beta^3},\\
&B^{(3)}_2=\sum_{\alpha=1}^3
\Bigl(
\frac{y_1y_2y_3}{y_\alpha^3}
+\frac{y_\alpha^3}{y_1y_2y_3}
\Bigr),\\
&B^{(3)}_3=
  \frac{(y_1 + y_2)y_1y_2}{y_3^3}
+ \frac{(y_1 + y_3)y_1 y_3}{y_2^3}
+ \frac{(y_2 + y_3)y_2y_3}{y_1^3}
+ \frac{y_1^3}{y_2^2y_3}
+ \frac{y_1^3}{y_2y_3^2}
+ \frac{y_2^3}{y_1^2y_3}
+ \frac{y_2^3}{y_1y_3^2}\nonumber \\
&\quad + \frac{y_3^3}{y_1^2y_2}
+ \frac{y_3^3}{y_1y_2^2}.
\end{align}
\end{subequations}

By setting $y_\alpha=1$ we also determined the higher order terms in $x_2$ as
{\fontsize{9pt}{1pt}\selectfont
\begin{subequations}
\label{listofFlm_l3y1higherorderonly}
\begin{align}
&F^{(l=3)}_2(y_\alpha=1;x_1;x_2)=
\cdots+(376+585 x_1+243 x_1^2+17 x_1^3)    (x_1x_2)^4
+(944+1657 x_1+820 x_1^2+107 x_1^3)    (x_1x_2)^5\nonumber \\
&\quad +(2236+4231 x_1+2457 x_1^2+425 x_1^3+9 x_1^4)    (x_1x_2)^6
+{\cal O}(x_2^7),\\
&F^{(l=3)}_3(y_\alpha=1;x_1;x_2)=
\cdots
+(601+1384 x_1+1988 x_1^2+1584 x_1^3+827 x_1^4+223 x_1^5+26 x_1^6)    (x_1x_2)^4\nonumber \\
&\quad
+(1772+4536 x_1+7120 x_1^2+6508 x_1^3+4032 x_1^4+1466 x_1^5+306 x_1^6+18 x_1^7)    (x_1x_2)^5\nonumber \\
&\quad +(4668+13240 x_1+22304 x_1^2+22776 x_1^3+15901 x_1^4+7090 x_1^5+1976 x_1^6+262 x_1^7+9 x_1^8) (x_1x_2)^6
+{\cal O}(x_2^7),\\
&F^{(l=3)}_4(y_\alpha=1;x_1;x_2)=
\cdots
+(726+1835 x_1+3455 x_1^2+4721 x_1^3+4822 x_1^4+3985 x_1^5+2480 x_1^6+1178 x_1^7+404 x_1^8\nonumber \\
&\quad\quad +79 x_1^9+9 x_1^{10})    (x_1x_2)^4\nonumber \\
&\quad +(2248+6643 x_1+13556 x_1^2+20322 x_1^3+23113 x_1^4+21456 x_1^5+15541 x_1^6+9019 x_1^7+4007 x_1^8+1269 x_1^9\nonumber \\
&\quad\quad +271 x_1^{10}+27 x_1^{11}) (x_1x_2)^5\nonumber \\
&\quad +(6496+21179 x_1+46497 x_1^{2}+74808 x_1^{3}+92773 x_1^{4}+93815 x_1^{5}+76173 x_1^{6}+50327 x_1^{7}+26614 x_1^{8}+10771 x_1^{9}\nonumber \\
&\quad\quad +3277 x_1^{10} +630 x_1^{11}+64 x_1^{12})(x_1x_2)^6
+{\cal O}(x_2^7),\\
&F^{(l=3)}_5(y_\alpha=1;x_1;x_2)=
\cdots
+(726+1960 x_1+4031 x_1^2+6539 x_1^3+8978 x_1^4+10041 x_1^5+9984 x_1^6+8481 x_1^7+6171 x_1^8\nonumber \\
&\quad\quad +3785 x_1^9+2005 x_1^{10} +828 x_1^{11}+260 x_1^{12}+62 x_1^{13}+9 x_1^{14})    (x_1x_2)^4\nonumber \\
&\quad +(2464+7552 x_1+17083 x_1^{2}+30123 x_1^{3}+44805 x_1^{4}+55370 x_1^{5}+60470 x_1^{6}+57241 x_1^{7}+47450 x_1^{8}+34065 x_1^{9}\nonumber \\
&\quad\quad +21459 x_1^{10}+11335 x_1^{11}+5042 x_1^{12}+1793 x_1^{13}+496 x_1^{14}+81 x_1^{15}+9 x_1^{16})(x_1x_2)^5\nonumber \\
&\quad +(7361+25536x_1+62274x_1^{2}+117946x_1^{3}+187159x_1^{4}+249877x_1^{5}+293312x_1^{6}+301691x_1^{7}+273969x_1^{8}\nonumber \\
&\quad\quad +219177x_1^{9}+155225x_1^{10}+95571x_1^{11}+50913x_1^{12}+23033x_1^{13}+8613x_1^{14}+2486x_1^{15}+530x_1^{16}+71x_1^{17}\nonumber \\
&\quad\quad +x_1^{18})(x_1x_2)^6
+{\cal O}(x_2^7),\\
&F^{(l=3)}_6(y_\alpha=1;x_1;x_2)=
\cdots
+(726+1960 x_1+4156 x_1^{2}+7115 x_1^{3}+10921 x_1^{4}+14548 x_1^{5}+17184 x_1^{6}+18397 x_1^{7}\nonumber \\
&\quad\quad +18010 x_1^{8}+15941 x_1^{9} +12657 x_1^{10} +9130 x_1^{11}+5902 x_1^{12}+3383 x_1^{13}+1655 x_1^{14}+684 x_1^{15}+243 x_1^{16}+62 x_1^{17}\nonumber \\
&\quad\quad +9 x_1^{18})    (x_1x_2)^{4}\nonumber \\
&\quad (2464+7768 x_1+18208 x_1^{2}+34299 x_1^{3}+56675 x_1^{4}+82020 x_1^{5}+105725 x_1^{6}+123878 x_1^{7}+132582 x_1^{8}+129910 x_1^{9}\nonumber \\
&\quad\quad +115915 x_1^{10}+95126 x_1^{11}+71098 x_1^{12}+48444 x_1^{13}+29543 x_1^{14}+16178 x_1^{15}+7806 x_1^{16}+3250 x_1^{17}+1109 x_1^{18}\nonumber \\
&\quad\quad +306 x_1^{19}+63 x_1^{20}+9 x_1^{21})(x_1x_2)^5\nonumber \\
&\quad +(7704+27137x_1+69205x_1^{2}+140279x_1^{3}+247098x_1^{4}+381244x_1^{5}+526202x_1^{6}+659483x_1^{7}+757072x_1^{8}\nonumber \\
&\quad\quad +798327x_1^{9} +773938x_1^{10}+692866x_1^{11}+571629x_1^{12}+433989x_1^{13}+301543x_1^{14}+191243x_1^{15}+109959x_1^{16}\nonumber \\
&\quad\quad +56552x_1^{17}+25682x_1^{18} +10070x_1^{19}+3345x_1^{20}+875x_1^{21}+170x_1^{22}+17x_1^{23}+x_1^{24})(x_1x_2)^6
+{\cal O}(x_2^7),\\
&F^{(l=3)}_7(y_\alpha=1;x_1;x_2)=
\cdots
+(726+1960 x_1+4156 x_1^{2}+7240 x_1^{3}+11497 x_1^{4}+16491 x_1^{5}+21816 x_1^{6}+25948 x_1^{7}\nonumber \\
&\quad\quad +29070 x_1^{8}+30317 x_1^{9} +29700 x_1^{10} +26965 x_1^{11}+22815 x_1^{12}+17914 x_1^{13}+13237 x_1^{14}+8934 x_1^{15}+5500 x_1^{16}\nonumber \\
&\quad\quad +3033 x_1^{17}+1511 x_1^{18} +667 x_1^{19}+243 x_1^{20} +62 x_1^{21}+9 x_1^{22})    (x_1x_2)^{4}\nonumber \\
&\quad +(2464+7768 x_1+18424 x_1^{2}+35424 x_1^{3}+61067 x_1^{4}+94539 x_1^{5}+134660 x_1^{6}+174740 x_1^{7}+212369 x_1^{8}\nonumber \\
&\quad\quad +240862 x_1^{9}+257138 x_1^{10}+256942 x_1^{11}+241560 x_1^{12}+213087 x_1^{13}+177511 x_1^{14}+138115 x_1^{15}+100613 x_1^{16}\nonumber \\
&\quad\quad +67968 x_1^{17}+42746 x_1^{18}+24710 x_1^{19}+13071 x_1^{20}+6103 x_1^{21}+2566 x_1^{22}+919 x_1^{23}+288 x_1^{24}+63 x_1^{25}\nonumber \\
&\quad\quad +9 x_1^{26})(x_1x_2)^5\nonumber \\
&\quad+(7704+27480x_1+71149x_1^{2}+148289x_1^{3}+273084x_1^{4}+450527x_1^{5}+681185x_1^{6}+943924x_1^{7}+1221033x_1^{8}\nonumber \\
&\quad\quad  +1477314x_1^{9}+1682392x_1^{10}+1802650x_1^{11}+1823359x_1^{12}+1741134x_1^{13}+1574032x_1^{14}+1342719x_1^{15}\nonumber \\
&\quad\quad +1080961x_1^{16} +818888x_1^{17}+583416x_1^{18}+389269x_1^{19}+242245x_1^{20}+139312x_1^{21}+73839x_1^{22}+35663x_1^{23}\nonumber \\
&\quad\quad +15519x_1^{24}+5938x_1^{25} +1933x_1^{26}+524x_1^{27}+116x_1^{28}+17x_1^{29}+x_1^{30})(x_1x_2)^6
+{\cal O}(x_2^7),\\
&F^{(l=3)}_8(y_\alpha=1;x_1;x_2)=
\cdots
+(726+1960 x_1+4156 x_1^{2}+7240 x_1^{3}+11622 x_1^{4}+17067 x_1^{5}+23759 x_1^{6}+30580 x_1^{7}\nonumber \\
&\quad\quad +36746 x_1^{8}
+41728 x_1^{9} +45220 x_1^{10} +46545 x_1^{11}+45662 x_1^{12}+42361 x_1^{13}+37173 x_1^{14}+31070 x_1^{15}+24719 x_1^{16}\nonumber \\
&\quad\quad +18567 x_1^{17}+13041 x_1^{18}+8532 x_1^{19} +5150 x_1^{20}+2889 x_1^{21}+1494 x_1^{22}+667 x_1^{23}+243 x_1^{24}+62 x_1^{25}+9 x_1^{26})    (x_1x_2)^{4}\nonumber \\
&\quad +(2464+7768 x_1+18424 x_1^{2}+35640 x_1^{3}+62192 x_1^{4}+98931 x_1^{5}+147395 x_1^{6}+204324 x_1^{7}+265516 x_1^{8}\nonumber \\
&\quad\quad +326472 x_1^{9}+381889 x_1^{10}+425794 x_1^{11}+453082 x_1^{12}+460026 x_1^{13}+445376 x_1^{14}+412884 x_1^{15}+365880 x_1^{16}\nonumber \\
&\quad\quad +309944 x_1^{17}+249820 x_1^{18}+191742 x_1^{19}+139377 x_1^{20}+96307 x_1^{21}+62718 x_1^{22}+38335 x_1^{23}+21692 x_1^{24}\nonumber \\
&\quad\quad +11368 x_1^{25}+5419 x_1^{26}+2376 x_1^{27}+901 x_1^{28}+288 x_1^{29}+63 x_1^{30}+9 x_1^{31})(x_1x_2)^5\nonumber \\
&\quad+(7704+27480x_1+71492x_1^{2}+150233x_1^{3}+281437x_1^{4}+477592x_1^{5}+754464x_1^{6}+1109330x_1^{7}+1532221x_1^{8}\nonumber \\
&\quad\quad  +2000720x_1^{9}+2485249x_1^{10}+2943015x_1^{11}+3331771x_1^{12}+3607864x_1^{13}+3741430x_1^{14}+3721944x_1^{15}\nonumber \\
&\quad\quad +3553779x_1^{16} +3256071x_1^{17}+2860475x_1^{18}+2408301x_1^{19}+1940890x_1^{20}+1496791x_1^{21}+1102369x_1^{22}\nonumber \\
&\quad\quad +773074x_1^{23}+514563x_1^{24} +324142x_1^{25}+192517x_1^{26}+107317x_1^{27}+55750x_1^{28}+26636x_1^{29}+11586x_1^{30}\nonumber \\
&\quad\quad +4535x_1^{31}+1582x_1^{32}+470x_1^{33}+116x_1^{34} +17x_1^{35}+x_1^{36})(x_1x_2)^6
+{\cal O}(x_2^7),
\end{align}
\end{subequations}
}
where the lower order terms $\cdots$ are obtained from \eqref{listofFlm_l3withy} by setting $y=1$.

\subsection{$l=4$}
\label{app:listofFlm_l4withy}

We find that $F^{(l=4)}_m(y_\alpha;x_1;x_2)$ are given as
{\fontsize{9pt}{1pt}\selectfont
\begin{subequations}
\label{listofFlm_l4withy}
\begin{align}
&F_1^{(l=4)}(y_\alpha;x_1;x_2)=
1+
(3 + A^{(4)}_1
)x_1x_2
+(9 + 3A^{(4)}_1 + B^{(4)}_2)(x_1x_2)^2
+(22+9A^{(4)}_1+3B^{(4)}_2+C^{(4)}_2)(x_1x_2)^3\nonumber \\
&\quad
+{\cal O}(x_2^4),\\
&F_2^{(l=4)}(y_\alpha;x_1;x_2)=
1+
\Bigl[3 + A^{(4)}_1
 + (3 + A^{(4)}_1)x_1  
\Bigr]x_1x_2\nonumber \\
&\quad +\Bigl[
15 + 6A^{(4)}_1 + A^{(4)}_2 + 2B^{(4)}_2 + C^{(4)}_2
 + (18 + 8A^{(4)}_1 + 3B^{(4)}_2 + C^{(4)}_2)x_1  
 + (6 + 2A^{(4)}_1 + B^{(4)}_2)        x_1^2
\Bigr](x_1x_2)^2\nonumber \\
&\quad +\Bigl[
50+25A^{(4)}_1+3A^{(4)}_2+11B^{(4)}_2+6C^{(4)}_2+B^{(4)}_3
+(81+40A^{(4)}_1+3A^{(4)}_2+18B^{(4)}_2+8C^{(4)}_2+B^{(4)}_3)x_1\nonumber \\
&\quad\quad
+(35+16A^{(4)}_1+7B^{(4)}_2+2C^{(4)}_2)x_1^2
+(4+A^{(4)}_1)x_1^3
\Bigr](x_1x_2)^3
+{\cal O}(x_2^4),\\
&F_3^{(l=4)}(y_\alpha;x_1;x_2)=
1+
\Bigl[3 + A^{(4)}_1
 + (3 + A^{(4)}_1)x_1  
 + (3 + A^{(4)}_1)x_1^2
\Bigr]x_1x_2\nonumber \\
&\quad +\Bigl[
15 + 6A^{(4)}_1 + A^{(4)}_2 + 2B^{(4)}_2 + C^{(4)}_2
 + (24 + 11A^{(4)}_1 + A^{(4)}_2 + 4B^{(4)}_2 + 2C^{(4)}_2) x_1  \nonumber \\
&\quad\quad
 + (30 + 13A^{(4)}_1 + A^{(4)}_2 + 5B^{(4)}_2 + 2C^{(4)}_2) x_1^2
 + (15 + 7A^{(4)}_1 + 3B^{(4)}_2 + C^{(4)}_2)         x_1^3
 + (6 + 2A^{(4)}_1 + B^{(4)}_2)                 x_1^4
\Bigr](x_1x_2)^2\nonumber \\
&\quad
+\Bigl[
60+31A^{(4)}_1+6A^{(4)}_2+A^{(4)}_3+14B^{(4)}_2+9C^{(4)}_2+2B^{(4)}_3+C^{(4)}_3+D^{(4)}_3\nonumber \\
&\quad\quad
+(131+71A^{(4)}_1+11A^{(4)}_2+35B^{(4)}_2+20C^{(4)}_2+4B^{(4)}_3+C^{(4)}_3+2D^{(4)}_3)x_1\nonumber \\
&\quad\quad
+(188+100A^{(4)}_1+13A^{(4)}_2+49B^{(4)}_2+26C^{(4)}_2+5B^{(4)}_3+C^{(4)}_3+2D^{(4)}_3)x_1^2\nonumber \\
&\quad\quad
+(147+77A^{(4)}_1+7A^{(4)}_2+39B^{(4)}_2+18C^{(4)}_2+3B^{(4)}_3+D^{(4)}_3)x_1^3\nonumber \\
&\quad\quad
+(77+38A^{(4)}_1+2A^{(4)}_2+18B^{(4)}_2+7C^{(4)}_2+B^{(4)}_3)x_1^4
+(20+9A^{(4)}_1+4B^{(4)}_2+C^{(4)}_2)x_1^5\nonumber \\
&\quad\quad
+(4+A^{(4)}_1)x_1^6
\Bigr](x_1x_2)^3
+{\cal O}(x_2^4),\\
&F_4^{(l=4)}(y_\alpha;x_1;x_2)=
1+
\Bigl[3 + A^{(4)}_1
 + (3 + A^{(4)}_1)x_1  
 + (3 + A^{(4)}_1)x_1^2
 + (3 + A^{(4)}_1)x_1^3
\Bigr]x_1x_2\nonumber \\
&\quad +\Bigl[
15 + 6A^{(4)}_1 + A^{(4)}_2 + 2B^{(4)}_2 + C^{(4)}_2
 + (24 + 11A^{(4)}_1 + A^{(4)}_2 + 4B^{(4)}_2 + 2C^{(4)}_2)  x_1  \nonumber \\
&\quad\quad
 + (36 + 16A^{(4)}_1 + 2A^{(4)}_2 + 6B^{(4)}_2 + 3C^{(4)}_2) x_1^2
 + (39 + 18A^{(4)}_1 + A^{(4)}_2 + 7B^{(4)}_2 + 3C^{(4)}_2)  x_1^3\nonumber \\
&\quad\quad
 + (27 + 12A^{(4)}_1 + A^{(4)}_2 + 5B^{(4)}_2 + 2C^{(4)}_2)  x_1^4
 + (15 + 7A^{(4)}_1 + 3B^{(4)}_2 + C^{(4)}_2)          x_1^5
 + (6 + 2A^{(4)}_1 + B^{(4)}_2)                  x_1^6
\Bigr](x_1x_2)^2\nonumber \\
&\quad
+\Bigl[
60+31A^{(4)}_1+6A^{(4)}_2+A^{(4)}_3+14B^{(4)}_2+9C^{(4)}_2+2B^{(4)}_3+C^{(4)}_3+D^{(4)}_3\nonumber \\
&\quad\quad
+(141+77A^{(4)}_1+14A^{(4)}_2+A^{(4)}_3+38B^{(4)}_2+23C^{(4)}_2+5B^{(4)}_3+2C^{(4)}_3+3D^{(4)}_3)x_1\nonumber \\
&\quad\quad
+(248+137A^{(4)}_1+24A^{(4)}_2+A^{(4)}_3+69B^{(4)}_2+41C^{(4)}_2+9B^{(4)}_3+3C^{(4)}_3+5D^{(4)}_3)x_1^2\nonumber \\
&\quad\quad
+(332+182A^{(4)}_1+28A^{(4)}_2+A^{(4)}_3+93B^{(4)}_2+52C^{(4)}_2+11B^{(4)}_3+3C^{(4)}_3+6D^{(4)}_3)x_1^3\nonumber \\
&\quad\quad
+(314+173A^{(4)}_1+24A^{(4)}_2+90B^{(4)}_2+48C^{(4)}_2+10B^{(4)}_3+2C^{(4)}_3+5D^{(4)}_3)x_1^4\nonumber \\
&\quad\quad
+(249+135A^{(4)}_1+16A^{(4)}_2+70B^{(4)}_2+35C^{(4)}_2+7B^{(4)}_3+C^{(4)}_3+3D^{(4)}_3)x_1^5\nonumber \\
&\quad\quad
+(146+76A^{(4)}_1+6A^{(4)}_2+39B^{(4)}_2+17C^{(4)}_2+3B^{(4)}_3+D^{(4)}_3)x_1^6\nonumber \\
&\quad\quad
+(62+31A^{(4)}_1+2A^{(4)}_2+15B^{(4)}_2+6C^{(4)}_2+B^{(4)}_3)x_1^7
+(20+9A^{(4)}_1+4B^{(4)}_2+C^{(4)}_2)x_1^8\nonumber \\
&\quad\quad
+(4+A^{(4)}_1)x_1^9
\Bigr](x_1x_2)^3
+{\cal O}(x_2^4),\\
&F_5^{(l=4)}(y_\alpha;x_1;x_2)=
1+
\Bigl[3 + A^{(4)}_1
 + (3 + A^{(4)}_1)x_1  
 + (3 + A^{(4)}_1)x_1^2
 + (3 + A^{(4)}_1)x_1^3
 + (3 + A^{(4)}_1)x_1^4
\Bigr]x_1x_2\nonumber \\
&\quad +
\Bigl[
15 + 6A^{(4)}_1 + A^{(4)}_2 + 2B^{(4)}_2 + C^{(4)}_2
 + (24 + 11A^{(4)}_1 + A^{(4)}_2 + 4B^{(4)}_2 + 2C^{(4)}_2)  x_1  \nonumber \\
&\quad\quad
 + (36 + 16A^{(4)}_1 + 2A^{(4)}_2 + 6B^{(4)}_2 + 3C^{(4)}_2) x_1^2
 + (45 + 21A^{(4)}_1 + 2A^{(4)}_2 + 8B^{(4)}_2 + 4C^{(4)}_2) x_1^3\nonumber \\
&\quad\quad
 + (51 + 23A^{(4)}_1 + 2A^{(4)}_2 + 9B^{(4)}_2 + 4C^{(4)}_2) x_1^4
 + (36 + 17A^{(4)}_1 + A^{(4)}_2 + 7B^{(4)}_2 + 3C^{(4)}_2)  x_1^5\nonumber \\
&\quad\quad
 + (27 + 12A^{(4)}_1 + A^{(4)}_2 + 5B^{(4)}_2 + 2C^{(4)}_2)  x_1^6
 + (15 + 7A^{(4)}_1 + 3B^{(4)}_2 + C^{(4)}_2)          x_1^7
 + (6 + 2A^{(4)}_1 + B^{(4)}_2)                  x_1^8
\Bigr](x_1x_2)^2\nonumber \\
&\quad
+\Bigl[
60+31A^{(4)}_1+6A^{(4)}_2+A^{(4)}_3+14B^{(4)}_2+9C^{(4)}_2+2B^{(4)}_3+C^{(4)}_3+D^{(4)}_3\nonumber \\
&\quad\quad
+(141+77A^{(4)}_1+14A^{(4)}_2+A^{(4)}_3+38B^{(4)}_2+23C^{(4)}_2+5B^{(4)}_3+2C^{(4)}_3+3D^{(4)}_3)x_1\nonumber \\
&\quad\quad
+(258+143A^{(4)}_1+27A^{(4)}_2+2A^{(4)}_3+72B^{(4)}_2+44C^{(4)}_2+10B^{(4)}_3+4C^{(4)}_3+6D^{(4)}_3)x_1^2\nonumber \\
&\quad\quad
+(392+219A^{(4)}_1+39A^{(4)}_2+2A^{(4)}_3+113B^{(4)}_2+67C^{(4)}_2+15B^{(4)}_3+5C^{(4)}_3+9D^{(4)}_3)x_1^3\nonumber \\
&\quad\quad
+(509+284A^{(4)}_1+48A^{(4)}_2+2A^{(4)}_3+147B^{(4)}_2+85C^{(4)}_2+19B^{(4)}_3+6C^{(4)}_3+11D^{(4)}_3)x_1^4\nonumber \\
&\quad\quad
+(518+291A^{(4)}_1+46A^{(4)}_2+A^{(4)}_3+154B^{(4)}_2+86C^{(4)}_2+19B^{(4)}_3+5C^{(4)}_3+11D^{(4)}_3)x_1^5\nonumber \\
&\quad\quad
+(479+267A^{(4)}_1+40A^{(4)}_2+A^{(4)}_3+141B^{(4)}_2+77C^{(4)}_2+17B^{(4)}_3+4C^{(4)}_3+9D^{(4)}_3)x_1^6\nonumber \\
&\quad\quad
+(378+209A^{(4)}_1+27A^{(4)}_2+111B^{(4)}_2+57C^{(4)}_2+12B^{(4)}_3+2C^{(4)}_3+6D^{(4)}_3)x_1^7\nonumber \\
&\quad\quad
+(248+134A^{(4)}_1+15A^{(4)}_2+70B^{(4)}_2+34C^{(4)}_2+7B^{(4)}_3+C^{(4)}_3+3D^{(4)}_3)x_1^8\nonumber \\
&\quad\quad
+(131+69A^{(4)}_1+6A^{(4)}_2+36B^{(4)}_2+16C^{(4)}_2+3B^{(4)}_3+D^{(4)}_3)x_1^9\nonumber \\
&\quad\quad
+(62+31A^{(4)}_1+2A^{(4)}_2+15B^{(4)}_2+6C^{(4)}_2+B^{(4)}_3)x_1^{10}
+(20+9A^{(4)}_1+4B^{(4)}_2+C^{(4)}_2)x_1^{11}\nonumber \\
&\quad\quad
+(4+A^{(4)}_1)x_1^{12}
\Bigr](x_1x_2)^3
+{\cal O}(x_2^4),\\
&F_6^{(l=4)}(y_\alpha;x_1;x_2)=
1+
\Bigl[3 + A^{(4)}_1
 + (3 + A^{(4)}_1)x_1  
 + (3 + A^{(4)}_1)x_1^2
 + (3 + A^{(4)}_1)x_1^3
 + (3 + A^{(4)}_1)x_1^4\nonumber \\
&\quad\quad
 + (3 + A^{(4)}_1)x_1^5
\Bigr]x_1x_2\nonumber \\
&\quad +
\Bigl[15 + 6A^{(4)}_1 + A^{(4)}_2 + 2B^{(4)}_2 + C^{(4)}_2
 + (24 + 11A^{(4)}_1 + A^{(4)}_2 + 4B^{(4)}_2 + 2C^{(4)}_2)   x_1   \nonumber \\
&\quad\quad
 + (36 + 16A^{(4)}_1 + 2A^{(4)}_2 + 6B^{(4)}_2 + 3C^{(4)}_2)  x_1^2
+ (45 + 21A^{(4)}_1 + 2A^{(4)}_2 + 8B^{(4)}_2 + 4C^{(4)}_2)  x_1^3 \nonumber \\
&\quad\quad
 + (57 + 26A^{(4)}_1 + 3A^{(4)}_2 + 10B^{(4)}_2 + 5C^{(4)}_2) x_1^4
 + (60 + 28A^{(4)}_1 + 2A^{(4)}_2 + 11B^{(4)}_2 + 5C^{(4)}_2) x_1^5\nonumber \\
&\quad\quad
 + (48 + 22A^{(4)}_1 + 2A^{(4)}_2 + 9B^{(4)}_2 + 4C^{(4)}_2)  x_1^6
 + (36 + 17A^{(4)}_1 + A^{(4)}_2 + 7B^{(4)}_2 + 3C^{(4)}_2)   x_1^7\nonumber \\
&\quad\quad
 + (27 + 12A^{(4)}_1 + A^{(4)}_2 + 5B^{(4)}_2 + 2C^{(4)}_2)   x_1^8
 + (15 + 7A^{(4)}_1 + 3B^{(4)}_2 + C^{(4)}_2)           x_1^9
 + (6 + 2A^{(4)}_1 + B^{(4)}_2)                   x_1^{10}
\Bigr](x_1x_2)^2\nonumber \\
&\quad
+\Bigl[
60+31A^{(4)}_1+6A^{(4)}_2+A^{(4)}_3+14B^{(4)}_2+9C^{(4)}_2+2B^{(4)}_3+C^{(4)}_3+D^{(4)}_3\nonumber \\
&\quad\quad
+(141+77A^{(4)}_1+14A^{(4)}_2+A^{(4)}_3+38B^{(4)}_2+23C^{(4)}_2+5B^{(4)}_3+2C^{(4)}_3+3D^{(4)}_3)x_1\nonumber \\
&\quad\quad
+(258+143A^{(4)}_1+27A^{(4)}_2+2A^{(4)}_3+72B^{(4)}_2+44C^{(4)}_2+10B^{(4)}_3+4C^{(4)}_3+6D^{(4)}_3)x_1^2\nonumber \\
&\quad\quad
+(402+225A^{(4)}_1+42A^{(4)}_2+3A^{(4)}_3+116B^{(4)}_2+70C^{(4)}_2+16B^{(4)}_3+6C^{(4)}_3+10D^{(4)}_3)x_1^3\nonumber \\
&\quad\quad
+(569+321A^{(4)}_1+59A^{(4)}_2+3A^{(4)}_3+167B^{(4)}_2+100C^{(4)}_2+23B^{(4)}_3+8C^{(4)}_3+14D^{(4)}_3)x_1^4\nonumber \\
&\quad\quad
+(713+402A^{(4)}_1+70A^{(4)}_2+3A^{(4)}_3+211B^{(4)}_2+123C^{(4)}_2+28B^{(4)}_3+9C^{(4)}_3+17D^{(4)}_3)x_1^5\nonumber \\
&\quad\quad
+(758+429A^{(4)}_1+73A^{(4)}_2+3A^{(4)}_3+228B^{(4)}_2+131C^{(4)}_2+30B^{(4)}_3+9C^{(4)}_3+18D^{(4)}_3)x_1^6\nonumber \\
&\quad\quad
+(743+421A^{(4)}_1+69A^{(4)}_2+2A^{(4)}_3+225B^{(4)}_2+127C^{(4)}_2+29B^{(4)}_3+8C^{(4)}_3+17D^{(4)}_3)x_1^7\nonumber \\
&\quad\quad
+(668+377A^{(4)}_1+58A^{(4)}_2+A^{(4)}_3+202B^{(4)}_2+111C^{(4)}_2+25B^{(4)}_3+6C^{(4)}_3+14D^{(4)}_3)x_1^8\nonumber \\
&\quad\quad
+(543+303A^{(4)}_1+43A^{(4)}_2+A^{(4)}_3+162B^{(4)}_2+86C^{(4)}_2+19B^{(4)}_3+4C^{(4)}_3+10D^{(4)}_3)x_1^9\nonumber \\
&\quad\quad
+(377+208A^{(4)}_1+26A^{(4)}_2+111B^{(4)}_2+56C^{(4)}_2+12B^{(4)}_3+2C^{(4)}_3+6D^{(4)}_3)x_1^{10}\nonumber \\
&\quad\quad
+(233+127A^{(4)}_1+15A^{(4)}_2+67B^{(4)}_2+33C^{(4)}_2+7B^{(4)}_3+C^{(4)}_3+3D^{(4)}_3)x_1^{11}\nonumber \\
&\quad\quad
+(131+69A^{(4)}_1+6A^{(4)}_2+36B^{(4)}_2+16C^{(4)}_2+3B^{(4)}_3+D^{(4)}_3)x_1^{12}\nonumber \\
&\quad\quad
+(62+31A^{(4)}_1+2A^{(4)}_2+15B^{(4)}_2+6C^{(4)}_2+B^{(4)}_3)x_1^{13}
+(20+9A^{(4)}_1+4B^{(4)}_2+C^{(4)}_2)x_1^{14}\nonumber \\
&\quad\quad
+(4+A^{(4)}_1)x_1^{15}
\Bigr](x_1x_2)^3
+{\cal O}(x_2^4),\\
&F_7^{(l=4)}(y_\alpha;x_1;x_2)=
1+
\Bigl[3 + A^{(4)}_1
 + (3 + A^{(4)}_1)x_1  
 + (3 + A^{(4)}_1)x_1^2
 + (3 + A^{(4)}_1)x_1^3
 + (3 + A^{(4)}_1)x_1^4
 + (3 + A^{(4)}_1)x_1^5\nonumber \\
&\quad\quad
 + (3 + A^{(4)}_1)x_1^6
\Bigr]x_1x_2\nonumber \\
&\quad +
\Bigl[15 + 6A^{(4)}_1 + A^{(4)}_2 + 2B^{(4)}_2 + C^{(4)}_2
 + (24 + 11A^{(4)}_1 + A^{(4)}_2 + 4B^{(4)}_2 + 2C^{(4)}_2)   x_1   \nonumber \\
&\quad\quad
 + (36 + 16A^{(4)}_1 + 2A^{(4)}_2 + 6B^{(4)}_2 + 3C^{(4)}_2)  x_1^2
+ (45 + 21A^{(4)}_1 + 2A^{(4)}_2 + 8B^{(4)}_2 + 4C^{(4)}_2)  x_1^3 \nonumber \\
&\quad\quad
 + (57 + 26A^{(4)}_1 + 3A^{(4)}_2 + 10B^{(4)}_2 + 5C^{(4)}_2) x_1^4
 + (66 + 31A^{(4)}_1 + 3A^{(4)}_2 + 12B^{(4)}_2 + 6C^{(4)}_2) x_1^5\nonumber \\
&\quad\quad
 + (72 + 33A^{(4)}_1 + 3A^{(4)}_2 + 13B^{(4)}_2 + 6C^{(4)}_2) x_1^6
 + (57 + 27A^{(4)}_1 + 2A^{(4)}_2 + 11B^{(4)}_2 + 5C^{(4)}_2) x_1^7 \nonumber \\
&\quad\quad
 + (48 + 22A^{(4)}_1 + 2A^{(4)}_2 + 9B^{(4)}_2 + 4C^{(4)}_2)  x_1^8
 + (36 + 17A^{(4)}_1 + A^{(4)}_2 + 7B^{(4)}_2 + 3C^{(4)}_2)   x_1^9 \nonumber \\
&\quad\quad
 + (27 + 12A^{(4)}_1 + A^{(4)}_2 + 5B^{(4)}_2 + 2C^{(4)}_2)   x_1^{10}
 + (15 + 7A^{(4)}_1 + 3B^{(4)}_2 + C^{(4)}_2)           x_1^{11}
 + (6 + 2A^{(4)}_1 + B^{(4)}_2)                   x_1^{12}
\Bigr](x_1x_2)^2\nonumber \\
&\quad
+\Bigl[
60+31A^{(4)}_1+6A^{(4)}_2+A^{(4)}_3+14B^{(4)}_2+9C^{(4)}_2+2B^{(4)}_3+C^{(4)}_3+D^{(4)}_3\nonumber \\
&\quad\quad +(141+77A^{(4)}_1+14A^{(4)}_2+A^{(4)}_3+38B^{(4)}_2+23C^{(4)}_2+5B^{(4)}_3+2C^{(4)}_3+3D^{(4)}_3)x_1\nonumber \\
&\quad\quad +(258+143A^{(4)}_1+27A^{(4)}_2+2A^{(4)}_3+72B^{(4)}_2+44C^{(4)}_2+10B^{(4)}_3+4C^{(4)}_3+6D^{(4)}_3)x_1^2\nonumber \\
&\quad\quad +(402+225A^{(4)}_1+42A^{(4)}_2+3A^{(4)}_3+116B^{(4)}_2+70C^{(4)}_2+16B^{(4)}_3+6C^{(4)}_3+10D^{(4)}_3)x_1^3\nonumber \\
&\quad\quad +(579+327A^{(4)}_1+62A^{(4)}_2+4A^{(4)}_3+170B^{(4)}_2+103C^{(4)}_2+24B^{(4)}_3+9C^{(4)}_3+15D^{(4)}_3)x_1^4\nonumber \\
&\quad\quad +(773+439A^{(4)}_1+81A^{(4)}_2+4A^{(4)}_3+231B^{(4)}_2+138C^{(4)}_2+32B^{(4)}_3+11C^{(4)}_3+20D^{(4)}_3)x_1^5\nonumber \\
&\quad\quad +(953+540A^{(4)}_1+97A^{(4)}_2+5A^{(4)}_3+285B^{(4)}_2+168C^{(4)}_2+39B^{(4)}_3+13C^{(4)}_3+24D^{(4)}_3)x_1^6\nonumber \\
&\quad\quad +(1022+583A^{(4)}_1+102A^{(4)}_2+4A^{(4)}_3+312B^{(4)}_2+181C^{(4)}_2+42B^{(4)}_3+13C^{(4)}_3+26D^{(4)}_3)x_1^7\nonumber \\
&\quad\quad +(1043+595A^{(4)}_1+103A^{(4)}_2+4A^{(4)}_3+319B^{(4)}_2+184C^{(4)}_2+43B^{(4)}_3+13C^{(4)}_3+26D^{(4)}_3)x_1^8\nonumber \\
&\quad\quad +(995+567A^{(4)}_1+94A^{(4)}_2+3A^{(4)}_3+306B^{(4)}_2+173C^{(4)}_2+40B^{(4)}_3+11C^{(4)}_3+24D^{(4)}_3)x_1^9\nonumber \\
&\quad\quad +(893+507A^{(4)}_1+81A^{(4)}_2+2A^{(4)}_3+273B^{(4)}_2+152C^{(4)}_2+35B^{(4)}_3+9C^{(4)}_3+20D^{(4)}_3)x_1^{10}\nonumber \\
&\quad\quad +(732+413A^{(4)}_1+61A^{(4)}_2+A^{(4)}_3+223B^{(4)}_2+120C^{(4)}_2+27B^{(4)}_3+6C^{(4)}_3+15D^{(4)}_3)x_1^{11}\nonumber \\
&\quad\quad +(542+302A^{(4)}_1+42A^{(4)}_2+A^{(4)}_3+162B^{(4)}_2+85C^{(4)}_2+19B^{(4)}_3+4C^{(4)}_3+10D^{(4)}_3)x_1^{12}\nonumber \\
&\quad\quad +(362+201A^{(4)}_1+26A^{(4)}_2+108B^{(4)}_2+55C^{(4)}_2+12B^{(4)}_3+2C^{(4)}_3+6D^{(4)}_3)x_1^{13}\nonumber \\
&\quad\quad +(233+127A^{(4)}_1+15A^{(4)}_2+67B^{(4)}_2+33C^{(4)}_2+7B^{(4)}_3+C^{(4)}_3+3D^{(4)}_3)x_1^{14}\nonumber \\
&\quad\quad +(131+69A^{(4)}_1+6A^{(4)}_2+36B^{(4)}_2+16C^{(4)}_2+3B^{(4)}_3+D^{(4)}_3)x_1^{15}\nonumber \\
&\quad\quad
+(62+31A^{(4)}_1+2A^{(4)}_2+15B^{(4)}_2+6C^{(4)}_2+B^{(4)}_3)x_1^{16}
+(20+9A^{(4)}_1+4B^{(4)}_2+C^{(4)}_2)x_1^{17}\nonumber \\
&\quad\quad
+(4+A^{(4)}_1)x_1^{18}
\Bigr](x_1x_2)^3
+{\cal O}(x_2^4),\\
&F_8^{(l=4)}(y_\alpha;x_1;x_2)=
1+
\Bigl[3 + A^{(4)}_1
 + (3 + A^{(4)}_1)x_1  
 + (3 + A^{(4)}_1)x_1^2
 + (3 + A^{(4)}_1)x_1^3
 + (3 + A^{(4)}_1)x_1^4
 + (3 + A^{(4)}_1)x_1^5\nonumber \\
&\quad\quad
 + (3 + A^{(4)}_1)x_1^6
+ (3 + A^{(4)}_1)x_1^7
\Bigr]x_1x_2\nonumber \\
&\quad +
\Bigl[15 + 6 A^{(4)}_1 + A^{(4)}_2 + 2 B^{(4)}_2 + C^{(4)}_2
 + (24 + 11 A^{(4)}_1 + A^{(4)}_2 + 4 B^{(4)}_2 + 2 C^{(4)}_2)    x_1    \nonumber \\
&\quad\quad
 + (36 + 16 A^{(4)}_1 + 2 A^{(4)}_2 + 6 B^{(4)}_2 + 3 C^{(4)}_2)  x_1^2
 + (45 + 21 A^{(4)}_1 + 2 A^{(4)}_2 + 8 B^{(4)}_2 + 4 C^{(4)}_2)  x_1^3  \nonumber \\
&\quad\quad
 + (57 + 26 A^{(4)}_1 + 3 A^{(4)}_2 + 10 B^{(4)}_2 + 5 C^{(4)}_2) x_1^4
 + (66 + 31 A^{(4)}_1 + 3 A^{(4)}_2 + 12 B^{(4)}_2 + 6 C^{(4)}_2) x_1^5\nonumber \\
&\quad\quad
 + (78 + 36 A^{(4)}_1 + 4 A^{(4)}_2 + 14 B^{(4)}_2 + 7 C^{(4)}_2) x_1^6
 + (81 + 38 A^{(4)}_1 + 3 A^{(4)}_2 + 15 B^{(4)}_2 + 7 C^{(4)}_2) x_1^7  \nonumber \\
&\quad\quad
 + (69 + 32 A^{(4)}_1 + 3 A^{(4)}_2 + 13 B^{(4)}_2 + 6 C^{(4)}_2) x_1^8
 + (57 + 27 A^{(4)}_1 + 2 A^{(4)}_2 + 11 B^{(4)}_2 + 5 C^{(4)}_2) x_1^9  \nonumber \\
&\quad\quad
 + (48 + 22 A^{(4)}_1 + 2 A^{(4)}_2 + 9 B^{(4)}_2 + 4 C^{(4)}_2)  x_1^{10}
 + (36 + 17 A^{(4)}_1 + A^{(4)}_2 + 7 B^{(4)}_2 + 3 C^{(4)}_2)    x_1^{11}\nonumber \\
&\quad\quad
 + (27 + 12 A^{(4)}_1 + A^{(4)}_2 + 5 B^{(4)}_2 + 2 C^{(4)}_2)    x_1^{12}
 + (15 + 7 A^{(4)}_1 + 3 B^{(4)}_2 + C^{(4)}_2)             x_1^{13} 
 + (6 + 2 A^{(4)}_1 + B^{(4)}_2)                      x_1^{14} 
\Bigr](x_1x_2)^2\nonumber \\
&\quad
+\Bigl[
60+31A^{(4)}_1+6A^{(4)}_2+A^{(4)}_3+14B^{(4)}_2+9C^{(4)}_2+2B^{(4)}_3+C^{(4)}_3+D^{(4)}_3\nonumber \\
&\quad\quad +(141+77A^{(4)}_1+14A^{(4)}_2+A^{(4)}_3+38B^{(4)}_2+23C^{(4)}_2+5B^{(4)}_3+2C^{(4)}_3+3D^{(4)}_3)x_1\nonumber \\
&\quad\quad +(258+143A^{(4)}_1+27A^{(4)}_2+2A^{(4)}_3+72B^{(4)}_2+44C^{(4)}_2+10B^{(4)}_3+4C^{(4)}_3+6D^{(4)}_3)x_1^2\nonumber \\
&\quad\quad +(402+225A^{(4)}_1+42A^{(4)}_2+3A^{(4)}_3+116B^{(4)}_2+70C^{(4)}_2+16B^{(4)}_3+6C^{(4)}_3+10D^{(4)}_3)x_1^3\nonumber \\
&\quad\quad +(579+327A^{(4)}_1+62A^{(4)}_2+4A^{(4)}_3+170B^{(4)}_2+103C^{(4)}_2+24B^{(4)}_3+9C^{(4)}_3+15D^{(4)}_3)x_1^4\nonumber \\
&\quad\quad +(783+445A^{(4)}_1+84A^{(4)}_2+5A^{(4)}_3+234B^{(4)}_2+141C^{(4)}_2+33B^{(4)}_3+12C^{(4)}_3+21D^{(4)}_3)x_1^5\nonumber \\
&\quad\quad +(1013+577A^{(4)}_1+108A^{(4)}_2+6A^{(4)}_3+305B^{(4)}_2+183C^{(4)}_2+43B^{(4)}_3+15C^{(4)}_3+27D^{(4)}_3)x_1^6\nonumber \\
&\quad\quad +(1217+694A^{(4)}_1+126A^{(4)}_2+6A^{(4)}_3+369B^{(4)}_2+218C^{(4)}_2+51B^{(4)}_3+17C^{(4)}_3+32D^{(4)}_3)x_1^7\nonumber \\
&\quad\quad +(1322+757A^{(4)}_1+136A^{(4)}_2+6A^{(4)}_3+406B^{(4)}_2+238C^{(4)}_2+56B^{(4)}_3+18C^{(4)}_3+35D^{(4)}_3)x_1^8\nonumber \\
&\quad\quad +(1370+785A^{(4)}_1+139A^{(4)}_2+6A^{(4)}_3+423B^{(4)}_2+246C^{(4)}_2+58B^{(4)}_3+18C^{(4)}_3+36D^{(4)}_3)x_1^9\nonumber \\
&\quad\quad +(1355+777A^{(4)}_1+135A^{(4)}_2+5A^{(4)}_3+420B^{(4)}_2+242C^{(4)}_2+57B^{(4)}_3+17C^{(4)}_3+35D^{(4)}_3)x_1^{10}\nonumber \\
&\quad\quad +(1280+733A^{(4)}_1+124A^{(4)}_2+4A^{(4)}_3+397B^{(4)}_2+226C^{(4)}_2+53B^{(4)}_3+15C^{(4)}_3+32D^{(4)}_3)x_1^{11}\nonumber \\
&\quad\quad +(1145+653A^{(4)}_1+106A^{(4)}_2+3A^{(4)}_3+354B^{(4)}_2+198C^{(4)}_2+46B^{(4)}_3+12C^{(4)}_3+27D^{(4)}_3)x_1^{12}\nonumber \\
&\quad\quad +(957+543A^{(4)}_1+84A^{(4)}_2+2A^{(4)}_3+294B^{(4)}_2+161C^{(4)}_2+37B^{(4)}_3+9C^{(4)}_3+21D^{(4)}_3)x_1^{13}\nonumber \\
&\quad\quad +(731+412A^{(4)}_1+60A^{(4)}_2+A^{(4)}_3+223B^{(4)}_2+119C^{(4)}_2+27B^{(4)}_3+6C^{(4)}_3+15D^{(4)}_3)x_1^{14}\nonumber \\
&\quad\quad +(527+295A^{(4)}_1+42A^{(4)}_2+A^{(4)}_3+159B^{(4)}_2+84C^{(4)}_2+19B^{(4)}_3+4C^{(4)}_3+10D^{(4)}_3)x_1^{15}\nonumber \\
&\quad\quad +(362+201A^{(4)}_1+26A^{(4)}_2+108B^{(4)}_2+55C^{(4)}_2+12B^{(4)}_3+2C^{(4)}_3+6D^{(4)}_3)x_1^{16}\nonumber \\
&\quad\quad +(233+127A^{(4)}_1+15A^{(4)}_2+67B^{(4)}_2+33C^{(4)}_2+7B^{(4)}_3+C^{(4)}_3+3D^{(4)}_3)x_1^{17}\nonumber \\
&\quad\quad +(131+69A^{(4)}_1+6A^{(4)}_2+36B^{(4)}_2+16C^{(4)}_2+3B^{(4)}_3+D^{(4)}_3)x_1^{18}\nonumber \\
&\quad\quad
+(62+31A^{(4)}_1+2A^{(4)}_2+15B^{(4)}_2+6C^{(4)}_2+B^{(4)}_3)x_1^{19}
+(20+9A^{(4)}_1+4B^{(4)}_2+C^{(4)}_2)x_1^{20}\nonumber \\
&\quad\quad
+(4+A^{(4)}_1)x_1^{21}
\Bigr](x_1x_2)^3
+{\cal O}(x_2^4),
\end{align}
\end{subequations}
}
where
\begin{subequations}
\label{ABCl4}
\begin{align}
&A^{(4)}_1=\sum_{\alpha\neq\beta}^4\frac{y_\alpha}{y_\beta},\quad
A^{(4)}_2=\sum_{\alpha\neq\beta}^4\frac{y_\alpha^2}{y_\beta^2},\quad
B^{(4)}_2=
\sum_{\alpha<\beta}
\sum_{\substack{\gamma<\delta\\ (\gamma,\delta\neq \alpha,\beta)}}
\frac{y_\alpha y_\beta}{y_\gamma y_\delta},\\
&C^{(4)}_2=\sum_{\alpha<\beta}^4\sum_{\gamma(\neq\alpha,\beta)}^4
\Bigl(
\frac{y_\alpha y_\beta}{y_\gamma^2}
+\frac{y_\gamma^2}{y_\alpha y_\beta}
\Bigr),\quad
A^{(4)}_3=\sum_{\alpha\neq\beta}\frac{y_\alpha^3}{y_\beta^3},\\
&B^{(4)}_3=\sum_\alpha \sum_{\beta(\neq\alpha)}\sum_{\gamma(\neq\alpha,\beta)}\sum_{\delta(\neq\alpha,\beta,\gamma)}\frac{y_\alpha^2 y_\beta}{y_\gamma^2 y_\delta},\\
&C^{(4)}_3=\sum_\alpha \sum_{\beta(\neq\alpha)}\sum_{\gamma(\neq\alpha,\beta)}\Bigl(\frac{y_\alpha^2 y_\beta}{y_\gamma^3}
+\frac{y_\gamma^3}{y_\alpha^2y_\beta}\Bigr),\quad
D^{(4)}_3=\sum_{\alpha} \Bigl(
\frac{y_\alpha^4}{y_1y_2y_3y_4}
+\frac{y_1y_2y_3y_4}{y_\alpha^4}
\Bigr).
\end{align}
\end{subequations}

\bibliographystyle{utphys}
\bibliography{ref}

\end{document}